\documentclass[10pt,journal,compsoc]{IEEEtran}

\ifCLASSOPTIONcompsoc
\usepackage[nocompress]{cite}
\else
\usepackage{cite}
\fi

\newcommand{\Model}{\texttt{DaL}} 

\usepackage[ruled, linesnumbered, vlined, commentsnumbered]{algorithm2e}
\usepackage{booktabs}
\usepackage{amsfonts}
\usepackage{amsmath}
\usepackage{eucal}
\usepackage{xcolor,colortbl}
\usepackage{multirow}
\usepackage{subcaption} 
\usepackage{csquotes}
\usepackage{adjustbox}
\usepackage{balance}
\usepackage{hyperref}
\usepackage{standalone}
\usepackage{tikz}
\usepackage{pgfplots}
\definecolor{beaublue}{rgb}{0.74, 0.83, 0.9}
\definecolor{steel}{rgb}{0, 0.2, 0.9}
\usepackage[most,listings,skins,breakable]{tcolorbox}

\newtcolorbox{quotebox}{colback=beaublue,boxrule=0.4pt,colframe=black,fonttitle=\bfseries,top=2pt,bottom=2pt}

\newcolumntype{P}[1]{>{\centering\arraybackslash}p{#1}}

\definecolor{one}{HTML}{2b7bba}
\definecolor{two}{HTML}{d52221}

 \newcommand{\markone}[4]{\begin{adjustbox}{max width=.1\textwidth}\begin{picture}(20,5)
    {\linethickness{0.2mm}\color{one}\put(#1,3){\line(1,0){#2}}\color{one}\put(#3,3){\circle*{4}}}\end{picture}\end{adjustbox}}

    \newcommand{\marktwo}[4]{\begin{adjustbox}{max width=.1\textwidth}\begin{picture}(20,5)
    {\linethickness{0.2mm}\color{two}\put(#1,3){\line(1,0){#2}}\color{two}\put(#3,0.5){\large$\star$}}\end{picture}\end{adjustbox}}

\DeclareMathAlphabet\mathbfcal{OMS}{cmsy}{b}{n}
\SetKwInput{kwDeclare}{Declare}

\makeatletter
\tcbset{
    myhbox/.style 2 args={%
        enhanced, 
        breakable,
        colback=white,
        colframe=steel!50!black,
        attach boxed title to top left={yshift*=-\tcboxedtitleheight}, 
        title={#2},
        boxed title size=title,
        boxed title style={%
            sharp corners, 
            rounded corners=northwest, 
            colback=tcbcolframe, 
            boxrule=0pt,
        },
        underlay boxed title={%
            \path[fill=tcbcolframe] (title.south west)--(title.south east) 
                to[out=0, in=180] ([xshift=5mm]title.east)--
                (title.center-|frame.east)
                [rounded corners=\kvtcb@arc] |- 
                (frame.north) -| cycle; 
        },
        #1
    }
}   
\makeatother

\ifCLASSINFOpdf
\else
\fi
\newcolumntype{P}[1]{>{\centering\arraybackslash}m{#1}}
\newcolumntype{Y}{>{\centering\arraybackslash}X}

\begin{document}
\bstctlcite{MyBSTcontrol}

	\title{Dividable Configuration Performance Learning}
	
	\author{
		Jingzhi Gong,~\IEEEmembership{}
		Tao Chen,~\IEEEmembership{}
		Rami Bahsoon%
		
		\IEEEcompsocitemizethanks{
		    \IEEEcompsocthanksitem Corresponding author: Tao Chen (email: t.chen@bham.ac.uk).
			\IEEEcompsocthanksitem Jingzhi Gong is with the Department of Computer Science and Engineering, University of Electronic Science and Technology of China, China, 610056, and the Department of Computer Science, Loughborough University, UK, LE11 3TU. This research was conducted when Jingzhi Going visited the University of Electronic Science and Technology of China.
			\IEEEcompsocthanksitem Tao Chen and Rami Bahsoon are with the School of Computer Science, University of Birmingham, UK, B15 2TT.
		}
		\thanks{}}

	
	\IEEEtitleabstractindextext{%
		\begin{abstract}

 Machine/deep learning models have been widely adopted to predict the configuration performance of software systems.
 However, a crucial yet unaddressed challenge is how to cater for the sparsity inherited from the configuration landscape: the influence of configuration options (features) and the distribution of data samples are highly sparse. In this paper, we propose a model-agnostic and sparsity-robust framework for predicting configuration performance, dubbed \Model, based on the new paradigm of dividable learning that builds a model via ``divide-and-learn''. To handle sample sparsity, the samples from the configuration landscape are divided into distant divisions, for each of which we build a sparse local model, e.g., regularized Hierarchical Interaction Neural Network, to deal with the feature sparsity. A newly given configuration would then be assigned to the right model of division for the final prediction. Further, \Model~adaptively determines the optimal number of divisions required for a system and sample size without any extra training or profiling. Experiment results from 12 real-world systems and five sets of training data reveal that, compared with the state-of-the-art approaches, \Model~performs no worse than the best counterpart on 44 out of 60 cases (within which 31 cases are significantly better) with up to $1.61\times$ improvement on accuracy; requires fewer samples to reach the same/better accuracy; and producing acceptable training overhead. In particular, the mechanism that adapted the parameter $d$ can reach the optimal value for 76.43\% of the individual runs. The result also confirms that the paradigm of dividable learning is more suitable than other similar paradigms such as ensemble learning for predicting configuration performance. Practically, \Model~considerably improves different global models when using them as the underlying local models, which further strengthens its flexibility. To promote open science, all the data, code, and supplementary materials of this work can be accessed at our repository: \texttt{\textcolor{blue}{\url{https://github.com/ideas-labo/DaL-ext}}}.
		\end{abstract}

		\begin{IEEEkeywords}
			performance engineering, configurable software systems, configuration learning, performance modeling, performance prediction, software configuration.
	\end{IEEEkeywords}}

	
	\maketitle
	
	\IEEEdisplaynontitleabstractindextext

	%
	\IEEEpeerreviewmaketitle




\IEEEraisesectionheading{\section{Introduction}
\label{sec:introduction}}


\IEEEPARstart{A}{lmost} every modern software system is configurable, hence configuration management has become one of the most important phases in software engineering, especially when considering its drastic impact on
software performance, such as latency, runtime, and energy consumption~\cite{DBLP:journals/tse/SayaghKAP20,DBLP:conf/wosp/0001BWY18,DBLP:journals/pacmse/0001L24,DBLP:journals/tosem/ChenL23a,DBLP:journals/corr/abs-2403-03322}. Indeed, while highly configurable software systems provide great flexibility, they also introduce potential risks that can hinder their performance due to the daunting number of configuration options. For example, \textsc{x264}, which is a video encoder, comes with 16 options to tune which can significantly influence its runtime. As such, understanding what performance can be obtained under a given configuration before the deployment is essential for satisfying performance requirements. Practically, the knowledge of configuration performance not only enables better decisions on configuration tuning~\cite{DBLP:conf/sigsoft/0001Chen21,DBLP:journals/tosem/ChenL23} but also reduces the efforts of configuration testing~\cite{DBLP:conf/kbse/ChenHL022} as well as renders runtime self-adaptation plausible~\cite{DBLP:journals/tosem/ChenLBY18,DBLP:conf/wcre/Chen22,DBLP:conf/seams/Chen22}.

To understand configuration performance, one naive solution is to directly profile the software system for all possible configurations when needed. This, however, is impractical, because (1) there might be simply too many configurations to measure~\cite{DBLP:conf/mascots/JamshidiC16,DBLP:conf/sigsoft/0001Chen21,DBLP:conf/sigsoft/SiegmundGAK15, DBLP:conf/icse/HaZ19}. For example, \textsc{HIPA$^{cc}$} (a compiler for image processing) has more than 10,000 possible configurations. (2) Even when such a number is small, measuring a single configuration can still be extremely expensive and time-consuming~\cite{DBLP:conf/mascots/JamshidiC16,DBLP:conf/sigsoft/0001Chen21,DBLP:conf/kbse/LiXCT20,DBLP:conf/icse/LiX0WT20}: Wang \textit{et al.}~\cite{DBLP:conf/sigsoft/WangHJK13} report that it could take weeks of running time to benchmark and profile even a simple system. Therefore, an accurate performance model that can cheaply predict the expected performance of a newly given configuration is in high demand. 

Modeling the correlation between configurations and performance is, nevertheless, not an easy task. This is because as the complexity of modern software systems increases, the number of configurable options continues to expand and the interactions between options become more complicated, leading to significant difficulty in building an accurate model~\cite{DBLP:conf/icse/SiegmundKKABRS12,DBLP:journals/tse/ChenB17}. The problem, namely configuration performance learning, fits well with the strength of machine learning in the Artificial Intelligence (AI) paradigm. Indeed, in the past decade, machine learning models have demonstrated more promising results for learning configuration performance compared with the analytical performance models, as they are capable of modeling the complex interplay between a large number of variables by observing hidden patterns from data~\cite{DBLP:conf/icse/WeberAS21,DBLP:conf/icse/VelezJSAK21, DBLP:conf/wosp/HanYP21,DBLP:conf/sigsoft/SiegmundGAK15,DBLP:conf/esem/ShuS0X20,DBLP:journals/sqj/SiegmundRKKAS12,DBLP:conf/icse/HaZ19}.

However, since machine learning modeling is data-driven, the characteristics and properties of the measured data for configurable software systems pose non-trivial challenges to the learning, primarily because it is known that the configuration landscapes of the systems do not follow a ``smooth'' shape~\cite{DBLP:conf/mascots/JamshidiC16}. For example, adjusting between different cache strategies can drastically influence the performance, but they are often represented as a single-digit change on the landscape~\cite{DBLP:conf/sigsoft/0001Chen21}. This leads to the notion of sparsity in two aspects:

\begin{itemize}
    \item Only a small number of configuration options can significantly influence the performance, hence there is a clear \textbf{\textit{feature sparsity}} involved~\cite{DBLP:conf/nips/HuangJYCMN10, DBLP:conf/sigsoft/SiegmundGAK15, DBLP:conf/icse/HaZ19, DBLP:conf/icse/VelezJSAK21}.
    \item The samples from the configuration landscape tend to form different divisions with diverse values of performance and configuration options, especially when the training data is limited due to expensive measurement---a typical case of \textbf{\textit{sample sparsity}}~\cite{DBLP:journals/corr/abs-2202-03354,DBLP:conf/icml/LiuCH20,DBLP:conf/icml/ShibagakiKHT16}. This is particularly true when not all configurations are valid~\cite{DBLP:conf/sigsoft/SiegmundGAK15}.
\end{itemize}

While prior work can handle feature sparsity through tree-liked model~\cite{DBLP:conf/kbse/GuoCASW13}, feature selection~\cite{DBLP:journals/fgcs/LiLTWHQD19, DBLP:conf/mascots/GrohmannEEKKM19,DBLP:journals/tse/ChenB17}, or regularizing deep learning~\cite{DBLP:conf/icse/HaZ19, DBLP:journals/jmlr/GlorotBB11,DBLP:conf/esem/ShuS0X20,DBLP:journals/tosem/ChengGZ23}, the sample sparsity has almost been ignored, which causes a major obstacle to the effectiveness of the machine learning-based performance model. For example, it is known that sparse data samples can easily force the model to focus and memorize too much on a particular region in the landscape of configuration data, leading to a serious issue of overfitting\footnote{Overfitting means a learned model fits well with the training data but works poorly on new data.}~\cite{DBLP:journals/corr/abs-2202-03354}.

To address the above gap, in this paper, we propose \Model, a framework for configuration performance learning via the concept of ``divide-and-learn''. \Model~comes under a newly proposed paradigm termed \textit{dividable learning}---the key that enables it to be naturally model-agnostic and can improve any existing models that learn configuration performance. The basic idea is that, to handle sample sparsity, we divide the samples (configurations and their performance) into different divisions, each of which is independently learned by a local model of any kind. In this way, the highly sparse configuration data can be separated into different regions of samples, which are locally smooth, and hence their patterns and feature sparsity can be more easily captured.

In a nutshell, our main contributions are:
\begin{enumerate}

    \item We formulate, on top of the regression problem of learning configuration performance, a new classification problem without explicit labels.
    \item We modify Classification and Regression Tree (\texttt{CART})~\cite{loh2011classification} as a clustering algorithm to ``divide'' the samples into different divisions with similar characteristics, for each of which we build a local model of any choice.
    
    \item Newly given configurations would be assigned into a division inferred by a Random Forest classifier~\cite{DBLP:conf/icdar/Ho95}, which is trained using the pseudo-labeled data from the \texttt{CART}. The local model of the assigned division would be used for the final prediction thereafter.
    \item Under 12 systems with diverse performance attributes, scale, and domains, as well as five different training sizes, we evaluate \Model~against the common state-of-the-art approaches, different underlying local models, and a number of its variants.

\end{enumerate}

\subsection{New Extensions}

It is worth noting that this work is a significant extension of our work published at FSE'23~\cite{DBLP:conf/sigsoft/JChen2023}, for which we make the following key additional contributions:

\begin{itemize}
    \item We provide a more systematic qualitative study to analyze the sparsity characteristics of configuration data, including both a literature review and an empirical study, hence better motivating our needs (Section~\ref{sec:background}).

    \item The contribution in our FSE'23 work requires manually tuning a crucial parameter $d$, which controls the number of divisions. This work proposes a novel adaptive mechanism that dynamically adapts the $d$ value to an appropriate level without additional training or profiling (Section~\ref{subsubsec:adapting_depth}). 

    \item To determine the optimal $d$ value in the adaptation, we proposed a new indicator $\mu$HV, extending from the standard HV that is widely used for multi-objective evaluation, which can better reflect the goodness and balance between the ability to handle sample sparsity and the amount of data for learning in the divided configuration data (Section~\ref{subsubsec:adapting_depth}). 

    \item We evaluate four additional systems that are of different characteristics and additionally compare \Model~with the most recent work from TOSEM'23~\cite{DBLP:journals/tosem/ChengGZ23}, which reports on a new state-of-the-art approach after our FSE'23 work and has shown compelling results, together with three extra research questions (\textbf{RQ3}, \textbf{RQ4}, and partial \textbf{RQ5}) that help to more thoroughly assess our contributions (Section~\ref{sec:evaluation}).
\end{itemize}

The experiment results are encouraging: compared with the best state-of-the-art approach, we demonstrate that \Model

\begin{itemize}
        \item achieves no worse accuracy on 44 out of 60 cases with 31 of them being significantly better. The improvements can be up to $1.61\times$ against the best counterpart; 
    \item uses fewer samples to reach the same/better accuracy.
    \item incurs acceptable training time considering the improvements in accuracy while the adaptation of $d$ has negligible overhead without requiring extra training. 
\end{itemize}

Interestingly, we also reveal that:

\begin{itemize}
    \item \Model~is model-agnostic, significantly improving the accuracy of a given local model for each division compared with using the model alone as a global model (which is used to learn the entire training dataset). However, \Model~using the hierarchical deep learning-based approach published at TOSEM'23~\cite{DBLP:journals/tosem/ChengGZ23} as the local model produces the most accurate results.
    \item Compared with ensemble learning, which is the other similar paradigm that shares information between local models, \Model, which follows the paradigm of dividable learning that completely isolates the local model, performs considerably better in dealing with the sample sparsity for configuration data with up to $28.50\times$ accuracy improvement over the second-best approach depending on the local model used.
    \item The tailored \texttt{CART}, the mechanism that adapts $d$, and the proposed $\mu$HV indicators can indeed individually contribute to the effectiveness of \Model.
    \item \Model's error tends to correlate upward and quadratically with its only parameter $d$ that sets the number of divisions. Yet, the optimal $d$ value indeed varies depending on the actual systems and training/testing data. Despite such, \Model~can adapt to the optimal $d$ value in 76.43\% of the individual runs while even when it misses hit, a promising $d$ value, which leads to generally marginal accuracy degradations to that of the optimal $d$, can still be selected.
    


\end{itemize}

\subsection{Organization}

This paper is organized as follows: Section~\ref{sec:background} introduces the problem formulation and the notions of sparsity in software performance learning. Section~\ref{sec:framework} delineates the tailored problem formulation and our detailed designs of \Model. Section~\ref{sec:setup} presents the research questions and the experiment design, followed by the analysis of results in Section~\ref{sec:evaluation}. The reasons that \Model~works, its strengths, limitations, and threats to validity are discussed in Section~\ref{sec:discussion}. Section~\ref{sec:related} and~\ref{sec:conclusion} present the related work and conclude the paper, respectively.

\section{Problem Formulation and Methodology}
\label{sec:background}

In this section, we introduce the essential problem formulation and the research methodology which leads to the key observations that motivate this work.


\subsection{Problem Formulation}


In the software engineering community, configuration performance learning has been most commonly tackled by using various machine learning models (or at least partially)~\cite{DBLP:journals/tse/YuWHH19,DBLP:conf/splc/TempleAPBJR19,DBLP:journals/tosem/ChenLBY18,DBLP:journals/corr/abs-1801-02175,DBLP:conf/esem/HanY16}. Such a data-driven process relies on observing the software’s actual behaviors and builds a statistical model to predict the configuration performance without heavy human intervention~\cite{DBLP:books/daglib/0020252}. 


Formally, modeling the performance of software with $n$ configuration options is a regression problem that builds:
\begin{equation}
    \mathcal{P} = f(\mathbfcal{S})\text{, } \mathcal{P}\in\mathbb{R}
    \label{eq:prob}
\end{equation}
whereby $\mathbfcal{S}$ denotes the training samples of configuration-performance pairs, such that $\mathbf{\overline{x}} \in \mathbfcal{S}$. $\mathbf{\overline{x}}$ is a configuration and $\mathbf{\overline{x}}=(x_{1},x_{2},\cdots,x_{n})$, where each configuration option $x_{i}$ is either binary or categorical/numerical. The corresponding performance is denoted as $\mathcal{P}$. 


The goal of machine learning-based modeling is to learn a regression function $f$ using all training data samples such that for newly given configurations, the predicted performance is as close to the actual performance as possible. 

\begin{table*}[t!]
\caption{Selected papers that contain domain knowledge and explicit statements related to the characteristics of sparsity in configuration data (sorted by the citation count from Google Scholar at the time of collecting paper metadata, i.e., Apr 2024). The key phrases are highlighted in bold. The completed list of reviewed papers can be found at: \texttt{\textcolor{blue}{\url{https://github.com/ideas-labo/DaL-ext/blob/main/supplementary_materials/SLR_full_list.xlsb}}}}.
\centering
\begin{adjustbox}{width=\textwidth,center}
\begin{tabular}{p{3.2cm}lllp{13cm}p{2.8cm}}
\toprule
\textbf{Authors/Reference} & \textbf{Year}  & \textbf{Venue} & \textbf{$\#$Citation} & \textbf{Description and/or definition related to sparsity} & \textbf{What is sparse?}\\
\midrule

Aken \textit{et al.}~\cite{DBLP:conf/sigmod/AkenPGZ17} & 2017 & SIGMOD & 640 & ``\textit{DBMSs can have hundreds of knobs, but \textbf{only a subset actually affect} the DBMS’s performance.}'' & Configuration knobs\\
\hline

Siegmund \textit{et al.}~\cite{DBLP:conf/sigsoft/SiegmundGAK15} & 2015 & ESEC/FSE & 305 & ``\textit{The ratio over all subject systems for OW with Plackett-Burman sampling is 0.31, indicating that \textbf{one third of the options significantly contribute} to the performance of a system.}'' & Configuration options\\
\hline

Sarkar \textit{et al.}~\cite{DBLP:conf/kbse/SarkarGSAC15} & 2015 & ASE & 180 & ``\textit{As a consequence, the performance of the system may vary substantially depending on whether \textbf{a certain feature} is selected or not.}'' & Features\\
\hline

Oh \textit{et al.}~\cite{DBLP:conf/sigsoft/OhBMS17} & 2017 & FSE & 155 & \textcolor{black}{``\textit{Stairs arise from discrete feature decisions; \textbf{some features are highly-influential} in performance while others have little or no impact.}''} & Features\\
\hline

Thrane \textit{et al.}~\cite{DBLP:journals/access/ThraneZC20} & 2020 & IEEE Access & 155 & ``\textit{The satellite images offer much information, and the entirety is not necessarily relevant for radio performance prediction, especially at lower frequencies. The use of such images results ultimately in a model that is harder to train since the \textbf{latent features obtained by the CNN might be sparse}.}'' & Features\\
\hline

Jamshidi and Casale~\cite{DBLP:conf/mascots/JamshidiC16} & 2016 & MASCOTS & 142 & ``\textit{More specifically, this means low-order interactions among \textbf{a few dominating factors} can explain the main changes in the response function observed in the experiments.}'' & Factors\\
\hline

Nair \textit{et al.}~\cite{DBLP:journals/corr/abs-1801-02175} & 2020 & TSE & 141 & ``\textit{We know of many software options where \textbf{a small change can lead to radically different} software performance.}'' & Configuration options\\
\hline

Marcus and Papaemmanoui~\cite{DBLP:journals/pvldb/MarcusP19} & 2019 & VLDB & 138 & ``\textit{The problem with this solution is \textbf{sparsity}:
if one has many different operator types, the vectors used to
represent them will have an increasingly \textbf{larger proportion
of zeros}.}'' & Vectors of variables\\
\hline

Jamshidi \textit{et al.}~\cite{DBLP:conf/kbse/JamshidiSVKPA17} & 2017 & ASE & 128 & ``\textit{\textbf{Only a subset of options is influential} which is largely preserved across all environment changes.}'' & Configuration options\\
\hline

Chen and Bahsoon~\cite{DBLP:journals/tse/ChenB17} & 2017 & TSE & 100 & ``\textit{Too limited inputs may not provide
enough information of relevance to the QoS (i.e., the information that drives the changes in QoS), which restricts the
model accuracy and applicability. On the other hand, too
many inputs can generate noise in the modeling, because it
introduces \textbf{irrelevant information and large redundancy} in
the inputs (i.e., the same information has been provided by
more than one selected primitives, thus it becomes noise),
this will downgrade the model accuracy and generate
unnecessary overhead.}'' & Configuration primitives and inputs\\
\hline

Ha and Zhang~\cite{DBLP:conf/icse/HaZ19} & 2019 & ICSE & 93 & ``\textit{It has been observed that the software \textbf{performance functions are usually very sparse} (i.e. only a small number of configuration variables and their interactions have significant impact on system performance).}'' & Variables of performance function\\
\hline

Zhou \textit{et al.}~\cite{DBLP:journals/pvldb/ZhouSLF20} & 2020 & VLDB & 68 & ``\textit{The
\textbf{performance-related features are sparsely scattered} in the
graph (e.g., for a matrix of 1000×1000, many rows only have
two 1s, meaning that the corresponding operator only has 2
directly related operators).}'' & Performance features\\
\hline

Kanellis \textit{et al.}~\cite{DBLP:conf/hotstorage/KanellisAV20} & 2020 & HotStorage & 49 & ``\textit{Surprisingly, we find that with YCSB
workload-A on Cassandra, tuning \textbf{just five knobs can achieve
99\% of the performance} achieved by the best configuration
that is obtained by tuning many knobs.}'' & Configuration knobs\\
\hline

Fekry \textit{et al.}~\cite{DBLP:conf/kdd/FekryCPRH20} & 2020 & KDD & 39 & ``\textit{However, we have observed that for each workload \textbf{only a small subset of those parameters has a significant impact} on overall performance.}'' & Configuration parameters\\
\hline

Chen \textit{et al.}~\cite{DBLP:conf/icse/0003XC021} & 2021 & ICSE & 38 & ``\textit{\textbf{Only a small number of optimization flags}, referred to as impactful optimizations, can have \textbf{noticeable impact} on the runtime performance of a specific program.}'' & Configuration flag\\
\hline

Zhang \textit{et al.}~\cite{DBLP:journals/pvldb/ZhangCLWTLC22} & 2022 & VLDB & 36 & ``\textit{Given a limited tuning budget, tuning over the configuration space with all the knobs is inefficient. It is recommended to \textbf{preselect important knobs} to prune the configuration space.}'' & Configuration knobs\\
\hline

Grohmann \textit{et al.}~\cite{DBLP:conf/middleware/GrohmannNIKL19} & 2019 & Middleware & 35 & ``\textit{We collect 1,040 platform metrics using the PCP monitoring tool as described in Section 3.1. 952 of these platform metrics consider the host, 88 are specific to service instances (i.e., containers) running on the host. As expected, \textbf{not all the metrics are relevant} for the machine learning model and in many cases metric preprocessing is required such that they can be useful or leveraged by the algorithm.}'' & Measured configuration metrics\\
\hline

Zhang~\cite{DBLP:conf/sigmod/ZhangW0T0022} & 2022 & SIGMOD & 34 & \textcolor{black}{``\textit{Modern DBMSs have hundreds of plan operators, encoding the plan operators could cause the ``dimensionality issue" when augmenting the \textbf{sparse and high-dimensional variables} to GP kernel.}''} & GP kernel\\
\hline

\hline

Ha and Zhang~\cite{DBLP:conf/icsm/Ha019} & 2019 & ICSME & 29 & ``\textit{For software performance functions, their \textbf{Fourier coefficients are always very sparse},
i.e. most of Fourier coefficients are zeros. The reason is that only a small number of configurations have significant impact on system performance.}'' & Configuration coefficients\\
\hline

\hline

Kanellis \textit{et al.}~\cite{DBLP:journals/pvldb/KanellisDKMCV22} & 2022 & VLDB & 22 & ``\textit{However, recent studies have shown that \textbf{tuning a handful of knobs can be sufficient} to achieve near-optimal performance and significantly reducing the number of knobs can accelerate the tuning process.}'' & Configuration knobs\\
\hline

Zhang \textit{et al.}~\cite{DBLP:journals/jsa/ZhangLWWZH18} & 2018 & JSA & 18 & ``\textit{\textbf{Not all the features have impact} on the performance evaluation result, we can find a way to confirm the real effective features and remove the redundancy ones.}'' & Configuration features\\
\hline


Gong and Chen~\cite{DBLP:conf/msr/GongC22} & 2022 & MSR & 12 & ``\textit{Unlike other domains, software configuration is often highly
sparse, leading to unusual data distributions. Specifically, \textbf{a few configuration options could have large influence} on the software
performance, while the others are trivial.}'' & Configuration options\\
\hline

Marker~\cite{DBLP:conf/kbse/MarkerBG14} & 2014 & ASE & 12 & \textcolor{black}{``\textit{The (implementation × performance) space of DLA is stair-stepped. Each stair is a set of implementations with very similar performance and (surprisingly) share \textbf{key design decision(s)}.}''} & \textcolor{black}{Design decisions}\\
\hline

Shu \textit{et al.}~\cite{DBLP:conf/esem/ShuS0X20} & 2020 & ESEM & 12 & ``\textit{This means
that only a small number of parameters have significant impact on the model. In other words, the \textbf{parameters of the neural network could be sparse}.}'' & Configuration parameters\\
\hline

Cheng \textit{et al.}~\cite{DBLP:journals/tosem/ChengGZ23} & 2021 & TOSEM & 9 & ``\textit{One important prior knowledge is that \textbf{only a small number of configuration options and their interactions have a significant impact} on system performance, implying sparsity on the parameters of the performance model (i.e., making parameters of many insignificant options and interactions equal to zero).}'' & Configuration options\\
\hline

Acher \textit{et al.}~\cite{DBLP:conf/splc/Acher0LBJKBP22} & 2022 & SPLC & 9 & ``\textit{Given these results, we can say that out of the thousands of options of Linux kernel, \textbf{only a few hundreds actually influence} its binary size.}'' & Configuration options\\
\hline

Xie \textit{et al.}~\cite{DBLP:conf/sc/XieTCCHLOVW19} & 2019 & PDSW & 8 & ``\textit{Our analysis shows that the write behaviors of the GPFS system are \textbf{dominated by the metadata load and load skew} within the supercomputer and the resources used in its filesystem.}'' (The metadata load and load skew here are configuration options.) & Configuration options\\
\hline

Malik \textit{et al.}~\cite{DBLP:conf/sc/MalikFP09} & 2013 & WORKS & 7 & ``\textit{\textbf{Not all the features actually have considerable impact} on the execution time, therefore the collected provenance data needs to be filtered and reduced.}'' & Configuration features\\
\hline

Gong and Chen~\cite{DBLP:conf/sigsoft/JChen2023} & 2024 & FSE & 6 & \textcolor{black}{``\textit{The division of samples and the trained local model handles the
\textbf{sample sparsity} while the rDNN deals with the \textbf{feature sparsity}.
}''} & \textcolor{black}{Feature \& sample distribution}\\
\hline

Trajkovic \textit{et al.}~\cite{DBLP:journals/tecs/TrajkovicKHZ22} & 2022 & TECS & 5 & ``\textit{To be able to generate a prediction model with high accuracy, we need to select features with a significant impact on the values of the targets. Including the \textbf{features that have little or no impact on the targets may result in overfitting}.}'' & Parameter features\\
\hline

Chen \textit{et al.}~\cite{chen2024mmo} & 2024 & TSE & 5 & \textcolor{black}{``\textit{They have been specifically tailored to cater for the key
properties of the tuning problem, e.g., \textbf{high sparsity} and expensive measurements.}''} & Configuration options\\
\hline

Gong and Chen~\cite{gong2024predicting} & 2024 & FSE & 2 & \textcolor{black}{``\textit{Noteworthy, SeMPL does not achieve the highest rank for Storm, which can be attributed to a combination of factors including the \textbf{sparsity of samples} (due to the nature of the system) and the presence of more diverse environments, primarily limited to hardware changes.}''} & \textcolor{black}{Sample distribution}\\



\bottomrule
\end{tabular}
\end{adjustbox}
\label{tb:papers}
\end{table*}

\subsection{Sparsity in Configuration Data: A Qualitative Study}
\label{sec:motivation}

\subsubsection{Literature review}

To confirm the known characteristics of configuration data with respect to sparsity and how it is tackled, we first conducted a literature review for papers related to the performance of software configuration\footnote{Notably, the main purpose of the review is not to exhaustively extract much different information from existing work on configuration performance learning, but to find the general ideas about how sparsity in configuration data is defined and considered when building a model.} published during the last decade, i.e., from 2013 to 2024. Specifically, we performed an automatic search over six highly influential indexing services, i.e., ACM Library, IEEE Xplore, Google Scholar, ScienceDirect, SpringerLink, and Wiley Online, by using the search string below:

\begin{displayquote}
\textit{(``software" OR ``system" \textcolor{black}{OR ``software product lines''}) AND (``configuration" \textcolor{black}{OR ``configurable''}) AND (``performance prediction" OR ``performance modeling" OR ``performance learning" OR ``configuration tuning") AND ``machine learning"} 
\end{displayquote}

The search result consists of 5,920 studies with duplication except for non-English documents. Next, we filtered out the duplication by examining the titles and also eliminated clearly irrelevant documents, e.g., those about students' learning performance. This has resulted in 422 highly relevant candidate studies. Through a detailed review of the candidate studies, we then applied various criteria to further extract a set of more representative works. In particular, a study is temporarily selected as a primary study if it meets all of the following inclusion criteria:

\begin{itemize}
\item The paper presents machine learning algorithm(s) for modeling configuration performance\footnote{\textcolor{black}{There have been diverse kinds of research related to configurable systems, including but not limited to, configuration performance learning, configuration debugging, and configuration tuning. We chose to include only papers that involve machine learning algorithms in learning configuration because our scope is to propose a new approach to better handle the data sparsity inherited in the configuration data; our literature review aims to solely serve the purpose of confirming how sparsity is currently handled in such a context. Otherwise, we might include works that do consider sparsity but from an optimization/search/tuning perspective, e.g., how to overcome the local optima caused by sparsity instead of learning it~\cite{bestconfig}.}}; This can also include search algorithm(s) for tuning the configurations to reach better performance in which some models need to be learned, e.g., different variants of Bayesian Optimization~\cite{DBLP:conf/icse/0003XC021,DBLP:journals/corr/abs-1801-02175}.
\item The paper presents a study or claims, either explicitly or implicitly, related to the sparsity of configuration data in the domain studied.
\item The goal of the paper is to predict, optimize, or analyze the performance of a software system.
\item The paper has at least one section that explicitly specifies the algorithm(s) used.
\item The paper contains quantitative evaluations with details abouts how the results were obtained.

\end{itemize}

Subsequently, we applied the following exclusion criteria on the previously included studies, which would be removed if they meet any below: 
\begin{itemize}
\item The paper is not software or system-related.
\item The paper is not published in a peer-reviewed venue.
\item The paper is a survey, review, or tutorial.
\item The paper reports a work-in-progress research, i.e., shorter than 8 double-column or 15 single-column pages.
\end{itemize}

We obtained 71 highly related candidate studies from the above procedures. Next, a snowballing systematic search was performed using Google Scholar. This involved a backward snowballing exploration of the references cited within the identified candidate studies, enabling us to uncover additional sources that are related to sparsity. Additionally, we systematically examined the studies that cited these candidate studies in a forward snowballing search to identify recent publications that mention the sparsity in configuration data. The results are then further filtered subject to the inclusion and exclusion criteria above. Throughout the reviewing process, the first author took the primary responsibility for conducting the search and the second author served as a quality checker of the outcomes. 
This has led to five additional studies, resulting in a total of 76 primary studies for data collection. 

All authors are part of the data collection process, which involves three iterations, as commonly used by the literature reviews in Software Engineering~\cite{li2020evaluate, zou2018practitioners}:



\begin{itemize}
    \item In the first iteration, each of the authors conducted initial data collection independently, read carefully throughout the 76 primary studies, extracted sentences that describe sparsity and subjects that are observed to be sparse (e.g., features, vectors, and GP kernel), and summarized the data into a table. 
    Notably, there were situations where an author was not sure how a statement was related to sparsity (e.g., statements about sparse matrix). In these cases, the corresponding data were marked for further investigation in the next iteration.
    \item In the second iteration, the authors reviewed and cross-checked each other's data summary tables, ultimately integrating them into a unified table. The unclear data that were marked in the previous iteration were rechecked and we arranged meetings to reach a common agreement for each item of which the collected data from the authors were different. This is achieved through discussions; further investigations from the existing literature; or consultation with external experts. 
    \item Finally, the goals for the third iteration are to summarize the statistics of the integrated table from the second iteration; count the number of studies for each type of sparsity; and aggregate similar sparsity definitions. 
 
\end{itemize}


The selected results have been shown in Table~\ref{tb:papers}. Albeit with diverse terminologies (e.g., knobs, metrics, or parameters), we see that almost all papers have identified evident sparse patterns in software features related to configuration options, therefore, existing work reveals high \textbf{\textit{feature sparsity}}, which refers to the fact that only a small number of configuration options are prominent to the performance. This is inevitable as the configuration space for software systems is generally rugged with respect to the configuration options~\cite{DBLP:conf/sigsoft/0001Chen21,DBLP:conf/kbse/JamshidiSVKPA17}.

\begin{figure*}[!t]
\centering
\footnotesize

\begin{subfigure}{.5\columnwidth}
  \centering
    \includegraphics[width=\linewidth]{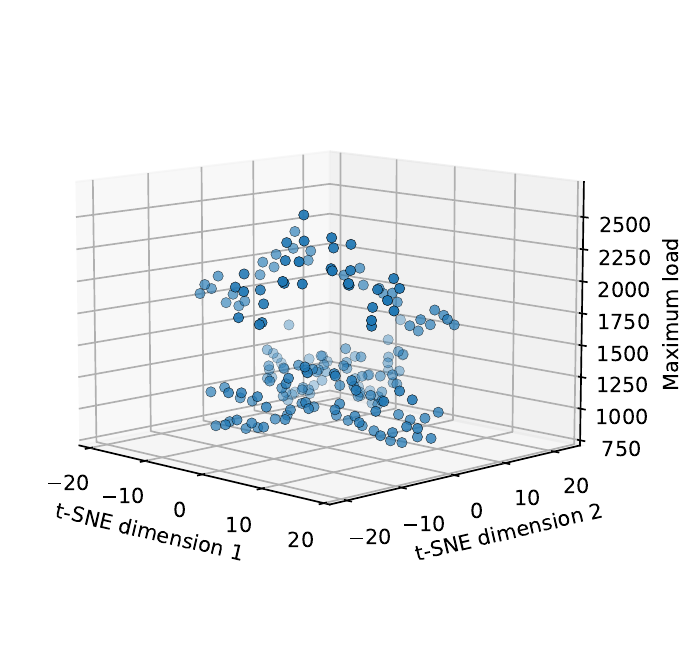} \vspace{-0.8cm}
  \caption{\textsc{Apache}}
  \label{fig:depth-Apache}
\end{subfigure}
~\hspace{-0.3cm}
\begin{subfigure}{.5\columnwidth}
  \centering
  \includegraphics[width=\linewidth]{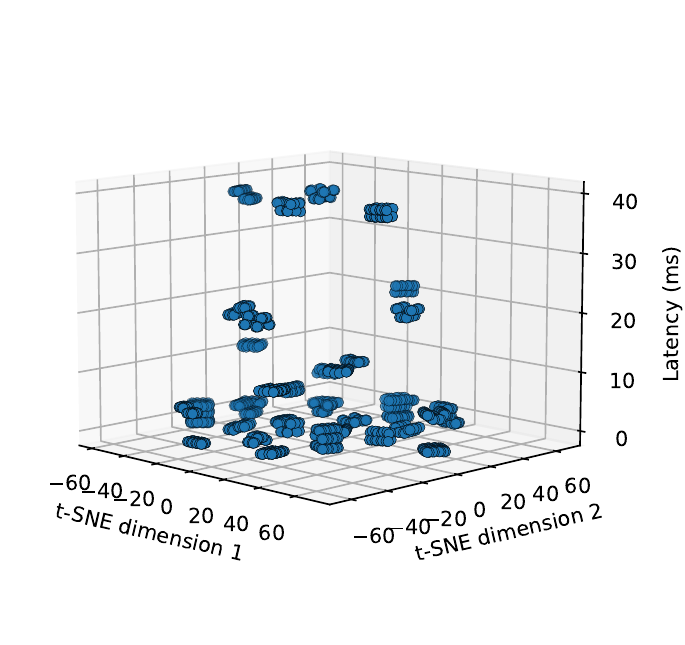}  \vspace{-0.8cm}
  \caption{\textsc{BDB-C}}
  \label{fig:depth-BDBC}
\end{subfigure}
~\hspace{-0.27cm}
\begin{subfigure}{.5\columnwidth}
  \centering
  \includegraphics[width=\linewidth]{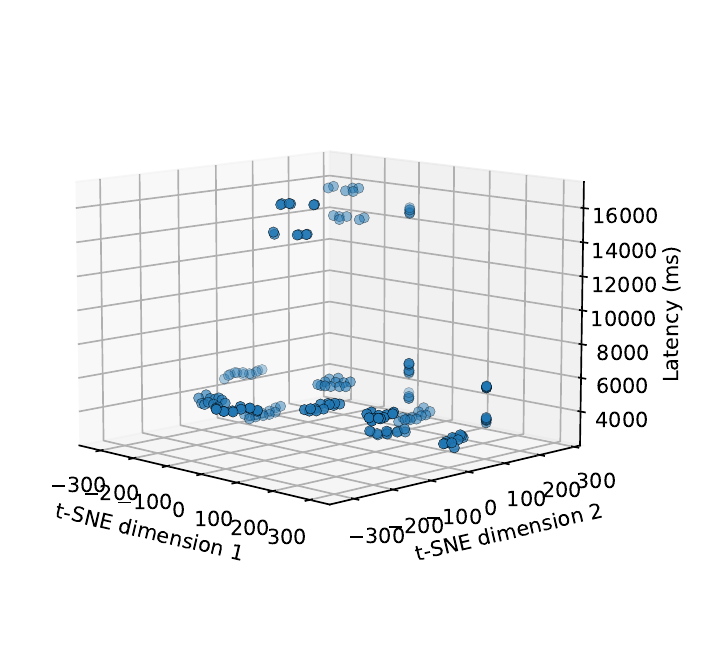} \vspace{-0.8cm}
  \caption{\textsc{BDB-J}}
  \label{fig:depth-BDBJ}
\end{subfigure}
~\hspace{-0.3cm}
\begin{subfigure}{.5\columnwidth}
  \centering
  \includegraphics[width=\linewidth]{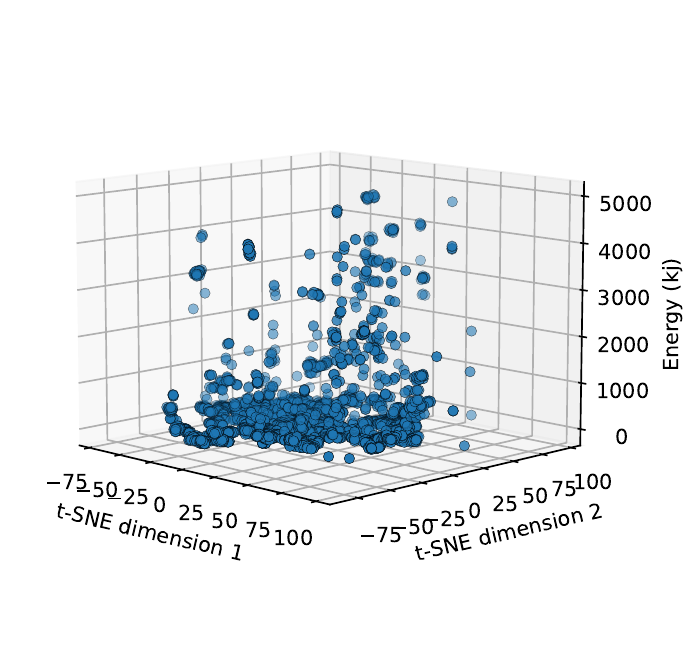}  \vspace{-0.8cm}
  \caption{\textsc{kanzi}}
  \label{fig:depth-kanzi}
\end{subfigure}

\begin{subfigure}{.5\columnwidth}
  \centering
  \includegraphics[width=\linewidth]{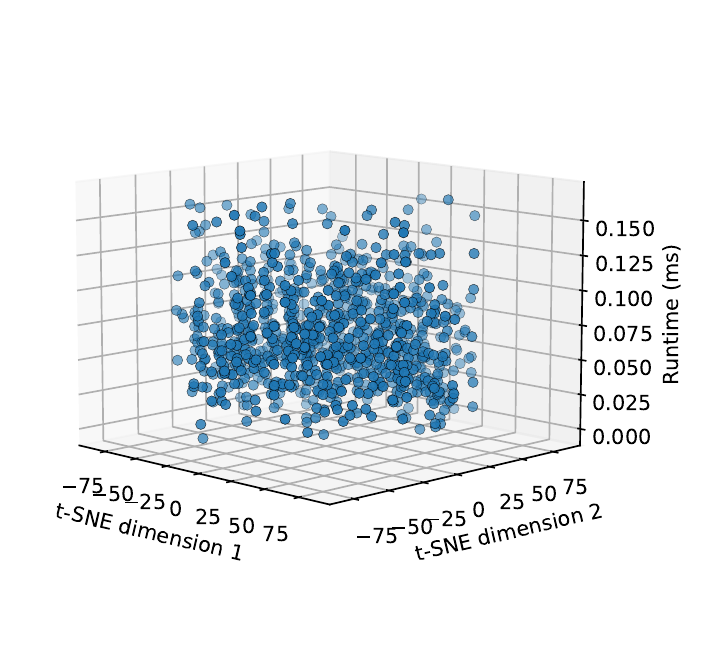} \vspace{-0.8cm}
  \caption{\textsc{SQLite}}
  \label{fig:depth-SQLite}
\end{subfigure}
~\hspace{-0.3cm}
\begin{subfigure}{.5\columnwidth}
  \centering
  \includegraphics[width=\linewidth]{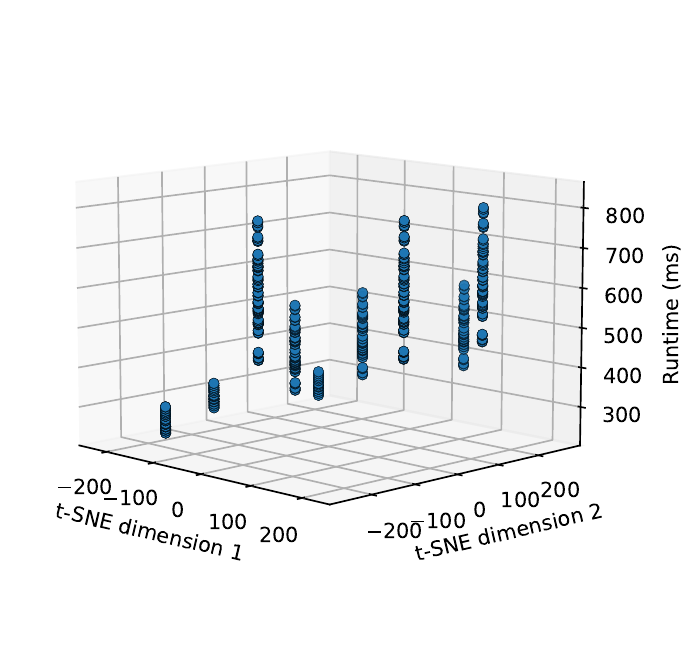} \vspace{-0.8cm}
  \caption{\textsc{x264}}
  \label{fig:depth-x264}
\end{subfigure}
~\hspace{-0.3cm}
\begin{subfigure}{.5\columnwidth}
  \centering
    \includegraphics[width=\linewidth]{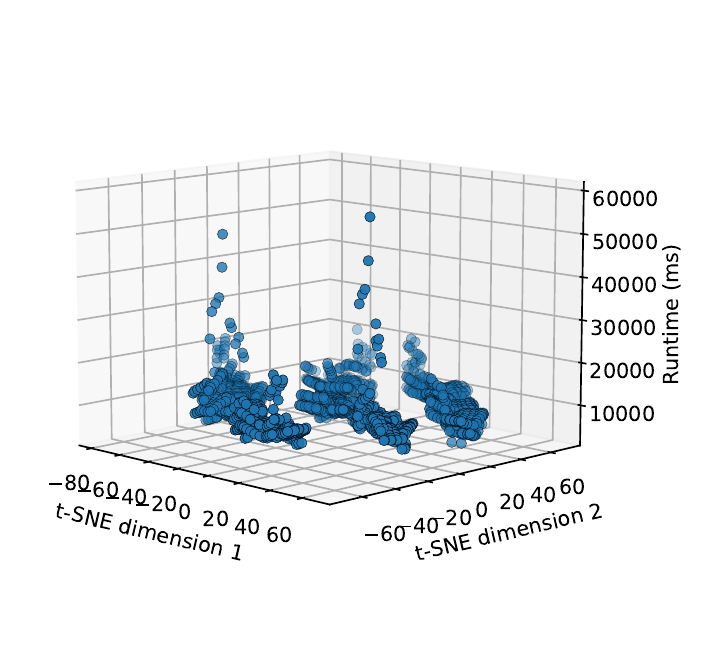} \vspace{-0.8cm}
  \caption{\textsc{Dune MGS}}
  \label{fig:depth-Dune}
\end{subfigure}
~\hspace{-0.3cm}
\begin{subfigure}{.5\columnwidth}
  \centering
 \includegraphics[width=\linewidth]{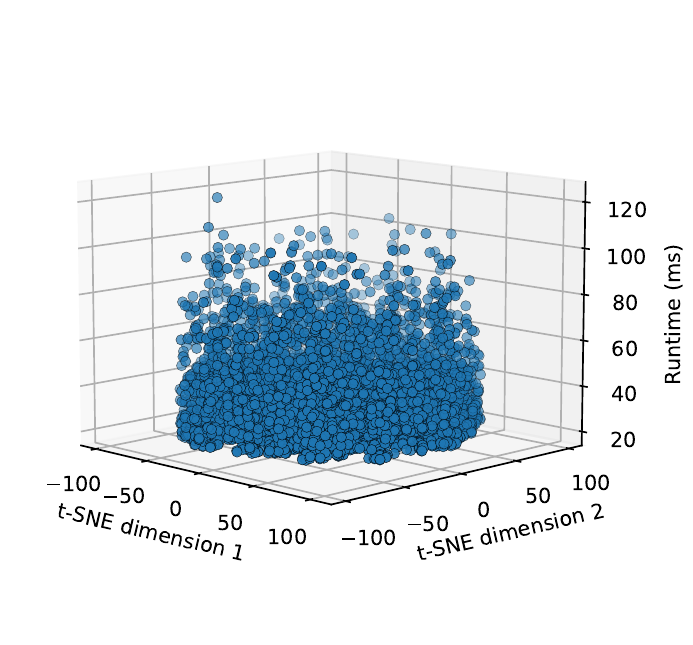} \vspace{-0.8cm}
  \caption{\textsc{HIPA$^{cc}$}}
  \label{fig:depth-HIPAcc}
\end{subfigure}

\begin{subfigure}{.5\columnwidth}
  \centering
    \includegraphics[width=\linewidth]{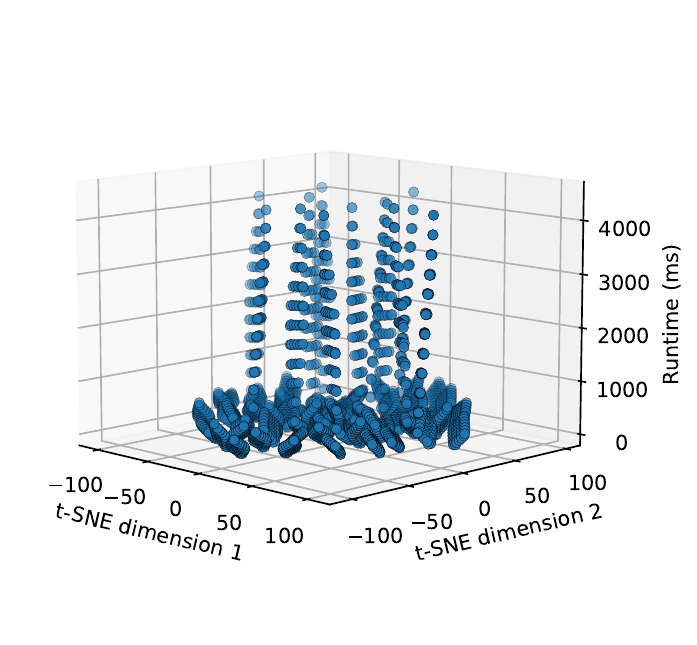}\vspace{-0.5cm} 
  \caption{\textsc{HSMGP}}
  \label{fig:depth-HSMGP}
\end{subfigure}
~\hspace{-0.3cm}
\begin{subfigure}{.5\columnwidth}
  \centering
    \includegraphics[width=\linewidth]{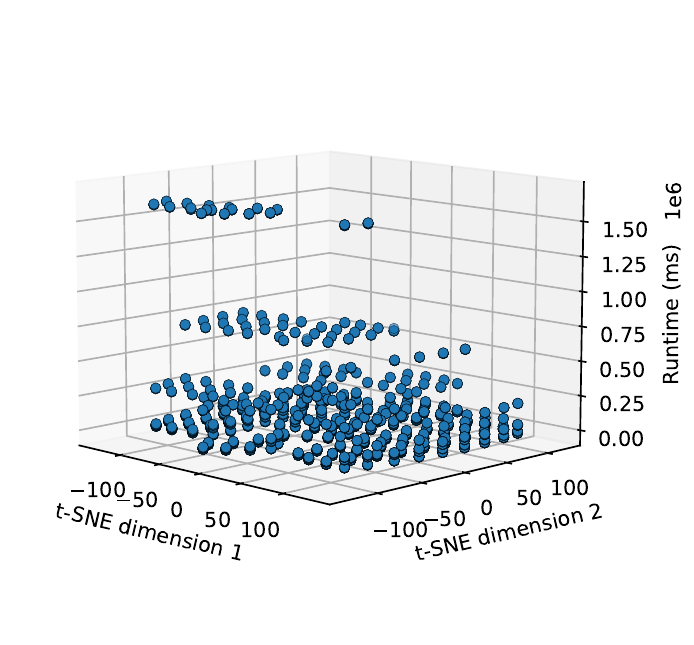} \vspace{-0.8cm}
  \caption{\textsc{Lrzip}}
  \label{fig:depth-Lrzip}
\end{subfigure}
~\hspace{-0.3cm}
\begin{subfigure}{.5\columnwidth}
  \centering
    \includegraphics[width=\linewidth]{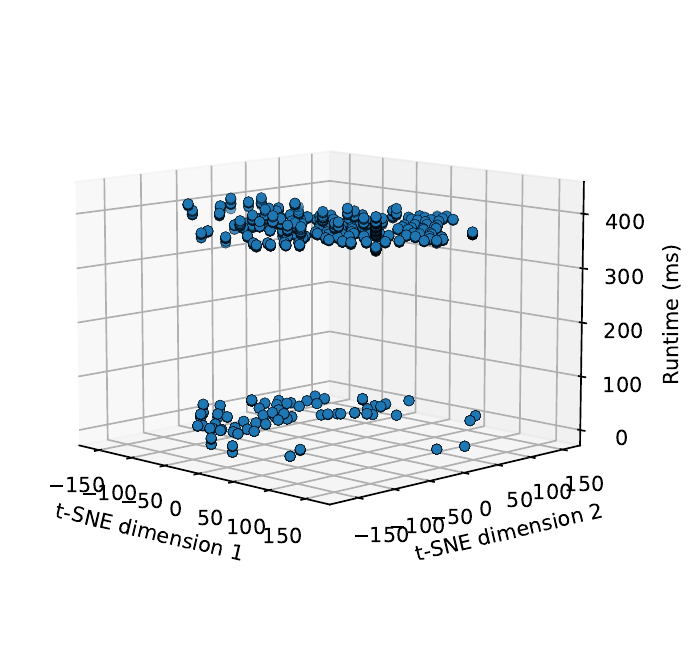} \vspace{-0.8cm}
  \caption{\textsc{nginx}}
  \label{fig:depth-nginx}
\end{subfigure}
~\hspace{-0.3cm}
\begin{subfigure}{.5\columnwidth}
  \centering
    \includegraphics[width=\linewidth]{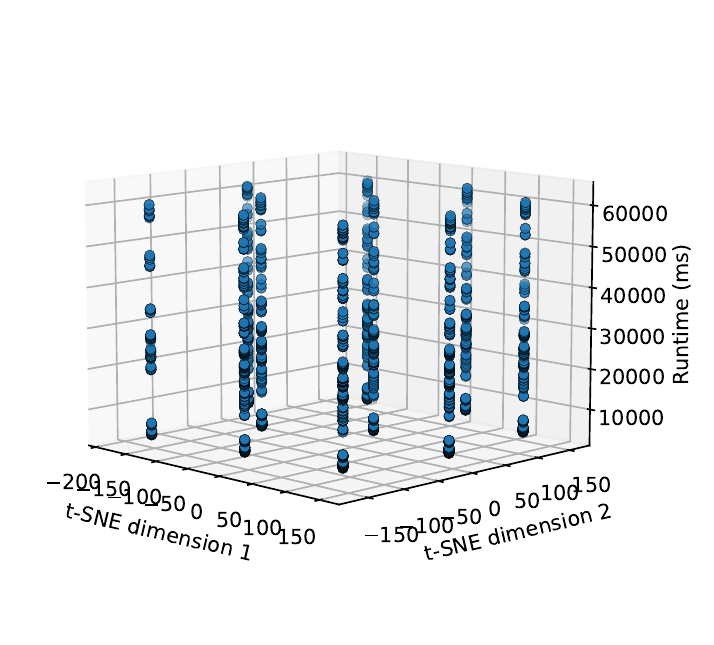}  \vspace{-0.8cm}
  \caption{\textsc{VP8}}
  \label{fig:depth-VP8}
\end{subfigure}

  \caption{Projection of configurations in the landscape using t-SNE \textcolor{black}{(Note that the t-SNE dimensions are extracted and newly emerged features that do not correspond to any actual configuration options of the systems)}.}
     \label{fig:3d-exp}
\end{figure*}

\subsubsection{Empirical study}
\label{sec:emprical}

Although the literature review has revealed evidence of sparsity in configuration data, it is difficult to ensure that all the major aspects/characteristics of sparsity in configuration data have been covered. As such, to better understand the sparse nature of configurable software systems, we empirically analyzed the data collected from the real-world systems studied in prior work. We do so by reviewing the systems used in the work identified from the literature review stage and selecting the ones according to the following criteria:

\begin{itemize}
\item The system is used by more than one paper.
\item There are clear articulations on how the data is collected under the systems, e.g., how many repeated measurements.
\item The data contains neither missing nor invalid measurements, such as ``NaN'' or other suspicious values like ``$-1$'' or ``9999''.
\item The data should express a full spectrum of all the valid configurations for the system.
\end{itemize}

The above criteria could still lead to many systems that are of similar categories and characteristics, but it would be unrealistic to study them all and hence further representatives need to be extracted. To that end, for each category of the software systems, (e.g., a video encoder or a compiler), we select the one(s) that have been overwhelmingly studied across most of the papers identified.


The extraction has led to 12 real-world configurable software systems in our analysis (their details will be discussed in Section~\ref{sec:setup}).

Since it is unrealistic to visualize the configuration landscape consisting of multi-dimensional configuration options, we process the configuration data using t-distributed Stochastic Neighbor Embedding (t-SNE)~\cite{DBLP:conf/nips/HintonR02}, which processes all options and produces two new, but reduced dimensions in the visualization. In a nutshell, t-SNE is an unsupervised algorithm wherein similar configurations are modeled by nearby points and dissimilar configurations are modeled by distant points with high probability. Note that t-SNE is often used for dimensionality reduction, where only the information of the most important features is preserved.

In Figure~\ref{fig:3d-exp}, we demonstrated the t-SNE processed results. Notably, we see that the systems all exhibit a consistent pattern---Our key discovery is therefore:




\begin{tcbitemize}[%
    raster columns=1, 
    raster rows=1
    ]
  \tcbitem[myhbox={}{Key Discovery}] \textit{Even with the key options that are the most influential to the performance (as represented in the reduced dimensions by t-SNE), the samples still do not exhibit a ``smooth'' distribution over the configuration landscape. Instead, those with similar characteristics can condense into local divisions, but globally, those from different local divisions might be spread sparsely.}
\end{tcbitemize}

This is a typical case of high \textbf{\textit{sample sparsity}}~\cite{DBLP:journals/corr/abs-2202-03354,DBLP:conf/icml/LiuCH20,DBLP:conf/icml/ShibagakiKHT16}. Indeed, it is difficult for precisely defining to what extent the sparsity is considered as high, but with the visualized results from our empirical study, there is strong evidence that the configuration data is certainly non-smooth and hence raising concerns of ``high sparsity smells''.


In general, such a high sample sparsity is caused by the following reasons: 

\begin{itemize}
    \item the inherited consequence of high feature sparsity. 
    \item the complex and nonlinear interaction between configuration options, especially those categorical ones that can sharply influence the performance, e.g., adjusting between different cache strategies can drastically influence the performance, but they are often represented as a single-digit change on the landscape~\cite{DBLP:conf/sigsoft/0001Chen21}. This leads to a highly rugged configuration landscape that leads to sparsity~\cite{DBLP:conf/sigsoft/0001Chen21,DBLP:conf/kbse/JamshidiSVKPA17}.
    \item the fact that not all configurations are valid because of the constraints (e.g., an option can be used only if another option has been turned on)~\cite{DBLP:conf/sigsoft/SiegmundGAK15}, thereby there are many “empty areas” in the configuration landscape.
\end{itemize}


\subsubsection{Discussion}
\label{background-sec-discussion}

The above findings reveal the key factors to consider for our problem: when using machine learning models to learn concepts from the configuration data, we need a model that: 

\begin{enumerate}
    \item handles the complex interactions between the configuration options with high feature sparsity while;
    \item captures the diverse characteristics of configuration samples over all divisions caused by the high sample sparsity, e.g., in Figure~\ref{fig:3d-exp}, where samples in different divisions have diverged performance ranges.
\end{enumerate}
For the former challenge, there have been some proposed approaches previously, such as \texttt{DeepPerf}~\cite{DBLP:conf/icse/HaZ19}, \texttt{Perf-AL}~\cite{DBLP:conf/esem/ShuS0X20}, and \texttt{HINNPerf}~\cite{DBLP:journals/tosem/ChengGZ23}. The latter, unfortunately, is often ignored in existing work for configuration performance learning as we observed from our literature review and Table~\ref{tb:papers}, causing a major obstacle for a model to learn and generalize the data for predicting the performance of the newly-given configuration. This is because those highly sparse samples increase the risk for models to overfit the training data, for instance by memorizing and biasing values in certain respective divisions~\cite{DBLP:journals/corr/abs-2202-03354}, especially considering that we can often have limited samples from the configuration landscape due to the expensive measurement of configurable systems. 

The above is the main motivation of this work, for which we ask: how can we improve the accuracy of predicting software configuration performance under such a high sample sparsity?




\section{Divide-and-Learn: Dividable Learning for Configuration Performance}
\label{sec:framework}


Drawing on our observations of the configuration data, we propose \Model---a framework of dividable learning that enables better prediction of the software configuration performance via ``divide-and-learn''. To mitigate the sample sparsity issue, the key idea of \Model~is that, since different divisions of configurations show drastically diverse characteristics, i.e., rather different performance values with distant values of key configuration options, we seek to independently learn a local model for each of those divisions that contain \textit{locally smooth} samples, thereby the learning can be more focused on the particular characteristics exhibited from the divisions and handle the feature sparsity. Yet, this requires us to formulate, on top of the original regression problem of predicting the performance value, a new classification problem without explicit labels. As such, we modify the original problem formulation (Equation~\ref{eq:prob}) as below: 
\begin{equation}
    \mathbfcal{D} = g(\mathbfcal{S}) 
\end{equation}
\begin{equation}
    \forall D_i \in \mathbfcal{D}\text{: } \mathcal{P} = f(D_i)\text{, } \mathcal{P}\in\mathbb{R}
\end{equation}
Overall, we aim to achieve three goals:

\begin{itemize}
    \item \textbf{Goal 1:} dividing the data samples into diverse yet more focused divisions $\mathbfcal{D}$ (building function $g$) and;
    \item \textbf{Goal 2:} training a dedicated local model for each division $D_i$ (building function $f$) while;
    \item \textbf{Goal 3:} assigning a newly coming configuration into the right model for prediction (using functions $g$ and $f$).
\end{itemize}

\begin{figure}[t!]
  \centering
  \includegraphics[width=\columnwidth]{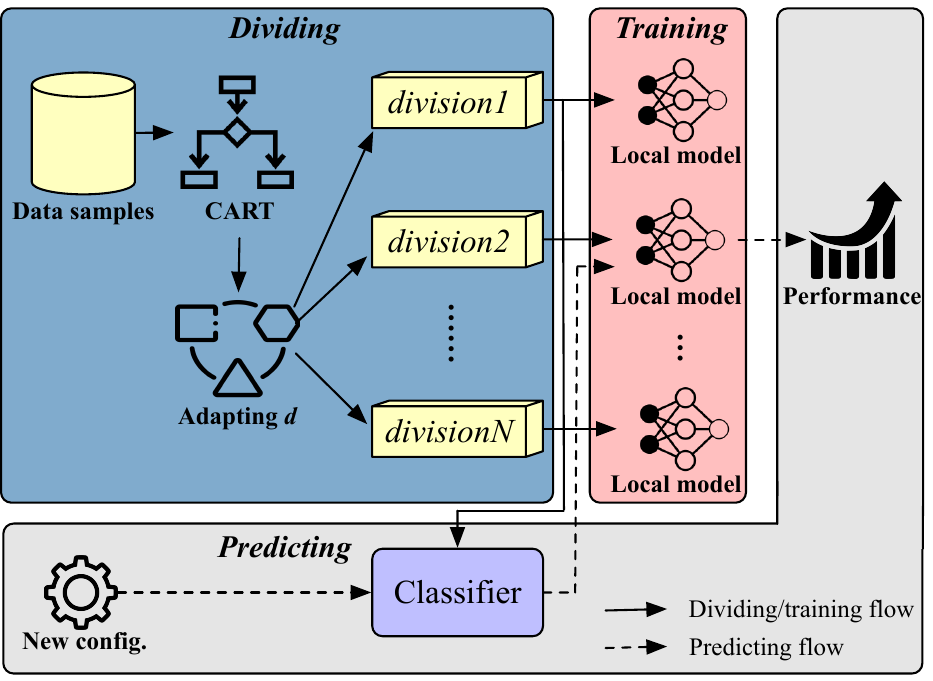}
  \caption{The architecture of \Model.}
  \label{fig:structure}
\end{figure}

Figure~\ref{fig:structure} illustrates the overall architecture of \Model, in which there are three core phases, namely \textit{Dividing}, \textit{Training}, and \textit{Predicting}. A pseudo-code can also be found in Algorithm~\ref{alg:dal-code}. In particular, the options' interactions at the \textit{Division} are handled by CART and a parameter $d$, i.e., only the top options are considered in the process of dividing the data samples. In the \textit{Training}, the interaction between all options is considered by the local model. Yet, since the local model is only responsible for one division of the data, the full interactions are split, each of which is handled separately and independently.

\subsection{Dividing}
\label{subsec:phase1_clustering}

The very first phase in \Model~is to appropriately divide the data into more focused divisions while doing so by considering both the configuration options and performance values. To that end, the key question we seek to address is: how to effectively cluster the configuration data with similar sample characteristics (\textbf{Goal 1})?

\subsubsection{Modifying \texttt{CART} for Generating Divisions}
\label{subsec:modeify-cart}

Indeed, for dividing data samples, it makes sense to consider various unsupervised clustering algorithms, such as \texttt{$k$Means}~\cite{macqueen1967some}, \texttt{Agglomerative clustering}~\cite{Inchoate:Ward63} or \texttt{DBSCAN}~\cite{DBLP:conf/kdd/EsterKSX96}. However, we found that they are ill-suited for our problem, because:
\begin{itemize}
    \item the distance metrics are highly system-dependent. For example, depending on the number of configuration options and whether they are binary/numeric options;
    \item it is difficult to combine the configuration options and performance value with appropriate discrimination; 
    \item and clustering algorithms are often non-interpretable.
\end{itemize}

\begin{algorithm}[t!]
	\DontPrintSemicolon
	\footnotesize
	
	\caption{Pseudo code of \Model}
	\label{alg:dal-code}
	\KwIn{A new configuration $\mathbf{\overline{c}}$ to be predicted}
     \KwOut{The predicted performance of $\mathbf{\overline{c}}$}
	
	\If{$\mathbfcal{M} = \emptyset$}
	{
	  \tcc{\textcolor{blue}{dividing phase.}}
      $\mathbfcal{S}\leftarrow$ randomly sample a set of configurations and their performance\\
      $\mathcal{T}\leftarrow$ \textsc{trainCART($\mathbfcal{S}$)}\\
      $d=$\textsc{adaptingDepth($\mathcal{T}$)}\\
      $d'=1$\\
      \While{$d'\leq d$}{
           \If{$d'<d$}
           {
             $\mathbfcal{D}\leftarrow$ extract all the leaf divisions of samples from $\mathcal{T}$ at the $d'$th depth\\
           }
           \Else{
              $\mathbfcal{D}\leftarrow$ extract all divisions of samples from $\mathcal{T}$ at the $d'$th depth\\
           }
           $d'=d'+1$\\
      }
      
      \tcc{\textcolor{blue}{training phase.}}
      \For{$\forall D_i \in \mathbfcal{D}$}  
      {
       $\mathbfcal{M}\leftarrow$ \textsc{trainLocalModel($D_i$)}\\
    
      }
    
    }
       
    \tcc{\textcolor{blue}{predicting phase.}}
    \If{$\mathcal{F}$ has not been trained}
	{
	$\mathbfcal{U}\leftarrow$ Removing performance data and labeling the configurations based on their divisions in $\mathbfcal{D}$\\
	$\mathbfcal{U'}\leftarrow$ \textsc{SMOTE($\mathbfcal{U}$)}\\
	$\mathcal{F}\leftarrow$ \textsc{trainRandomForest($\mathbfcal{U'}$)}
	}
	
    $D_i=$  \textsc{predict($\mathcal{F}$,$\mathbf{\overline{c}}$)}\\
    $\mathcal{M}=$ get the model from $\mathbfcal{M}$ that corresponds to the predicted division $D_i$\\

    \Return \textsc{predict($\mathcal{M}$,$\mathbf{\overline{c}}$)}\\
	
\end{algorithm}

As a result, in \Model, we extend Classification and Regression Tree (\texttt{CART}) as the clustering algorithm (lines 3-12 in Algorithm~\ref{alg:dal-code}) since (1) it is simple with interpretable/analyzable structure; (2) it ranks the important options as part of training (good for dealing with the feature sparsity issue), and (3) it does not suffer the issues above~\cite{DBLP:conf/kbse/SarkarGSAC15,DBLP:journals/ese/GuoYSASVCWY18,DBLP:journals/ase/NairMSA18,DBLP:conf/icse/Chen19b,DBLP:journals/tse/ChenB17,DBLP:journals/corr/abs-1801-02175,DBLP:conf/kbse/GuoCASW13}. In Section~\ref{subsec:component}, we will experimentally demonstrate that \texttt{CART} allows \Model~to obtain much more superior results than the state-of-the-art clustering algorithms in dividing our configuration data.

\begin{figure}[t!]
  \centering
  \includegraphics[width=\columnwidth]{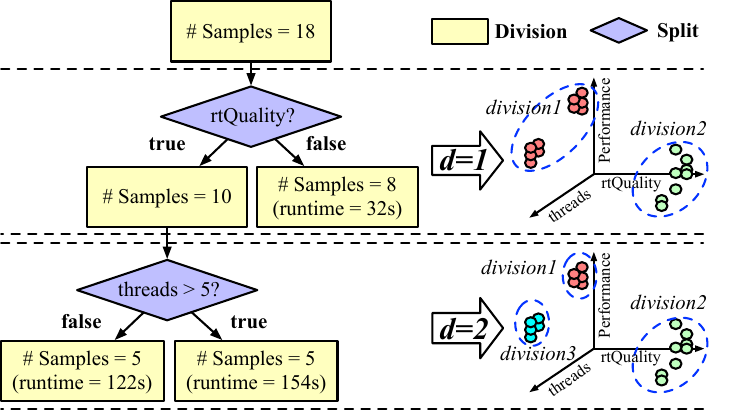}
  \caption{Projection of \texttt{CART} for \textsc{VP8} showing the possible divisions with different colors under alternative depth $d$.}
  \label{fig:DT_example}
\end{figure}

As illustrated in Figure~\ref{fig:DT_example}, \texttt{CART} is originally a supervised and binary tree-structured model, which recursively splits some, if not all, configuration options and the corresponding data samples based on tuned thresholds. A split would result in two divisions, each of which can be further split. In this work, we at first train the \texttt{CART} on the available samples of configurations and performance values, during which we use the most common mean performance of all samples for each division $D_i$ as the prediction~\cite{DBLP:journals/ese/GuoYSASVCWY18,DBLP:conf/kbse/GuoCASW13}:
\begin{equation}
    \overline{y}_{D_i}={{1}\over{|D_i|}} {\sum_{y_j \in D_i}y_j}
    \label{eq:seg}
\end{equation}
in which $y_j$ is a performance value. For example, Figure~\ref{fig:DT_example} shows a projected example, in which the configuration that satisfies ``\texttt{rtQuality=true}'' and ``\texttt{threads=3}'' would lead to an inferred runtime of 122 seconds, which is calculated over all the 5 samples involved using Equation~\ref{eq:seg}.

By choosing/ranking options that serve as the splits and tuning their thresholds, in \Model, we seek to minimize the following overall loss function during the \texttt{CART} training:
\begin{equation}
   \mathcal{L}= {1\over{|D_l|}}{\sum_{y_j \in D_l}{(y_j - \overline{y}_{D_l})}^2} + {1\over{|D_r|}}{\sum_{y_j \in D_r}{(y_j - \overline{y}_{D_r})}^2}
    \label{eq:loss}
\end{equation}
where $D_l$ and $D_r$ denote the left and right division from a split, respectively. This ensures that the divisions would contain data samples with similar performance values while they are formed with respect to the similar values of the key configuration options as determined by the splits/thresholds (i.e., as from the \textit{Key Discovery} in Section~\ref{sec:emprical}), i.e., the more important options would appear on the higher level of the tree with excessive splitting.

However, here we do not use \texttt{CART} to generalize prediction directly on new data once it is trained as it has been shown that the splits and simple average of performance values in the division alone can still fail to handle the complex interactions between the options, leading to insufficient accuracy~\cite{DBLP:conf/icse/HaZ19}. Further, with our loss function in Equation~\ref{eq:loss}, \texttt{CART} is prone to be overfitting, especially for software quality data~\cite{DBLP:journals/ese/KhoshgoftaarA01}. This exacerbates the issue of sample sparsity~\cite{DBLP:journals/corr/abs-2202-03354} under a small amount of data samples which is not uncommon for configurable software systems~\cite{DBLP:conf/icse/HaZ19,DBLP:conf/esem/ShuS0X20}. 


Instead, what we are interested in are the (branch and/or leaf) divisions generated therein (with respect to the training data), which enable us to use further dedicated and more focused local models for better generalizing to the new data (lines 7-12 in Algorithm~\ref{alg:dal-code}). As such, the final prediction is no longer a simple average while we do not care about the \texttt{CART} overfitting itself as long as it fits the training data well. This is similar to the case of unsupervised clustering for which the clustering is guided by implicit labels (via the loss function at Equation~\ref{eq:loss}). Specifically, in \Model~we extract the data samples according to the divisions made by the $d$th depth of the \texttt{CART}, including all the leaf divisions with depth smaller than $d$. An example can be seen from Figure~\ref{fig:DT_example}, where $d$ is a controllable parameter given by the software engineers. In this way, \Model~divides the data into a range of $[d+1,2^{d}]$ divisions\footnote{At depth $d$, there can be a minimum of $d$+1 divisions (where only one parent node at the $d-1$ depth is further divided into two nodes), and a maximum of $2^d$ divisions (where all parent nodes at the $d-1$ depth are further divided into two nodes). Hence, the range of the number of divisions is [d+1, $2^d$].} ($d \geq 0$), each of which will be captured by a local model learned thereafter\footnote{When $d=0$, \Model~is basically equivalent to a single local model.}. Note that when the number of data samples in the division is less than the minimum amount required by a model, we merge the two divisions of the same parent node.



As a concrete example, from Figure~\ref{fig:DT_example}, we see that there are two depths: when $d=1$ there would be two divisions (one branch and one leaf) with 10 and 8 samples respectively; similarly, when $d=2$ there would be three leaf divisions: two of each have 5 samples and one is the division with 8 samples from $d=1$ as it is a leaf. In this case, \texttt{CART} has detected that the \texttt{rtQuality} is a more important (binary) option to impact the performance, and hence it should be considered at a higher level in the tree. Note that for numeric options, e.g., \texttt{threads}, the threshold of splitting (\texttt{threads} $>5$) is also tuned as part of the training process of \texttt{CART}.

It is intuitive to understand that $d$ is such a critical parameter for \Model, thereby manually setting the right value for it can be time-consuming and error-prone. In what follows, we will delineate what role the $d$ plays in \Model~and propose an adaptive mechanism in \Model~that allows dynamic adaptation of the $d$ value during training. 


\subsubsection{The Role of Depth $d$ in \Model}
\label{subsubsec:play}

Since more divisions mean that the sample space is separated into more loosely related regions for dealing with the sample sparsity, one may expect that the accuracy will be improved, or at least, stay similar, thereby we should use the maximum possible $d$ from \texttt{CART} in the \textit{Dividing} phase. This, however, only exists in the ``utopia case'' where there is an infinite set of configuration data samples.

In essence, with the design of \Model, the depth $d$ will manage two conflicting objectives that influence its accuracy:

\begin{enumerate}
    \item greater ability to handle sample sparsity by separating the distant samples into divisions, each of which is learned by an isolated local model;
    \item and a larger amount of data samples in each division for the local model to be able to generalize.
\end{enumerate}

Clearly, a larger $d$ may benefit the ability to handle sample sparsity but it will inevitably reduce the data size per division for a local model to generalize since it is possible for \texttt{CART} to generate divisions with imbalanced sample sizes. From this perspective, we see $d$ as a value that controls the trade-off between the two objectives, and neither a too small nor too large $d$ would be ideal, as the former would lose the ability to deal with sample sparsity while the latter would leave too little data for a local model to learn, hence produce negative noises that harm the overall prediction. This is the key reason that setting $d$ for a given system is crucial when using \Model. 

\begin{figure}[!t]
  \centering
   \begin{subfigure}[t]{0.48\columnwidth}
        \centering
\includegraphics[width=\columnwidth]{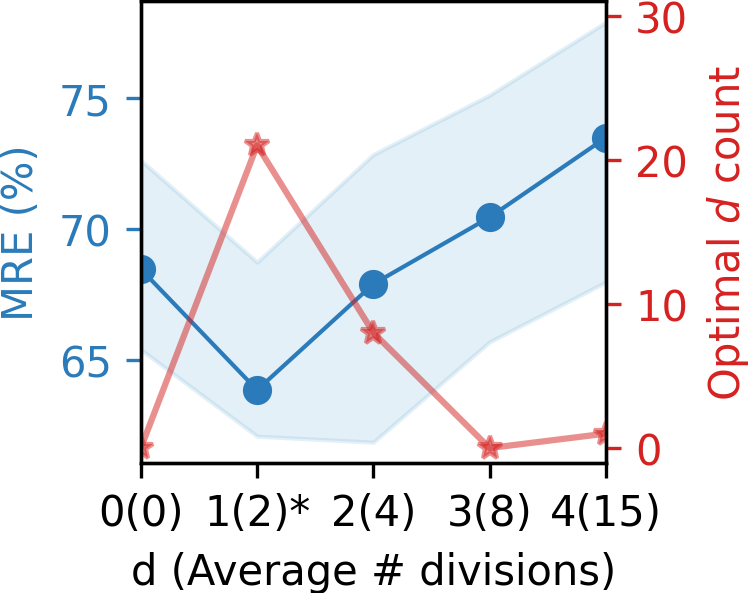}
    \subcaption{\textsc{SQLite}}
   \end{subfigure}
~\hfill
      \begin{subfigure}[t]{0.46\columnwidth}
        \centering
\includegraphics[width=\columnwidth]{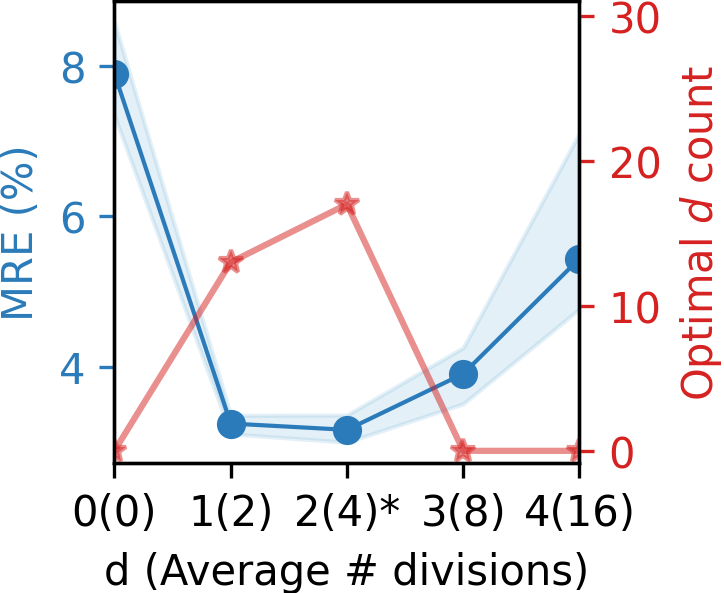}
 \subcaption{\textsc{Lrzip}}
   \end{subfigure}

    \caption{The changing optimal $d$ on \Model~depending on the software systems being modeled and the training/testing data across 30 runs.}
       \label{fig:optimal-d}
  \end{figure}

Unfortunately, finding the appropriate $d$ value is not straightforward, as it often requires repeatedly profiling different values following in a trial-and-error manner, which can be rather expensive, especially when the optimal $d$, which leads to the best mean relative error (MRE), varies depending on the system and even the training/testing data. For example, Figure~\ref{fig:optimal-d} illustrates the optimal $d$ for \Model~on two real-world software systems (the full coverage will be discussed in Section~\ref{subsec:sen}), from which it is clear that the  $d=1$ would lead to the best overall MRE for \textsc{SQLite}, but this becomes $d=2$ for \textsc{Lrzip}. If we consider the individual runs that involve different training and testing configuration data, the optimal $d$ could also differ (red lines). Notably, we see that some $d$ values, e.g., $d=4$ for \textsc{SQLite}, can be dramatically harmful to the accuracy, leading to a result that is even worse than the case when \Model~is absent (i.e., $d=0$).

\subsubsection{Adapting the Depth $d$}
\label{subsubsec:adapting_depth}

To overcome the above, in \Model, we design an adaptive mechanism as part of the \textit{Dividing} phase that is able to dynamically find the $d$ such that the two aforementioned objectives can be optimized and balanced (line 5 in Algorithm~\ref{alg:dal-code}).

Specifically, we use the following functions $h$ and $z$ to measure the ability to handle sample sparsity and the amount of information for a division $D_i$, respectively:
\begin{equation}
\begin{cases}
h(D_i)={1\over{|D_i|}}{\sum_{y_j \in D_i}{(y_j - \overline{y}_{D_i})}^2}\\
z(D_i)=-n_{D_i}\\
\end{cases}
\label{eq:adapt}
\end{equation}
whereby $z$ is the additive inverse of the assigned sample size, denoted as $n_{D_i}$. $h$ is basically the mean square error (or performance variance of the samples in a division) taken from the loss function (Equation~\ref{eq:loss}) that splits the divisions with respect to the important configuration options. As such, the samples in a division that are generally closer to each other in terms of the performance value, after being divided according to the importance of key configuration options, will more likely to be beneficial for a local model to learn. Intuitively, both $h$ and $z$ need to be minimized. Given a maximum number of $d_{max}$ divisions generated by \texttt{CART} under the training data, our purpose is to find the $d$ value ($0\leq d\leq d_{max}$) that leads to the overall best and most balanced divisions (a.k.a. knee points) in the objective space of $h$ and $z$ from those generated by all possible $d$ values. In essence, from a multi-objective optimization perspective, knee points represent the solutions that, when changed, can only marginally improve one objective by largely comprising the other~\cite{DBLP:journals/tosem/ChenLBY18,DBLP:journals/tec/YuMJDLZ22}, hence achieving a well-balanced trade-off.

To this end, we gain inspiration from a widely used quality indicator for evaluating multi-objective solution sets, namely Hypervolume (HV)~\cite{Zitzler1998}. In a nutshell, HV measures the volume between all nondominated points\footnote{A solution $\mathbf{\overline{a}}$ is dominated by $\mathbf{\overline{b}}$ if all objectives of $\mathbf{\overline{b}}$ are better or equivalent to those of $\mathbf{\overline{a}}$ while there is at least one objective of $\mathbf{\overline{b}}$ performs better than that of $\mathbf{\overline{a}}$. A solution is said nondominated if it cannot be dominated by any other solutions in the set.} in the objective space and a reference point (usually a nadir point); the larger the volume, the better convergence and diversity that the set achieves. The HV for a solution set $\mathbfcal{A}$ can be computed as:
\begin{equation}
	 \text{HV}(\mathbfcal{A},\mathbf{\overline{r}}) = \lambda(\bigcup_{\mathbf{\overline{a}}\in \mathbfcal{A}} \{\mathbf{\overline{x}}|\mathbf{\overline{a}} \prec \mathbf{\overline{x}} \prec \mathbf{\overline{r}}\})
	\label{eq:hv}
\end{equation}
where $\lambda$ is the Lebesgue measure\textcolor{black}{, which takes into account the trade-offs between the objectives and the spread of the non-dominated solutions in the objective space. It is a default metric in HV used to calculate the volume of the region covered by a set of non-dominated solutions therein. More details can be found in the work of Zitzler and Thiele~\cite{Zitzler1998}}; $\mathbf{\overline{r}}$ is the reference nadir point, which is often defined as 1.1 times the range of the nondominated set~\cite{li2020evaluate}. Li \textit{et al.}~\cite{li2020evaluate} show that, with an appropriate setting of the reference point, HV can well reflect the preference for balanced/knee solutions, which fits precisely with our needs. However, we cannot directly adopt the original HV due to the fact that it can completely omit the contributions of divisions based on their relative domination relations, i.e., a division does not contribute to the HV value at all if it is dominated by the other division under a certain $d$. Indeed, this makes sense in conventional multi-objective evaluation scenarios, but in our case of dividable learning for configuration performance with \Model, the contribution of each division counts, even if it has been dominated, since the local model under such a division could still impact the result when the newly given configuration falls therein. As a result, the original HV might misjudge the true effectiveness of a $d$ value.

\begin{figure*}[!t]
  \centering
  \hspace{-0.2cm}
   \begin{subfigure}[t]{0.19\textwidth}
        \centering
\includegraphics[width=\columnwidth]{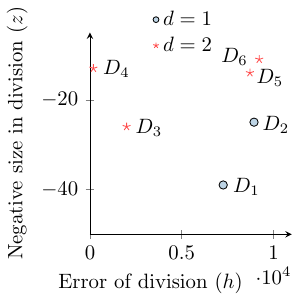}
    \subcaption{All divisions}
   \end{subfigure}
~\hfill
      \begin{subfigure}[t]{0.19\textwidth}
        \centering
\includegraphics[width=\columnwidth]{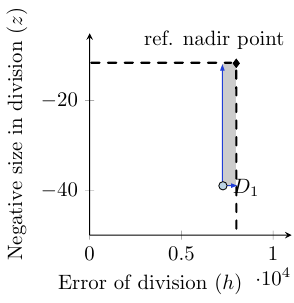}
 \subcaption{HV for $d=1$}
   \end{subfigure}
~\hfill
      \begin{subfigure}[t]{0.19\textwidth}
        \centering
\includegraphics[width=\columnwidth]{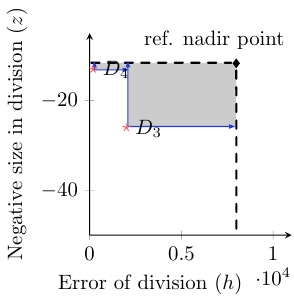}
  \subcaption{HV for $d=2$}
   \end{subfigure}
~\hfill
      \begin{subfigure}[t]{0.19\textwidth}
        \centering
\includegraphics[width=\columnwidth]{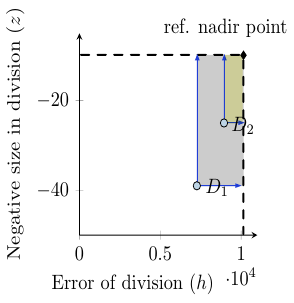}
     \subcaption{$\mu$HV for $d=1$}
   \end{subfigure}
~\hfill
      \begin{subfigure}[t]{0.19\textwidth}
        \centering
\includegraphics[width=\columnwidth]{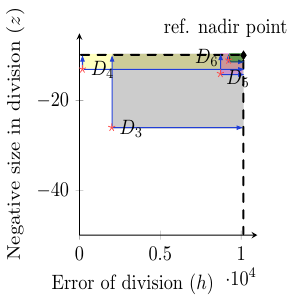}
    \subcaption{$\mu$HV for $d=2$}
   \end{subfigure}

    \caption{Comparing the differences between HV and $\mu$HV for divisions with $d=1$ and $d=2$ under system \textsc{x264}. The distinctly colored area indicates the individual HV value calculated based on one (for $\mu$HV) or more divisions (for HV). (a) shows that $D_1$ is the nondominated division for $d=1$ while $D_3$ and $D_4$ are the nondominated ones for $d=2$. In (b), the original HV value equals the area between $D_1$ alone and the reference nadir point, i.e., HV$=19825.43$. In (c), the original HV value equals the non-overlapped area from $D_3$ and $D_4$ to the reference nadir point, i.e., HV$=88025.51$. In (d), $\mu$HV is the mean over the HV value of the area between $D_1$ and the reference nadir point together with that of the area between $D_2$ and the reference nadir point, i.e., $\mu$HV$=(83951.55+18216.65) /2=51084.10$. In (e), $\mu$HV is the mean over the HV value for the area of each of the divisions $D_3$, $D_4$, $D_5$, and $D_6$, i.e., $\mu$HV$=(131212.91+30873.60+5862.67+1014.70)/4=42240.97$. With the original HV, $d=2$ would be chosen while using our proposed $\mu$HV, $d=1$ would be chosen. In fact, here $d=1$ tends to be better in general considering all the possible divisions in terms of $h$ and $z$. The actual validation also confirms that $d=1$ leads to $2.28\times$ more accurate result than that of $d=2$. }
       \label{fig:ahv-hv}
  \end{figure*}

Instead of leveraging the original HV, in this work, we propose a modified HV, dubbed averaging HV ($\mu$HV), that evaluates the average quality of the HV value that takes each division into account with respect to $h$ and $z$ regardless of the domination relations---a typical specialization of a generic concept to cater for our needs. Formally, the $\mu$HV for all the divisions in a set $\mathbfcal{D}$ under a particular $d$ value is calculated as below:
\begin{equation}
\begin{aligned}
&\mu \text{HV}(\mathbfcal{D},\mathbf{\overline{r}})={1\over{|\mathbfcal{D}|}}{\sum_{D_i \in \mathbfcal{D}}{(|h_r - h(D_i)|) \times (|z_r - z(D_i)|)}} \\ 
&\text{s.t. } \mathbf{\overline{r}}=\langle h_r,z_r\rangle
\end{aligned}
\label{eq:ahv}
\end{equation}
whereby $h_r$ and $z_r$ are the objective values for mean square error and data size of the reference nadir point $\mathbf{\overline{r}}$, respectively. Since there are only two objectives to consider in our case, each individual HV value within the $\mu$HV is essentially the area between the corresponding division and the reference nadir point in the objective space. With the above, we realize the following specializations to the original generic HV:
\begin{itemize}
    \item \textbf{Specialization 1:} Through averaging the HV value for each individual division's objectives, $\mu$HV would take into account the contribution made by a division that is dominated by the other. This, conversely, is not the case in the original HV as it works only on the nondominated ones.
    \item \textbf{Specialization 2:} Unlike the original HV that uses $1.1$ times the range of the nondominated divisions as the reference nadir point, $\mu$HV takes $1.1$ times the range of each of the worst objective values\footnote{This would be $0.9$ times for the objective $z$ as it is always a negative number.}. This is necessary to account for the divisions being dominated by others.
\end{itemize}

Considering the example illustrated in Figure~\ref{fig:ahv-hv}, in which it is possible to use $d=1$ or $d=2$ for \textsc{x264} (Figure~\ref{fig:ahv-hv}a). From Figures~\ref{fig:ahv-hv}b and~\ref{fig:ahv-hv}c, if the original HV is used directly, then we would have $\text{HV}_1=19825.43$ and $\text{HV}_2=88025.51$ for $d=1$ and $d=2$, respectively. Clearly, $d=2$ would be chosen here since $\text{HV}_1 < \text{HV}_2$. When using $\mu$HV, in Figures~\ref{fig:ahv-hv}d and~\ref{fig:ahv-hv}e, there are $\mu \text{HV}_1=51084.10$ and $\mu \text{HV}_2=42240.97$ for $d=1$ and $d=2$, respectively. As such, $d=1$ would be chosen since $\mu \text{HV}_1 > \mu \text{HV}_2$. In fact, $d=1$ is a better setting than $d=2$ as the latter contains $D_5$ and $D_6$, which are two much more inferior/less balanced divisions than those in $d=1$, leading to a higher probability of more severe negative impacts on the accuracy. Hence, $d=1$ tends to be more desired in general considering all the possible divisions in terms of $h$ and $z$. Indeed, the experiment validation confirms that it leads to $2.28\times$ more accurate results than that of $d=2$.

The above demonstrates the limitation of the original HV for adapting $d$ in dividable learning for configuration performance: because of the way it is computed, those divisions that are dominated by the others, e.g., $D_2$ for $d=1$ and $D_5$ for $d=2$, would have no contribution to the final value. This is a serious issue for our case given that the local model trained on those divisions is still important since \Model~naturally allows any local model to be used in isolation. For example, if upon prediction, there are newly given configurations belonging to the local models for $D_2$ or $D_5$, then these models would be used for predicting the performance independently. However, the original HV would have not been able to take that into account since the contributions of the corresponding divisions were ruled out, providing likely misleading guidance. $\mu$HV addresses such a shortcoming by averaging the individual HV value for all divisions under a $d$ value, hence taking the contribution of those being dominated by the other into account (\textbf{Specialization 1}). In this way, $\mu$HV can better assist in identifying the overall better $d$ value and its corresponding divisions with respect to $h$ and $z$. Yet, this requires changing the reference nadir point to consider not only the nondominated divisions (as in the original HV) but also anyone in the space (\textbf{Specialization 2}).

Algorithm~\ref{alg:adapt-code} illustrates the proposed adaptive mechanism with the following key steps:

\begin{enumerate}

    \item Compute the objective values of $h_i$ and $z_i$ for all division $D_i$ under the possible $d$. The set of divisions $\mathbfcal{D}_j$ and its corresponding $d_j$ are kept as a vector $\langle \mathbfcal{D}_j,d_j \rangle$.
    \item Calculate the $\mu$HV value for all the divisions under each $d$ up to the maximum $d$ possible as determined by \texttt{CART}. Here, $d=0$ is omitted since in such a case \Model~is essentially the same as the single local model.
    \item Return the $d$ with the largest $\mu$HV value.
\end{enumerate}

\textcolor{black}{With such an adaptive mechanism of $d$, we strike more balanced results for the aforementioned objectives while requiring neither additional training nor profiling.} The computational overhead of computing $\mu$HV is also ignorable, as we will show in Section~\ref{subsec:overhead}.

\begin{algorithm}[t!]
	\DontPrintSemicolon
	\footnotesize
	
	\caption{Pseudo code of \textsc{AdaptingDepth}}
	\label{alg:adapt-code}
	\KwIn{A trained CART model $\mathcal{T}$}
     \KwOut{The most balanced $d$ value}
      $d_j=1$\\
      
       \tcc{\textcolor{blue}{step 1.}}
     \While{$d_j\leq d_{max}$ from $\mathcal{T}$}{
      $d'=1$\\
      $\mathbfcal{D}_j\leftarrow \emptyset$\\
      \While{$d'\leq d_j$ from $\mathcal{T}$}{
          
           \If{$d'<d_j$}
           {
             $\mathbfcal{D}_j\leftarrow$ extract all the leaf divisions of samples from $\mathcal{T}$ at the $d'$th depth\\
           }
           \Else{
              $\mathbfcal{D}_j\leftarrow$ extract all divisions of samples from $\mathcal{T}$ at the $d'$th depth\\
           }
            $d'=d'+1$\\
      }

            \For{$\forall D_i \in \mathbfcal{D}_j$} {
             
            \If{$ \langle h_i,z_i \rangle \notin \langle \mathbf{\overline{h}},\mathbf{\overline{z}} \rangle$}{
            
            $\langle \mathbf{\overline{h}},\mathbf{\overline{z}} \rangle\leftarrow$ calculate $h_i$ and $z_i$ for $D_i$ according to Equation~\ref{eq:adapt}\\
            
            }

            }

            
                $\mathbfcal{N}\leftarrow \langle \mathbfcal{D}_j,d_j \rangle$\\
                  $d_j=d_j+1$\\
      }
      
        \tcc{\textcolor{blue}{step 2.}}
        $\mathbf{\overline{r}}=\langle h_r,z_r \rangle\leftarrow$ find the nadir reference point from $\langle \mathbf{\overline{h}},\mathbf{\overline{z}} \rangle$\\
        \For{$\forall \langle \mathbfcal{D}_j,d_j \rangle \in \mathbfcal{N}$} {
         $v_j=\mu \text{HV}(\mathbfcal{D}_j,\mathbf{\overline{r}})$ according to Equation~\ref{eq:ahv}\\
         $\mathbfcal{F}\leftarrow \langle v_j,d_j \rangle$\\
        }

      \tcc{\textcolor{blue}{step 3.}}
       \Return the $d_j$ with the largest $v_j$ such that $\langle v_j,d_j \rangle \in \mathbfcal{F}$\\

\end{algorithm}

\subsection{Training}

Given the divisions generated by the \textit{Dividing} phase, we train a local model for the samples from each division identified as part of \textbf{Goal 2} (lines 13-15 in Algorithm~\ref{alg:dal-code}). 
Theoretically, we can pair them with any model given the generic concept of dividable learning. However, we adopt the state-of-the-art regularized Hierarchical Interaction Neural Network~\cite{DBLP:journals/tosem/ChengGZ23} (namely \texttt{HINNPerf}) to serve as the default in this work, as Cheng \textit{et al.} provide compelling empirical evidence on its effectiveness for handling feature sparsity and the interaction therein well for configurable software, even with very small training sample size. Section~\ref{sec:evaluation} reports on our use and observation of pairing \Model~with \texttt{HINNPerf}.


In this work, we adopt exactly the same structure and training procedure as those used by Cheng \textit{et al.}~\cite{DBLP:journals/tosem/ChengGZ23}, hence we kindly refer interested readers to their work for the training details. Since the local models of the divisions are independent, we utilize parallel training as part of \Model. 

It is worth stressing that, as we will show in Section~\ref{subsec:local}, \Model~is a model-agnostic framework and can improve a wide range of local models compared with the case when they are used to learn the entire configuration data space.

\subsection{Predicting}

When a new configuration arrives for prediction, \Model~chooses a local model of division trained previously to infer its performance. Therefore, the question is: how to assign the new configuration to the right model (\textbf{Goal 3})? A naive solution is to directly feed the configuration into the \texttt{CART} from the \textit{Dividing} phase and check which divisions it associates with. Yet, since the performance of the new configuration is unforeseen from the \texttt{CART}'s training data, this solution requires \texttt{CART} to generalize accurately, which, as mentioned, can easily lead to poor results because \texttt{CART} is overfitting-prone when directly working on new data~\cite{DBLP:journals/ese/KhoshgoftaarA01}.


Instead, by using the divided samples from the \textit{Dividing} phase (which serves as pseudo labeled data), we train a Random Forest---a widely used classifier and is resilient to overfitting ~\cite{DBLP:conf/splc/ValovGC15,DBLP:conf/oopsla/QueirozBC16,DBLP:conf/icse/0003XC021}---to generalize the decision boundary and predict which division that the new configuration should be better assigned to (lines 16-22 in Algorithm~\ref{alg:dal-code}). Again, in this way, we are less concerned about the overfitting issue of \texttt{CART} as long as it matches the patterns of training data well. This now becomes a typical classification problem but there are only pseudo labels to be used in the training. Using the example from Figure~\ref{fig:DT_example} again, if $d=1$ is chosen as the best value then the configurations in the 10 sample set would have a label \textit{``division1''}; similarly, those in the 8 sample set would result in a label \textit{``division2''}.

However, one issue we experienced is that, even with $d=1$, the sample size of the two divisions can be rather imbalanced, which severely harms the quality of the classifier trained. For example, when training \textsc{BDB-C} with 18 samples, the first split in \texttt{CART} can lead to two divisions with 14 and 4 samples, respectively.
Therefore, before training the classifier we use Synthetic Minority Oversampling Technique (SMOTE)~\cite{DBLP:journals/bmcbi/BlagusL13a} to pre-process the pseudo label data, hence the division(s) with much fewer data (minority) can be more repeatedly sampled.


Finally, the classifier predicts a division whose local model would infer the performance of the new configuration.


\section{Experiment Setup}
\label{sec:setup}

Here, we delineate the settings of our evaluation. In this work, \Model~is implemented based on \texttt{Tensorflow} and \texttt{scikit-learn}. All experiments were carried out on a server with 64-core Intel Xeon 2.6GHz and 256G DDR RAM.


\subsection{Research Questions}

In this work, we comprehensively assess \Model~by answering the following research questions (RQ):

\begin{itemize}
    \item \textbf{RQ1:} How accurate is \Model~compared with the state-of-the-art approaches (i.e., \texttt{HINNPerf}, \texttt{DeepPerf}, \texttt{Perf-AL}, \texttt{DECART}, and \texttt{SPLConqueror}), \textcolor{black}{and other models from the machine learning community that share a similar concept (i.e., \texttt{IBMB}, \texttt{M5}, \texttt{PILOT}, and \texttt{MOB})}, for configuration performance learning?

    \item \textbf{RQ2:} To what extent \Model~can improve different generic models (i.e., \texttt{HINNPerf}, regularized Deep Neural Network, \texttt{CART}, Linear Regression, and Support Vector Regression) when they are used locally therein for predicting configuration performance?

    \item \textbf{RQ3:} How do \Model~perform compared with the existing ensemble approaches such as \texttt{Bagging} and \texttt{Boosting}?

    \item \textbf{RQ4:} What is the benefit of the components in \Model? This consists of three sub-questions:

    \begin{itemize}
        \item \textbf{RQ4.1:} What is the benefit of using \texttt{CART} for dividing in \Model~over the standard clustering algorithms?
        \item \textbf{RQ4.2:} What is the benefit of the mechanism that adapts $d$ compared with the variant that relies on pre-tuned fixed $d$ as what was used in our previous FSE work?
        \item \textbf{RQ4.3:} What is the benefit of $\mu$HV over the original HV in adapting $d$?
    \end{itemize}
    

    \item \textbf{RQ5:} What is the sensitivity of \Model~to a fixed $d$ and how well does the adaptive mechanism perform in finding the optimal $d$ for each individual run?
    
    \item \textbf{RQ6:} What is the model building time for \Model?

\end{itemize}

We ask \textbf{RQ1} to assess the effectiveness of \Model~under different sample sizes against the state-of-the-art approaches \textcolor{black}{and other models that share a similar concept}. Since \Model~is naturally model-agnostic and can be paired with different local models, we study \textbf{RQ2} to examine how the concept of dividable learning can benefit any given local model. Since \Model~is similar to the ensemble approaches, e.g., \texttt{Bagging}~\cite{breiman1996bagging} and \texttt{Boosting}~\cite{schapire2003boosting}, with \textbf{RQ3}, we seek to examine how it performs against those. \textbf{RQ4} is mainly an ablation analysis of \Model, consisting of confirming the necessity of using \texttt{CART} to determine the divisions over unsupervised clustering algorithms; verifying the effectiveness of adapting $d$ compared with its counterpart under a fixed best $d$ value obtained via trial-and-error; and assessing the benefit of using the newly proposed $\mu$HV in quantifying the usefulness of the divisions in a $d$ value. In \textbf{RQ5}, we examine, under an extensively increased training sample size, how the depth of division ($d$) can impact the performance of \Model~and how well the mechanism that adapts $d$ can locate the optimal setting in individual runs, which is the ground truth. Finally, we examine the overall overhead of \Model~in \textbf{RQ6}.

\subsection{Subject Systems and Sample Sizes}
\label{subsec:subject_system}

Leveraging the criteria and procedure mentioned in Section~\ref{sec:emprical}, we use the same datasets of all valid configurations from real-world systems as widely used in the literature~\cite{DBLP:conf/esem/ShuS0X20,DBLP:conf/icse/HaZ19,DBLP:conf/sigsoft/SiegmundGAK15, DBLP:journals/ese/GuoYSASVCWY18, DBLP:journals/corr/abs-2106-02716}. The 12 configurable software systems studied in this work are specified in Table~\ref{tb:subject-system}. As can be seen, these software systems come with diverse domains, scales, and performance concerns. Some of them contain only binary configuration options (e.g., \textsc{x264}) while the others involve mixed options (binary and numeric), e.g., \textsc{HSMGP}, leading to configuration data that can be more difficult to model and generalize~\cite{DBLP:conf/icse/HaZ19}. Note that the key performance-related configuration options have been identified in existing studies, hence although some systems have mixed options (e.g., \textsc{Apache}), they are still binary in the dataset as those are key options that can influence the performance.


The configuration data of all the systems are collected by prior studies using the standard benchmarks with repeated measurements~\cite{DBLP:conf/esem/ShuS0X20,DBLP:conf/icse/HaZ19,DBLP:conf/sigsoft/SiegmundGAK15, DBLP:journals/ese/GuoYSASVCWY18, DBLP:journals/corr/abs-2106-02716}. For example, the configurations of \textsc{Apache}---a popular Web server---are benchmarked using the tools \texttt{Autobench} and \texttt{Httperf}, where workloads are generated and increased until reaching the point before the server crashes, and then the maximum load is marked as the performance value~\cite{DBLP:journals/ese/GuoYSASVCWY18}. The process repeats a few times for each configuration to ensure reliability. When there exist multiple measured datasets for the same software system, we use the one with the largest size.


To ensure the generalizability of the results, for each system, we follow the protocol used by existing work~\cite{DBLP:journals/sqj/SiegmundRKKAS12,DBLP:conf/icse/HaZ19,DBLP:conf/esem/ShuS0X20} to obtain five sets of training sample size in the evaluation: 

\begin{itemize}
    \item \textbf{Binary systems:} We randomly sample $n$, $2n$, $3n$, $4n$, and $5n$ configurations and their measurements, where $n$ is the number of configuration options~\cite{DBLP:conf/icse/HaZ19,DBLP:conf/esem/ShuS0X20}.
    \item \textbf{Mixed systems:} We leverage the sizes suggested by \texttt{SPLConqueror}~\cite{DBLP:journals/sqj/SiegmundRKKAS12} (a state-of-the-art approach) depending on the amount of budget. 
    
\end{itemize}

The results have been illustrated in Table~\ref{tb:sizes}. All the remaining samples in the dataset are used for testing.

 \begin{table}[t!]

\centering
\footnotesize
\caption{Details of the subject systems. ($|\mathbfcal{B}|$/$|\mathbfcal{N}|$) denotes the number of binary/numerical options, and $|\mathbfcal{C}|$ denotes the number of valid configurations (full sample size).}
\begin{adjustbox}{width=\columnwidth,center}
\footnotesize
\setlength{\tabcolsep}{1mm}
\begin{tabular}{llllrl}
\toprule
\textbf{System} & \textbf{$|\mathbfcal{B}|$/$|\mathbfcal{N}|$} & \textbf{Performance} & \textbf{Description} & \textbf{$|\mathbfcal{C}|$} & \textbf{Used by}\\ 
\midrule
\textsc{Apache} & 9/0  & Maximum load & Web server              & 192 & \cite{DBLP:conf/icse/HaZ19,DBLP:journals/ese/GuoYSASVCWY18,DBLP:conf/esem/ShuS0X20} \\
\textsc{BDB-C}  & 16/0 & Latency (ms) & Database (C)    & 2560 & \cite{DBLP:conf/icse/HaZ19,DBLP:journals/ese/GuoYSASVCWY18,DBLP:conf/esem/ShuS0X20} \\
\textsc{BDB-J}  & 26/0 & Latency (ms) & Database (Java) & 180 & \cite{DBLP:conf/icse/HaZ19,DBLP:journals/ese/GuoYSASVCWY18,DBLP:conf/esem/ShuS0X20} \\
\textsc{kanzi}  & 31/0 & Energy (kj) & Compression tool & 3202 & \cite{DBLP:conf/icse/WeberKSAS23, 10172849} \\
\textsc{SQLite}  & 14/0 & Runtime (ms) & Database & 1000 & \cite{DBLP:journals/tse/KrishnaNJM21, DBLP:conf/icse/HaZ19, DBLP:journals/jss/CaoBWZLZ23} \\
\textsc{x264}   & 16/0 & Runtime (ms)  & Video encoder & 1152 & \cite{DBLP:conf/icse/HaZ19,DBLP:journals/ese/GuoYSASVCWY18,DBLP:conf/esem/ShuS0X20} \\
\textsc{Dune}   & 8/3 & Runtime (ms)  & Multi-grid solver           & 2304 & \cite{DBLP:conf/icse/HaZ19,DBLP:journals/ese/GuoYSASVCWY18,DBLP:conf/esem/ShuS0X20} \\
\textsc{HIPA$^{cc}$}   & 31/2 & Runtime (ms) & Image processing                & 13485 & \cite{DBLP:conf/icse/HaZ19,DBLP:conf/sigsoft/SiegmundGAK15} \\
\textsc{HSMGP}   & 11/3 & Runtime (ms) & Stencil-grid solver                & 3456 & \cite{DBLP:conf/icse/HaZ19,DBLP:conf/sigsoft/SiegmundGAK15} \\
\textsc{Lrzip}   & 9/3 & Runtime (ms) & Compression tool                & 5184 & \cite{DBLP:journals/corr/abs-2106-02716, DBLP:conf/icse/WeberKSAS23, 10172849} \\
\textsc{nginx}  & 12/2 & Runtime (ms) & Web server & 4416 & \cite{DBLP:conf/icse/WeberKSAS23, DBLP:conf/sigsoft/WangWHSSZFLSJ22} \\
\textsc{VP8}   & 9/4 & Runtime (ms) & Video encoder                & 2736 & \cite{DBLP:journals/corr/abs-2106-02716, DBLP:conf/icse/WeberKSAS23} \\
\bottomrule
\end{tabular}
\end{adjustbox}
\label{tb:subject-system}
\end{table}

\begin{table}[t!]
\caption{The training sample sizes used. $n$ denotes the number of configuration options in a binary system.}
\centering
\begin{adjustbox}{width=\columnwidth,center}

\begin{tabular}{lrrrrr}
\toprule
\textbf{System} & \textbf{Size 1 ($S_1$)} & \textbf{Size 2 ($S_2$)} & \textbf{Size 3 ($S_3$)} & \textbf{Size 4 ($S_4$)} & \textbf{Size 5 ($S_5$)} \\ 
\midrule
\textsc{Apache} & $n$  & $2n$ & $3n$  & $4n$ & $5n$\\
\textsc{BDB-C}  & $n$  & $2n$ & $3n$  & $4n$ & $5n$\\
\textsc{BDB-J}  & $n$  & $2n$ & $3n$  & $4n$ & $5n$\\
\textsc{kanzi}  & $n$  & $2n$ & $3n$  & $4n$ & $5n$\\
\textsc{SQLite}  & $n$  & $2n$ & $3n$  & $4n$ & $5n$\\
\textsc{x264}   & $n$  & $2n$ & $3n$  & $4n$ & $5n$\\
\textsc{Dune} & 224 & 692 & 1000 & 1365 & 1612\\
\textsc{HIPA$^{cc}$}  & 261 & 528 & 736 & 1281 & 2631\\
\textsc{HSMGP} & 77 & 173 & 384 & 480 & 864\\
\textsc{Lrzip} & 127 & 295 & 386 & 485 & 907\\
\textsc{nginx}  & 228  & 468 & 814  & 1012 & 1352\\
\textsc{VP8} & 121 & 273 & 356 & 467 & 830 \\
\bottomrule
\end{tabular}

\end{adjustbox}
\label{tb:sizes}
\end{table}

\subsection{Metric and Statistical Validation}

\subsubsection{Accuracy}

For all the experiments, mean relative error (MRE) is used as the evaluation metric for prediction accuracy, since it provides an intuitive indication of the error and has been widely used in the domain of software performance prediction~\cite{DBLP:conf/icse/HaZ19, DBLP:conf/esem/ShuS0X20, DBLP:journals/ese/GuoYSASVCWY18}. Formally, the MRE is computed as:
\begin{equation}
     MRE = {{1} \over {k}} \times {\sum^k_{t=1} {{|A_t - P_t|} \over {A_t}}} \times 100\%
\end{equation}
\noindent whereby $A_t$ and $P_t$ denote the $t$th actual and predicted performance, respectively. To mitigate bias, all experiments are repeated for 30 runs via bootstrapping without replacement.

\subsubsection{Statistical Test}




    
    




Since our evaluation commonly involves comparing more than two approaches, we apply Scott-Knott test~\cite{DBLP:journals/tse/MittasA13} to evaluate their statistical significance on the difference of MRE over 30 runs, as recommended by Mittas and Angelis~\cite{DBLP:journals/tse/MittasA13}. In a nutshell, Scott-Knott sorts the list of treatments (the approaches that model the system) by their median values of the MRE. Next, it splits the list into two sub-lists with the largest expected difference~\cite{xia2018hyperparameter}. For example, suppose that we compare $A$, $B$, and $C$, a possible split could be $\{A, B\}$, $\{C\}$, with the rank ($r$) of 1 and 2, respectively. This means that, in the statistical sense, $A$ and $B$ perform similarly, but they are significantly better than $C$. Formally, Scott-Knott test aims to find the best split by maximizing the difference $\Delta$ in the expected mean before and after each split:
\begin{equation}
    \Delta = \frac{|l_1|}{|l|}(\overline{l_1} - \overline{l})^2 + \frac{|l_2|}{|l|}(\overline{l_2} - \overline{l})^2
\end{equation}
whereby $|l_1|$ and $|l_2|$ are the sizes of two sub-lists ($l_1$ and $l_2$) from list $l$ with a size $|l|$. $\overline{l_1}$, $\overline{l_2}$, and $\overline{l}$ denote their mean MRE.

During the splitting, we apply a statistical hypothesis test $H$ to check if $l_1$ and $l_2$ are significantly different. This is done by using bootstrapping and $\hat{A}_{12}$~\cite{Vargha2000ACA}. If that is the case, Scott-Knott recurses on the splits. In other words, we divide the approaches into different sub-lists if both bootstrap sampling and effect size test suggest that a split is statistically significant (with a confidence level of 99\%) and with a good effect $\hat{A}_{12} \geq 0.6$. The sub-lists are then ranked based on their mean MRE. 



\begin{table*}[t!]
\caption{The median and interquartile range of MRE, denoted as Med-IQR, for \Model~and the state-of-the-art approaches for all the subject systems and training sizes over 30 runs. For each case, \setlength{\fboxsep}{1.5pt}\colorbox{green!20}{green cells} mean \Model~has the best median MRE; or \setlength{\fboxsep}{1.5pt}\colorbox{red!20}{red cells} otherwise. The one(s) with the best rank ($r$) from the Scott-Knott test is highlighted in bold.}
\resizebox{\textwidth}{!}{ 
\begin{tabular}{p{0.9cm}lp{0.1cm}p{1.8cm}|p{0.1cm}p{1.8cm}|p{0.1cm}p{1.8cm}|p{0.1cm}p{2.1cm}|p{0.1cm}p{1.8cm}|p{0.1cm}p{1.95cm}|p{0.1cm}p{1.8cm}|p{0.1cm}p{1.8cm}|p{0.1cm}p{1.96cm}|p{0.1cm}p{1.95cm}} \toprule
\multirow{2}{*}{System} & \multicolumn{1}{c}{\multirow{2}{*}{Size}} & \multicolumn{2}{c}{\texttt{DaL}} & \multicolumn{2}{c}{\texttt{HINNPerf}} & \multicolumn{2}{c}{\texttt{DeepPerf}} & \multicolumn{2}{c}{\texttt{Perf-AL}} & \multicolumn{2}{c}{\texttt{DECART}} & \multicolumn{2}{c}{\texttt{SPLConqueror}} & \multicolumn{2}{c}{\texttt{IBMB}} & \multicolumn{2}{c}{\texttt{M5}} & \multicolumn{2}{c}{\texttt{PILOT}} & \multicolumn{2}{c}{\texttt{MOB}} \\ \cline{3-22}
 & \multicolumn{1}{c}{} & $r$ & Med (IQR) & $r$ & Med (IQR) & $r$ & Med (IQR) & $r$ & Med (IQR) & $r$ & Med (IQR) & $r$ & Med (IQR) & $r$ & Med (IQR) & $r$ & Med (IQR) & $r$ & Med (IQR) & $r$ & Med (IQR) \\ \hline
\multirow{5}{*}{\textsc{Apache}} & $S_1$ & 3 & 21.86 (7.36) & \cellcolor{red!20}3 & \cellcolor{red!20}19.44 (10.53) & 2 & 20.19 (6.34) & 8 & 33.57 (11.34) & \cellcolor{red!20}\textbf{1} & \cellcolor{red!20}\textbf{19.44 (6.48)} & 7 & 28.50 (11.14) & 4 & 25.09 (4.77) & 5 & 26.14 (4.51) & 7 & 28.50 (11.14) & 6 & 27.92 (8.28) \\
 & $S_2$ & 6 & 12.89 (7.87) & \cellcolor{red!20}\textbf{1} & \cellcolor{red!20}\textbf{8.93 (3.89)} & 2 & 9.79 (4.56) & 8 & 32.97 (6.24) & 3 & 9.77 (5.36) & 5 & 18.33 (2.96) & 6 & 25.62 (2.37) & 6 & 23.91 (3.24) & 4 & 17.61 (2.41) & 7 & 28.88 (3.88) \\
 & $S_3$ & \cellcolor{green!20}\textbf{1} & \cellcolor{green!20}\textbf{6.56 (1.92)} & \textbf{1} & \textbf{7.01 (1.74)} & 2 & 7.98 (1.91) & 7 & 31.73 (4.66) & 2 & 8.07 (1.05) & 3 & 16.16 (1.41) & 5 & 25.92 (2.57) & 4 & 23.53 (2.09) & 3 & 16.16 (1.41) & 6 & 30.11 (3.62) \\
 & $S_4$ & 2 & 6.87 (1.95) & \cellcolor{red!20}\textbf{1} & \cellcolor{red!20}\textbf{6.58 (1.04)} & 2 & 6.85 (1.60) & 7 & 30.67 (6.84) & 3 & 7.47 (0.72) & 4 & 16.12 (0.78) & 6 & 25.90 (3.04) & 5 & 23.14 (0.81) & 4 & 16.12 (0.78) & 7 & 30.30 (4.08) \\
 & $S_5$ & \cellcolor{green!20}\textbf{1} & \cellcolor{green!20}\textbf{5.78 (1.08)} & \textbf{1} & \textbf{5.85 (0.99)} & 2 & 6.66 (1.61) & 7 & 30.45 (4.91) & 3 & 7.14 (0.89) & 4 & 16.12 (1.37) & 6 & 26.55 (2.26) & 5 & 23.04 (0.82) & 4 & 16.12 (1.37) & 7 & 30.79 (3.13) \\ \hline
\multirow{5}{*}{\textsc{BDB-C}} & $S_1$ & 4 & 83.60 (208.75) & 4 & 167.55 (144.47) & \cellcolor{red!20}\textbf{1} & \cellcolor{red!20}\textbf{43.66 (42.88)} & 3 & 82.38 (57.27) & 2 & 51.88 (45.91) & 7 & 772.44 (576.45) & 5 & 389.74 (289.18) & 6 & 505.90 (163.69) & 7 & 772.44 (579.59) & 6 & 539.31 (358.57) \\
 & $S_2$ & 3 & 21.92 (17.91) & 2 & 21.45 (13.61) & 2 & 18.17 (8.11) & 4 & 73.99 (22.51) & \cellcolor{red!20}\textbf{1} & \cellcolor{red!20}\textbf{13.37 (7.25)} & 6 & 526.69 (131.89) & 5 & 438.62 (165.58) & 6 & 541.59 (179.45) & 6 & 526.69 (131.89) & 7 & 626.96 (187.51) \\
 & $S_3$ & 2 & 8.31 (7.52) & \cellcolor{red!20}\textbf{1} & \cellcolor{red!20}\textbf{6.50 (9.20)} & 3 & 12.47 (4.08) & 3 & 69.74 (4.33) & \textbf{1} & \textbf{7.56 (5.07)} & 5 & 465.51 (88.42) & 4 & 437.64 (109.19) & 6 & 498.86 (126.41) & 5 & 465.51 (89.84) & 7 & 632.90 (112.44) \\
 & $S_4$ & \cellcolor{green!20}2 & \cellcolor{green!20}3.60 (4.50) & \textbf{1} & \textbf{4.86 (2.18)} & 2 & 9.06 (9.76) & 3 & 69.50 (2.49) & \textbf{1} & \textbf{5.17 (5.07)} & 5 & 438.47 (85.57) & 4 & 399.19 (152.10) & 6 & 503.49 (109.97) & 5 & 443.62 (86.20) & 7 & 594.25 (164.09) \\
 & $S_5$ & \cellcolor{green!20}\textbf{1} & \cellcolor{green!20}\textbf{1.88 (1.24)} & 3 & 4.04 (1.24) & 4 & 7.08 (8.03) & 5 & 69.29 (0.69) & 2 & 3.54 (1.92) & 7 & 461.53 (85.82) & 6 & 418.43 (143.13) & 8 & 486.14 (76.71) & 7 & 457.33 (85.98) & 9 & 618.53 (147.33) \\ \hline
\multirow{5}{*}{\textsc{BDB-J}} & $S_1$ & 3 & 5.37 (4.74) & 2 & 3.58 (1.74) & 3 & 2.98 (3.34) & 5 & 37.45 (2.45) & \cellcolor{red!20}\textbf{1} & \cellcolor{red!20}\textbf{2.36 (0.75)} & 5 & 43.23 (5.84) & 4 & 33.26 (3.52) & 4 & 33.71 (7.73) & 5 & 43.23 (6.80) & 6 & 51.89 (16.34) \\
 & $S_2$ & \cellcolor{green!20}\textbf{1} & \cellcolor{green!20}\textbf{1.83 (0.45)} & \textbf{1} & \textbf{1.96 (0.36)} & 2 & 1.91 (0.68) & 5 & 37.90 (1.67) & \textbf{1} & \textbf{1.84 (0.17)} & 5 & 38.30 (5.35) & 4 & 32.86 (3.31) & 3 & 32.51 (5.51) & 5 & 38.35 (6.13) & 6 & 49.16 (10.89) \\
 & $S_3$ & \cellcolor{green!20}\textbf{1} & \cellcolor{green!20}\textbf{1.58 (0.26)} & 2 & 1.75 (0.41) & 3 & 2.01 (1.10) & 6 & 37.19 (1.70) & \textbf{1} & \textbf{1.64 (0.24)} & 7 & 38.16 (3.88) & 4 & 32.04 (1.20) & 5 & 32.27 (3.41) & 7 & 38.20 (3.75) & 8 & 49.72 (5.45) \\
 & $S_4$ & \cellcolor{green!20}\textbf{1} & \cellcolor{green!20}\textbf{1.45 (0.22)} & 2 & 1.52 (0.24) & 3 & 1.70 (0.37) & 6 & 37.74 (2.59) & \textbf{1} & \textbf{1.47 (0.12)} & 6 & 36.88 (2.95) & 4 & 31.83 (2.75) & 5 & 31.73 (4.46) & 6 & 37.00 (3.58) & 7 & 50.87 (6.77) \\
 & $S_5$ & \textbf{1} & \textbf{1.40 (0.34)} & 2 & 1.48 (0.32) & 3 & 1.69 (0.48) & 6 & 35.76 (4.47) & \cellcolor{red!20}\textbf{1} & \cellcolor{red!20}\textbf{1.37 (0.26)} & 7 & 37.23 (3.13) & 4 & 32.07 (3.14) & 5 & 31.81 (4.87) & 7 & 37.14 (2.97) & 8 & 51.11 (5.46) \\ \hline
\multirow{5}{*}{\textsc{kanzi}} & $S_1$ & \cellcolor{green!20}\textbf{1} & \cellcolor{green!20}\textbf{196.15 (213.73)} & 2 & 262.27 (175.07) & \textbf{1} & \textbf{196.36 (122.61)} & 5 & 1434.70 (1328.18) & 2 & 276.50 (272.43) & 6 & 2528.40 (662.30) & 2 & 216.42 (270.56) & 3 & 718.11 (341.73) & 7 & 3106.10 (1285.5) & 4 & 898.25 (391.69) \\
 & $S_2$ & 2 & 95.31 (51.34) & \cellcolor{red!20}\textbf{1} & \cellcolor{red!20}\textbf{80.53 (51.36)} & 3 & 107.12 (43.75) & 7 & 961.11 (877.82) & 3 & 126.50 (142.62) & 8 & 1197.63 (349.62) & 4 & 243.08 (214.68) & 5 & 700.77 (211.53) & 8 & 1198.90 (354.37) & 6 & 970.59 (585.52) \\
 & $S_3$ & \cellcolor{green!20}2 & \cellcolor{green!20}59.30 (38.47) & \textbf{1} & \textbf{62.20 (14.07)} & 2 & 68.62 (30.41) & 5 & 280.93 (147.45) & 3 & 74.08 (32.14) & 7 & 1059.46 (166.51) & 4 & 208.36 (88.01) & 6 & 782.13 (180.76) & 7 & 1059.46 (166.90) & 7 & 1002.12 (392.28) \\
 & $S_4$ & \cellcolor{green!20}\textbf{1} & \cellcolor{green!20}\textbf{42.93 (18.75)} & \textbf{1} & \textbf{46.48 (19.58)} & 2 & 62.65 (26.54) & 5 & 266.53 (179.59) & 3 & 67.83 (15.79) & 7 & 1030.40 (166.98) & 4 & 160.99 (57.60) & 6 & 762.99 (208.45) & 7 & 1030.62 (163.83) & 7 & 1028.27 (269.47) \\
 & $S_5$ & \textbf{1} & \textbf{35.29 (17.13)} & \cellcolor{red!20}\textbf{1} & \cellcolor{red!20}\textbf{34.27 (11.87)} & 2 & 49.42 (9.99) & 4 & 169.12 (64.60) & 3 & 67.27 (17.95) & 7 & 1012.03 (261.66) & 5 & 211.48 (104.22) & 6 & 749.61 (119.62) & 7 & 1011.45 (264.24) & 7 & 942.95 (295.12) \\ \hline
\multirow{5}{*}{\textsc{SQLite}} & $S_1$ & \textbf{1} & \textbf{73.52 (20.40)} & \cellcolor{red!20}\textbf{1} & \cellcolor{red!20}\textbf{70.76 (27.86)} & 2 & 79.49 (25.36) & 6 & 573.26 (340.38) & \textbf{1} & \textbf{74.14 (17.98)} & 5 & 157.09 (70.92) & 6 & 199.96 (242.69) & 3 & 101.24 (26.56) & 4 & 151.86 (65.00) & 2 & 87.23 (2.08) \\
 & $S_2$ & \textbf{1} & \textbf{76.41 (16.66)} & 2 & 76.69 (23.32) & 3 & 77.75 (22.47) & 6 & 386.74 (323.30) & \cellcolor{red!20}2 & \cellcolor{red!20}76.20 (22.74) & 4 & 84.41 (23.00) & 4 & 81.65 (21.46) & 5 & 115.69 (29.35) & 4 & 84.41 (23.67) & 3 & 86.26 (1.95) \\
 & $S_3$ & \textbf{1} & \textbf{71.63 (11.58)} & \cellcolor{red!20}\textbf{1} & \cellcolor{red!20}\textbf{71.02 (17.30)} & \textbf{1} & \textbf{75.01 (15.58)} & 5 & 424.32 (311.00) & \textbf{1} & \textbf{74.16 (20.12)} & 2 & 75.72 (24.82) & 2 & 72.53 (21.01) & 4 & 116.19 (29.50) & 2 & 75.72 (24.83) & 3 & 85.79 (1.46) \\
 & $S_4$ & \cellcolor{green!20}\textbf{1} & \cellcolor{green!20}\textbf{66.94 (10.61)} & 2 & 70.84 (10.38) & 3 & 72.45 (12.11) & 6 & 363.18 (415.63) & 2 & 71.57 (14.72) & 3 & 69.61 (13.23) & 2 & 68.25 (13.72) & 5 & 120.45 (23.67) & 3 & 69.61 (13.80) & 4 & 85.61 (1.13) \\
 & $S_5$ & \cellcolor{green!20}\textbf{1} & \cellcolor{green!20}\textbf{64.20 (13.91)} & 2 & 70.15 (9.71) & 3 & 72.87 (8.86) & 6 & 212.19 (295.12) & 2 & 68.51 (8.54) & 2 & 67.78 (9.68) & 2 & 66.01 (10.76) & 5 & 122.04 (17.73) & 2 & 67.78 (9.68) & 4 & 85.43 (1.26) \\ \hline
\multirow{5}{*}{\textsc{x264}} & $S_1$ & 3 & 10.21 (4.35) & 2 & 8.44 (2.69) & \cellcolor{red!20}\textbf{1} & \cellcolor{red!20}\textbf{8.04 (3.06)} & 7 & 37.00 (9.43) & 2 & 9.27 (2.11) & 4 & 12.70 (4.19) & 5 & 32.14 (0.49) & 5 & 32.21 (0.36) & 4 & 12.70 (4.19) & 6 & 34.59 (3.08) \\
 & $S_2$ & \cellcolor{green!20}2 & \cellcolor{green!20}2.72 (1.52) & \textbf{1} & \textbf{2.77 (1.07)} & 2 & 3.17 (1.00) & 8 & 34.98 (7.81) & 3 & 6.29 (1.72) & 4 & 8.01 (1.19) & 6 & 32.06 (0.76) & 5 & 31.67 (0.26) & 4 & 7.98 (1.32) & 7 & 34.43 (2.82) \\
 & $S_3$ & 2 & 1.57 (1.52) & \cellcolor{red!20}\textbf{1} & \cellcolor{red!20}\textbf{1.37 (0.46)} & 2 & 2.23 (0.90) & 8 & 36.25 (7.73) & 3 & 4.57 (1.03) & 4 & 7.61 (0.81) & 6 & 32.06 (0.40) & 5 & 31.54 (0.13) & 4 & 7.61 (0.88) & 7 & 34.57 (2.06) \\
 & $S_4$ & 2 & 1.30 (1.17) & \cellcolor{red!20}\textbf{1} & \cellcolor{red!20}\textbf{0.81 (0.28)} & 3 & 1.74 (0.98) & 8 & 36.41 (6.47) & 4 & 3.66 (1.42) & 5 & 7.31 (0.49) & 7 & 32.10 (0.36) & 6 & 31.51 (0.16) & 5 & 7.34 (0.45) & 8 & 34.67 (1.64) \\
 & $S_5$ & 2 & 0.96 (0.79) & \cellcolor{red!20}\textbf{1} & \cellcolor{red!20}\textbf{0.58 (0.15)} & 3 & 1.44 (0.90) & 8 & 36.20 (3.75) & 4 & 2.15 (1.05) & 5 & 7.11 (0.60) & 7 & 32.04 (0.30) & 6 & 31.50 (0.11) & 5 & 7.11 (0.60) & 8 & 34.37 (1.63) \\ \hline
\multirow{5}{*}{\textsc{Dune}} & $S_1$ & 3 & 9.37 (1.15) & 2 & 7.98 (0.98) & \cellcolor{red!20}\textbf{1} & \cellcolor{red!20}\textbf{7.86 (0.93)} & 8 & 55.73 (0.12) & 9 & \texttimes & 4 & 13.03 (1.74) & 6 & 20.42 (2.38) & 7 & 22.71 (0.85) & 4 & 13.12 (1.91) & 5 & 18.39 (0.70) \\
 & $S_2$ & \cellcolor{green!20}\textbf{1} & \cellcolor{green!20}\textbf{5.95 (0.54)} & 2 & 6.15 (0.40) & \textbf{1} & \textbf{6.01 (0.21)} & 7 & 55.72 (0.53) & 8 & \texttimes & 3 & 13.09 (1.02) & 5 & 22.39 (2.06) & 6 & 22.89 (0.97) & 3 & 13.09 (0.99) & 4 & 18.52 (0.58) \\
 & $S_3$ & \cellcolor{green!20}\textbf{1} & \cellcolor{green!20}\textbf{4.81 (0.48)} & 2 & 5.48 (0.39) & 2 & 5.52 (0.34) & 6 & 55.53 (1.69) & 7 & \texttimes & 3 & 13.17 (0.57) & 5 & 22.91 (2.02) & 5 & 22.95 (0.85) & 3 & 13.18 (0.59) & 4 & 18.48 (0.59) \\
 & $S_4$ & \cellcolor{green!20}\textbf{1} & \cellcolor{green!20}\textbf{4.23 (0.35)} & 2 & 5.12 (0.38) & 3 & 5.26 (0.31) & 8 & 54.58 (1.48) & 9 & \texttimes & 4 & 13.15 (0.61) & 7 & 24.04 (1.39) & 6 & 22.85 (0.82) & 4 & 13.16 (0.61) & 5 & 18.75 (0.64) \\
 & $S_5$ & \cellcolor{green!20}\textbf{1} & \cellcolor{green!20}\textbf{3.98 (0.19)} & 2 & 4.98 (0.64) & 3 & 5.15 (0.34) & 8 & 34.53 (21.14) & 9 & \texttimes & 4 & 13.17 (0.55) & 7 & 23.86 (1.77) & 6 & 22.96 (0.78) & 4 & 13.19 (0.52) & 5 & 18.49 (0.47) \\ \hline
\multirow{5}{*}{\textsc{HIPA$^{cc}$}} & $S_1$ & \textbf{1} & \textbf{7.50 (1.42)} & \cellcolor{red!20}\textbf{1} & \cellcolor{red!20}\textbf{7.40 (1.05)} & 2 & 9.70 (1.28) & 7 & 31.99 (0.06) & 8 & \texttimes & 3 & 20.43 (1.64) & 5 & 26.18 (0.68) & 4 & 25.01 (0.19) & 3 & 20.57 (1.85) & 6 & 33.07 (2.31) \\
 & $S_2$ & \cellcolor{green!20}\textbf{1} & \cellcolor{green!20}\textbf{4.48 (0.38)} & 2 & 4.53 (0.48) & 3 & 6.89 (1.44) & 7 & 31.98 (0.06) & 9 & \texttimes & 4 & 19.73 (0.89) & 6 & 26.12 (0.49) & 5 & 24.97 (0.07) & 4 & 19.76 (0.86) & 8 & 33.79 (1.24) \\
 & $S_3$ & \cellcolor{green!20}\textbf{1} & \cellcolor{green!20}\textbf{3.43 (0.34)} & 2 & 3.67 (0.28) & 3 & 4.68 (0.90) & 7 & 31.99 (0.06) & 9 & \texttimes & 4 & 19.33 (0.70) & 6 & 26.00 (0.46) & 5 & 24.95 (0.05) & 4 & 19.32 (0.85) & 8 & 33.73 (0.87) \\
 & $S_4$ & \cellcolor{green!20}\textbf{1} & \cellcolor{green!20}\textbf{2.60 (0.15)} & 2 & 2.70 (0.18) & 3 & 3.58 (1.00) & 7 & 31.98 (0.06) & 9 & \texttimes & 4 & 18.99 (0.61) & 6 & 25.90 (0.32) & 5 & 24.92 (0.06) & 4 & 18.99 (0.60) & 8 & 33.66 (0.79) \\
 & $S_5$ & \cellcolor{green!20}\textbf{1} & \cellcolor{green!20}\textbf{2.11 (0.08)} & 2 & 2.23 (0.23) & 3 & 2.82 (0.57) & 7 & 31.97 (0.14) & 9 & \texttimes & 4 & 18.79 (0.42) & 6 & 25.73 (0.22) & 5 & 24.91 (0.12) & 4 & 18.79 (0.35) & 8 & 33.58 (0.66) \\ \hline
\multirow{5}{*}{\textsc{HSMGP}} & $S_1$ & \cellcolor{green!20}\textbf{1} & \cellcolor{green!20}\textbf{4.14 (1.44)} & 2 & 5.03 (2.56) & 3 & 7.09 (3.04) & 8 & 66.70 (1294.05) & 9 & \texttimes & 4 & 56.76 (18.43) & 5 & 65.36 (30.25) & 6 & 95.18 (11.51) & 4 & 56.76 (18.02) & 7 & 124.61 (36.63) \\
 & $S_2$ & \cellcolor{green!20}\textbf{1} & \cellcolor{green!20}\textbf{2.06 (0.19)} & 2 & 3.13 (0.49) & 3 & 3.69 (0.61) & 8 & 66.67 (0.11) & 9 & \texttimes & 4 & 50.74 (12.37) & 5 & 67.20 (15.34) & 6 & 89.54 (8.16) & 4 & 50.85 (10.53) & 7 & 121.43 (23.60) \\
 & $S_3$ & \cellcolor{green!20}\textbf{1} & \cellcolor{green!20}\textbf{1.42 (0.13)} & 2 & 2.34 (0.38) & 2 & 2.28 (0.37) & 6 & 66.63 (0.18) & 7 & \texttimes & 3 & 51.10 (8.01) & 4 & 62.29 (11.54) & 5 & 84.97 (4.24) & 3 & 51.73 (8.30) & 6 & 119.16 (16.57) \\
 & $S_4$ & \cellcolor{green!20}\textbf{1} & \cellcolor{green!20}\textbf{1.33 (0.08)} & 2 & 2.19 (0.23) & 3 & 2.23 (0.39) & 6 & 66.63 (0.14) & 9 & \texttimes & 4 & 49.45 (9.49) & 5 & 62.38 (9.96) & 7 & 84.76 (5.45) & 4 & 50.41 (8.05) & 8 & 117.78 (15.89) \\
 & $S_5$ & \cellcolor{green!20}\textbf{1} & \cellcolor{green!20}\textbf{1.19 (0.08)} & 2 & 2.09 (0.25) & 3 & 1.94 (0.63) & 6 & 66.59 (0.20) & 9 & \texttimes & 4 & 49.57 (3.15) & 5 & 58.63 (4.99) & 7 & 83.34 (3.06) & 4 & 49.58 (3.25) & 8 & 118.10 (7.24) \\ \hline
\multirow{5}{*}{\textsc{Lrzip}} & $S_1$ & \cellcolor{green!20}\textbf{1} & \cellcolor{green!20}\textbf{31.88 (10.96)} & 3 & 41.97 (12.02) & 2 & 35.40 (16.59) & 4 & 58.45 (0.12) & 9 & \texttimes & 6 & 353.37 (126.02) & 5 & 198.33 (88.16) & 7 & 416.66 (119.43) & 6 & 369.45 (117.97) & 8 & 556.01 (110.73) \\
 & $S_2$ & \cellcolor{green!20}\textbf{1} & \cellcolor{green!20}\textbf{10.51 (4.39)} & 3 & 24.10 (19.98) & 2 & 21.46 (4.53) & 4 & 58.46 (0.18) & 9 & \texttimes & 6 & 309.08 (60.71) & 5 & 144.88 (72.46) & 7 & 403.42 (63.75) & 6 & 315.26 (62.54) & 8 & 517.67 (80.56) \\
 & $S_3$ & \cellcolor{green!20}\textbf{1} & \cellcolor{green!20}\textbf{8.04 (2.37)} & 3 & 23.71 (16.49) & 2 & 18.68 (3.20) & 4 & 58.44 (0.21) & 9 & \texttimes & 6 & 317.52 (47.97) & 5 & 152.06 (92.00) & 7 & 389.89 (44.37) & 6 & 319.07 (50.92) & 8 & 531.18 (104.18) \\
 & $S_4$ & \cellcolor{green!20}\textbf{1} & \cellcolor{green!20}\textbf{5.95 (0.99)} & 3 & 19.82 (16.76) & 2 & 15.55 (1.71) & 4 & 58.48 (0.21) & 9 & \texttimes & 6 & 318.37 (54.99) & 5 & 151.04 (54.46) & 7 & 385.47 (29.42) & 6 & 325.49 (47.06) & 8 & 531.05 (73.06) \\
 & $S_5$ & \cellcolor{green!20}\textbf{1} & \cellcolor{green!20}\textbf{4.03 (0.26)} & 3 & 10.39 (2.92) & 2 & 10.17 (0.91) & 4 & 58.39 (0.21) & 9 & \texttimes & 6 & 325.68 (31.40) & 5 & 150.17 (20.91) & 7 & 383.70 (27.71) & 6 & 325.68 (30.81) & 8 & 531.16 (36.15) \\ \hline
\multirow{5}{*}{\textsc{nginx}} & $S_1$ & \cellcolor{green!20}\textbf{1} & \cellcolor{green!20}\textbf{4.11 (1.45)} & 2 & 8.08 (2.89) & 3 & 8.76 (4.10) & 7 & 1553.44 (965.63) & 8 & \texttimes & 4 & 574.23 (25.74) & 6 & 1218.33 (58.83) & 5 & 1042.65 (46.12) & 4 & 574.23 (25.74) & 7 & 1406.58 (65.16) \\
 & $S_2$ & \cellcolor{green!20}\textbf{1} & \cellcolor{green!20}\textbf{2.60 (1.14)} & 2 & 6.95 (1.74) & 2 & 6.15 (4.17) & 7 & 1637.60 (430.94) & 8 & \texttimes & 3 & 566.48 (20.63) & 5 & 1200.34 (76.00) & 4 & 1046.84 (41.65) & 3 & 568.39 (20.34) & 6 & 1379.87 (80.17) \\
 & $S_3$ & \cellcolor{green!20}\textbf{1} & \cellcolor{green!20}\textbf{2.14 (0.40)} & 3 & 6.46 (1.41) & 2 & 4.42 (1.44) & 8 & 1692.49 (305.78) & 9 & \texttimes & 4 & 567.42 (18.59) & 6 & 1206.44 (53.10) & 5 & 1045.37 (32.55) & 4 & 569.35 (15.63) & 7 & 1379.17 (61.17) \\
 & $S_4$ & \cellcolor{green!20}\textbf{1} & \cellcolor{green!20}\textbf{1.99 (0.47)} & 3 & 6.01 (1.68) & 2 & 4.09 (2.27) & 8 & 1784.76 (274.62) & 9 & \texttimes & 4 & 568.66 (19.43) & 6 & 1203.30 (54.38) & 5 & 1046.87 (29.85) & 4 & 571.25 (18.43) & 7 & 1376.25 (59.62) \\
 & $S_5$ & \cellcolor{green!20}\textbf{1} & \cellcolor{green!20}\textbf{1.98 (0.21)} & 3 & 7.27 (2.31) & 2 & 3.17 (1.72) & 8 & 1579.37 (515.54) & 9 & \texttimes & 4 & 564.37 (24.70) & 6 & 1199.31 (52.15) & 5 & 1039.58 (28.93) & 4 & 564.62 (23.57) & 7 & 1367.38 (60.70) \\ \hline
\multirow{5}{*}{\textsc{VP8}} & $S_1$ & \cellcolor{green!20}\textbf{1} & \cellcolor{green!20}\textbf{1.56 (0.18)} & \textbf{1} & \textbf{1.72 (0.25)} & 2 & 4.68 (3.27) & 4 & 60.05 (2.03) & 8 & \texttimes & 3 & 44.06 (4.85) & 6 & 92.98 (4.52) & 5 & 85.03 (4.01) & 3 & 44.06 (4.20) & 7 & 110.78 (6.00) \\
 & $S_2$ & \cellcolor{green!20}\textbf{1} & \cellcolor{green!20}\textbf{1.15 (0.05)} & 2 & 1.27 (0.15) & 3 & 2.39 (1.21) & 5 & 60.04 (0.24) & 9 & \texttimes & 4 & 42.10 (3.17) & 7 & 91.46 (5.56) & 6 & 85.68 (4.11) & 4 & 42.36 (3.42) & 8 & 108.46 (6.09) \\
 & $S_3$ & \cellcolor{green!20}\textbf{1} & \cellcolor{green!20}\textbf{1.08 (0.04)} & 2 & 1.25 (0.14) & 3 & 1.93 (0.72) & 5 & 60.02 (0.29) & 9 & \texttimes & 4 & 42.94 (2.82) & 7 & 91.76 (3.49) & 6 & 84.88 (2.66) & 4 & 43.02 (2.62) & 8 & 108.23 (5.29) \\
 & $S_4$ & \cellcolor{green!20}\textbf{1} & \cellcolor{green!20}\textbf{0.98 (0.05)} & 2 & 1.15 (0.11) & 3 & 1.54 (0.40) & 5 & 59.99 (0.27) & 9 & \texttimes & 4 & 42.05 (3.59) & 7 & 91.99 (3.14) & 6 & 84.78 (1.95) & 4 & 41.99 (3.12) & 8 & 108.47 (4.25) \\
 & $S_5$ & \cellcolor{green!20}\textbf{1} & \cellcolor{green!20}\textbf{0.90 (0.06)} & 2 & 1.12 (0.11) & 3 & 1.45 (0.23) & 5 & 59.95 (0.68) & 9 & \texttimes & 4 & 42.87 (2.39) & 7 & 92.21 (3.02) & 6 & 84.53 (2.44) & 4 & 42.66 (2.11) & 8 & 108.43 (3.71) \\ \hline
\multicolumn{2}{c}{Average $r$} & 1.47 &  & 1.88 &  & 2.37 &  & 6.1 &  & 5.38 &  & 4.67 &  & 5.13 &  & 5.45 &  & 4.65 &  & 6.62 & 
\\
\bottomrule
\end{tabular}
}

\label{tb:vsSOTA}
\end{table*}

\section{Evaluation}
\label{sec:evaluation}

We now present and discuss the experimental results.

\subsection{Comparing with the State-of-the-art Approaches}
\label{subsec:rq1}

\subsubsection{Method}

To understand how \Model~performs compared with the state-of-the-art approaches, we assess its accuracy against both the standard works for configuration performance learning that rely on statistical learning together with recent deep learning-based ones:

\begin{itemize}

  \item\texttt{SPLConqueror}~\cite{DBLP:journals/sqj/SiegmundRKKAS12}: linear regression with joint terms.

  \item\texttt{DECART}~\cite{DBLP:journals/ese/GuoYSASVCWY18}: an improved \texttt{CART} with hyperparameter tuning.

  \item\texttt{DeepPerf}~\cite{DBLP:conf/icse/HaZ19}: a single global regularized deep neural network.

  \item\texttt{Perf-AL}~\cite{DBLP:conf/esem/ShuS0X20}: an adversarial learning method. 

  \item\texttt{HINNPerf}~\cite{DBLP:journals/tosem/ChengGZ23}: a hierarchical deep neural network with embedding.

\end{itemize}

Additionally, we examine four approaches from the machine learning community that share a similar basic concept to \Model: 

\begin{itemize}
  \item\texttt{IBMB}~\cite{DBLP:conf/icml/Quinlan93}: a model that combines predictions from instance-based learning and model-based learning techniques.

  \item\texttt{M5}~\cite{Quinlan1992LearningWC}: a tree that divides to minimize the variance of each subset and trains a linear model for each leaf node.

  \item\texttt{MOB}~\cite{Zeileis2009partyWT}: a model that divides based on parametric models and parameter instability tests. 

  \item\texttt{PILOT}~\cite{raymaekers2023fast}: a model divides similarly to \texttt{CART} but without pruning and trains linear models at the leaf nodes.
\end{itemize}

All approaches can be used for any type of system except for \texttt{DECART}, which works on binary systems only. 
\textcolor{black}{The same randomly generated training and testing samples are used for all models, which are selected by using random sampling.}
Since there are 12 systems and 5 sets of sample sizes each, we obtained 60 cases to compare in total. The MREs and ranks from the Scott-Knott test are reported for all cases. To ensure consistency, we use the implementations published by their authors with the same parameter settings. We use the systems, training sizes, and statistical tests from Section~\ref{sec:setup}. All experiments are repeated for 30 runs. 



\begin{table*}[h!]
\centering
\caption{\Model~under different local models against using them as the global models. The format is the same as Table~\ref{tb:vsSOTA}.}
\setlength{\tabcolsep}{1.2mm}
\renewcommand\arraystretch{1.1}
\resizebox{\textwidth}{!}{ 
\begin{tabular}{p{1cm}lp{0.2cm}p{1.8cm}|p{0.2cm}p{1.8cm}|p{0.2cm}p{1.8cm}|p{0.2cm}p{1.8cm}|p{0.2cm}p{1.8cm}|p{0.2cm}p{1.8cm}|p{0.2cm}p{1.95cm}|p{0.2cm}p{2.1cm}|p{0.2cm}p{1.8cm}|p{0.2cm}p{1.95cm}} \toprule
\multirow{2}{*}{System} & \multicolumn{1}{c}{\multirow{2}{*}{Size}} & \multicolumn{2}{c}{\texttt{DaL}} & \multicolumn{2}{c}{\texttt{HINNPerf}} & \multicolumn{2}{c}{\texttt{DaL$_{DNN}$}} & \multicolumn{2}{c}{\texttt{DNN}} & \multicolumn{2}{c}{\texttt{DaL$_{CART}$}} & \multicolumn{2}{c}{\texttt{CART}} & \multicolumn{2}{c}{\texttt{DaL$_{LR}$}} & \multicolumn{2}{c}{\texttt{LR}} & \multicolumn{2}{c}{\texttt{DaL$_{SVR}$}} & \multicolumn{2}{c}{\texttt{SVR}} \\ \cline{3-22}
 & \multicolumn{1}{c}{} & $r$ & Med (IQR) & $r$ & Med (IQR) & $r$ & Med (IQR) & $r$ & Med (IQR) & $r$ & Med (IQR) & $r$ & Med (IQR) & $r$ & Med (IQR) & $r$ & Med (IQR) & $r$ & Med (IQR) & $r$ & Med (IQR) \\ \hline
\multirow{5}{*}{\textsc{Apache}} & $S_1$ & 2 & 21.86 (7.36) & \cellcolor{red!20}3 & \cellcolor{red!20}19.44 (10.53) & \textbf{1} & \textbf{21.02 (6.64)} & 2 & 20.19 (6.34) & 2 & 21.61 (6.69) & 2 & 20.85 (12.99) & \textbf{1} & \textbf{19.71 (6.80)} & 4 & 27.72 (4.98) & 3 & 23.30 (4.08) & 3 & 22.67 (3.66) \\
 & $S_2$ & 4 & 12.89 (7.87) & \cellcolor{red!20}\textbf{1} & \cellcolor{red!20}\textbf{8.93 (3.89)} & 3 & 9.82 (6.00) & 2 & 9.79 (4.56) & 3 & 10.12 (5.05) & 2 & 10.06 (3.05) & 3 & 10.77 (7.55) & 4 & 17.61 (2.41) & 4 & 17.59 (3.76) & 4 & 20.31 (3.18) \\
 & $S_3$ & \cellcolor{green!20}\textbf{1} & \cellcolor{green!20}\textbf{6.56 (1.92)} & \textbf{1} & \textbf{7.01 (1.74)} & 2 & 7.17 (2.67) & 3 & 7.98 (1.91) & 4 & 9.10 (1.31) & 4 & 9.03 (1.13) & 2 & 7.42 (2.14) & 6 & 16.16 (1.41) & 5 & 15.92 (2.45) & 7 & 18.50 (1.54) \\
 & $S_4$ & 3 & 6.87 (1.95) & 2 & 6.58 (1.04) & 3 & 6.59 (1.77) & 3 & 6.85 (1.60) & 5 & 7.97 (1.35) & 4 & 7.81 (1.35) & \cellcolor{red!20}\textbf{1} & \cellcolor{red!20}\textbf{6.39 (1.16)} & 7 & 16.12 (0.78) & 6 & 13.89 (2.35) & 8 & 18.57 (1.58) \\
 & $S_5$ & 3 & 5.78 (1.08) & 3 & 5.85 (0.99) & 2 & 5.96 (2.07) & 4 & 6.66 (1.61) & 5 & 7.63 (1.00) & 5 & 7.71 (1.03) & \cellcolor{red!20}\textbf{1} & \cellcolor{red!20}\textbf{5.75 (0.55)} & 7 & 16.12 (1.37) & 6 & 12.63 (1.76) & 8 & 17.82 (1.78) \\ \hline
\multirow{5}{*}{\textsc{BDB-C}} & $S_1$ & 4 & 83.60 (208.75) & 4 & 167.55 (144.47) & 2 & 41.77 (32.54) & 2 & 43.66 (42.88) & \textbf{1} & \textbf{47.20 (19.75)} & \cellcolor{red!20}\textbf{1} & \cellcolor{red!20}\textbf{36.17 (9.11)} & 3 & 65.09 (37.30) & 5 & 698.12 (713.51) & 3 & 65.38 (22.26) & 4 & 154.71 (52.45) \\
 & $S_2$ & 2 & 21.92 (17.91) & \textbf{1} & \textbf{21.45 (13.61)} & 2 & 17.22 (15.67) & \textbf{1} & \textbf{18.17 (8.11)} & \cellcolor{red!20}\textbf{1} & \cellcolor{red!20}\textbf{16.81 (8.59)} & \textbf{1} & \textbf{19.61 (7.19)} & 3 & 49.04 (47.11) & 4 & 545.48 (149.70) & 3 & 70.56 (64.18) & 3 & 117.87 (49.95) \\
 & $S_3$ & 3 & 8.31 (7.52) & 2 & 6.50 (9.20) & \cellcolor{red!20}2 & \cellcolor{red!20}5.74 (6.84) & 4 & 12.47 (4.08) & 3 & 6.72 (5.82) & \textbf{1} & \textbf{7.44 (3.70)} & 5 & 29.00 (78.81) & 7 & 465.51 (91.79) & 6 & 65.69 (155.89) & 5 & 97.22 (52.60) \\
 & $S_4$ & \cellcolor{green!20}3 & \cellcolor{green!20}3.60 (4.50) & \textbf{1} & \textbf{4.86 (2.18)} & \textbf{1} & \textbf{3.68 (2.78)} & 3 & 9.06 (9.76) & \textbf{1} & \textbf{4.15 (2.78)} & 2 & 5.74 (3.97) & 4 & 24.55 (4.18) & 6 & 441.04 (85.57) & 5 & 51.04 (29.64) & 5 & 93.39 (41.17) \\
 & $S_5$ & \cellcolor{green!20}\textbf{1} & \cellcolor{green!20}\textbf{1.88 (1.24)} & 3 & 4.04 (1.24) & 2 & 2.15 (1.52) & 4 & 7.08 (8.03) & 2 & 3.24 (3.09) & 2 & 3.32 (2.50) & 5 & 22.88 (3.78) & 8 & 456.02 (78.01) & 6 & 46.18 (19.08) & 7 & 105.14 (56.61) \\ \hline
\multirow{5}{*}{\textsc{BDB-J}} & $S_1$ & 2 & 5.37 (4.74) & 2 & 3.58 (1.74) & 2 & 3.14 (2.90) & \cellcolor{red!20}2 & \cellcolor{red!20}2.98 (3.34) & \cellcolor{red!20}\textbf{1} & \cellcolor{red!20}\textbf{2.98 (0.81)} & \textbf{1} & \textbf{3.03 (0.99)} & 3 & 5.31 (2.96) & 5 & 43.23 (6.15) & 4 & 22.77 (17.72) & 3 & 12.93 (3.37) \\
 & $S_2$ & \cellcolor{green!20}\textbf{1} & \cellcolor{green!20}\textbf{1.83 (0.45)} & \textbf{1} & \textbf{1.96 (0.36)} & 4 & 1.90 (0.37) & 3 & 1.91 (0.68) & 2 & 2.05 (0.33) & \textbf{1} & \textbf{1.99 (0.29)} & 4 & 3.77 (0.49) & 7 & 37.85 (6.27) & 6 & 15.28 (12.28) & 5 & 10.54 (0.97) \\
 & $S_3$ & \cellcolor{green!20}\textbf{1} & \cellcolor{green!20}\textbf{1.58 (0.26)} & 2 & 1.75 (0.41) & 2 & 1.61 (0.31) & 3 & 2.01 (1.10) & 2 & 1.92 (0.27) & 2 & 1.87 (0.35) & 3 & 3.41 (0.46) & 6 & 37.61 (3.49) & 5 & 15.03 (10.36) & 4 & 11.03 (0.89) \\
 & $S_4$ & \cellcolor{green!20}\textbf{1} & \cellcolor{green!20}\textbf{1.45 (0.22)} & 2 & 1.52 (0.24) & 3 & 1.61 (0.29) & 5 & 1.70 (0.37) & 4 & 1.67 (0.29) & 4 & 1.69 (0.36) & 6 & 3.57 (0.66) & 8 & 37.57 (3.48) & 7 & 9.20 (5.36) & 7 & 11.25 (1.35) \\
 & $S_5$ & \cellcolor{green!20}\textbf{1} & \cellcolor{green!20}\textbf{1.40 (0.34)} & 2 & 1.48 (0.32) & 2 & 1.46 (0.26) & 4 & 1.69 (0.48) & 3 & 1.67 (0.43) & 3 & 1.68 (0.39) & 5 & 3.23 (0.65) & 8 & 37.48 (2.86) & 6 & 8.21 (3.87) & 7 & 11.51 (2.21) \\ \hline
\multirow{5}{*}{\textsc{kanzi}} & $S_1$ & 2 & 196.15 (213.73) & 4 & 262.27 (175.07) & 2 & 166.22 (122.05) & 2 & 196.36 (122.61) & 2 & 178.67 (123.16) & 2 & 163.60 (121.50) & 5 & 551.03 (1244.41) & 6 & 3106.09 (1514.60) & 3 & 253.06 (150.61) & \cellcolor{red!20}\textbf{1} & \cellcolor{red!20}\textbf{129.01 (61.27)} \\
 & $S_2$ & 3 & 95.31 (51.34) & \cellcolor{red!20}\textbf{1} & \cellcolor{red!20}\textbf{80.53 (51.36)} & 2 & 90.27 (63.53) & 4 & 107.12 (43.75) & \textbf{1} & \textbf{86.55 (35.63)} & \textbf{1} & \textbf{91.65 (38.71)} & 6 & 383.06 (493.33) & 7 & 1191.03 (348.39) & 5 & 190.01 (134.20) & 2 & 105.79 (34.34) \\
 & $S_3$ & \cellcolor{green!20}2 & \cellcolor{green!20}59.30 (38.47) & \textbf{1} & \textbf{62.20 (14.07)} & 3 & 64.56 (62.71) & 2 & 68.62 (30.41) & 2 & 62.50 (21.15) & 2 & 60.13 (21.42) & 6 & 177.35 (384.90) & 5 & 1058.91 (166.05) & 4 & 198.38 (146.32) & 3 & 99.24 (33.82) \\
 & $S_4$ & \textbf{1} & \textbf{42.93 (18.75)} & \textbf{1} & \textbf{46.48 (19.58)} & \cellcolor{red!20}\textbf{1} & \cellcolor{red!20}\textbf{41.83 (15.47)} & 2 & 62.65 (26.54) & \textbf{1} & \textbf{48.69 (10.31)} & \textbf{1} & \textbf{49.06 (14.98)} & 6 & 134.99 (522.74) & 5 & 1030.25 (180.09) & 4 & 186.29 (80.16) & 3 & 89.93 (27.29) \\
 & $S_5$ & \textbf{1} & \textbf{35.29 (17.13)} & \cellcolor{red!20}\textbf{1} & \cellcolor{red!20}\textbf{34.27 (11.87)} & 4 & 38.57 (20.46) & 3 & 49.42 (9.99) & 2 & 44.15 (8.53) & 2 & 43.96 (10.22) & 7 & 605.42 (1213.79) & 7 & 1010.78 (264.34) & 6 & 152.57 (97.14) & 5 & 87.82 (16.52) \\ \hline
\multirow{5}{*}{\textsc{SQLite}} & $S_1$ & 3 & 73.52 (20.40) & \cellcolor{red!20}2 & \cellcolor{red!20}70.76 (27.86) & \textbf{1} & \textbf{73.62 (16.43)} & 3 & 79.49 (25.36) & 4 & 82.77 (20.99) & 4 & 83.67 (21.57) & 5 & 105.75 (83.62) & 6 & 151.86 (65.00) & 2 & 79.85 (19.78) & \textbf{1} & \textbf{74.80 (19.37)} \\
 & $S_2$ & \cellcolor{green!20}\textbf{1} & \cellcolor{green!20}\textbf{76.41 (16.66)} & 2 & 76.69 (23.32) & 3 & 77.81 (24.37) & 3 & 77.75 (22.47) & 4 & 86.76 (20.84) & 4 & 89.07 (18.89) & 5 & 161.86 (135.45) & 4 & 84.41 (23.67) & 2 & 79.52 (21.38) & 2 & 77.78 (11.49) \\
 & $S_3$ & \textbf{1} & \textbf{71.63 (11.58)} & \cellcolor{red!20}\textbf{1} & \cellcolor{red!20}\textbf{71.02 (17.30)} & 2 & 74.86 (15.95) & \textbf{1} & \textbf{75.01 (15.58)} & 3 & 87.65 (24.29) & 3 & 89.12 (23.09) & 4 & 103.75 (29.66) & 2 & 75.72 (24.83) & 2 & 80.75 (14.40) & 2 & 79.66 (10.31) \\
 & $S_4$ & \cellcolor{green!20}\textbf{1} & \cellcolor{green!20}\textbf{66.94 (10.61)} & 2 & 70.84 (10.38) & 2 & 70.41 (12.22) & 3 & 72.45 (12.11) & 6 & 86.47 (20.30) & 6 & 87.46 (19.46) & 6 & 82.25 (23.16) & 2 & 69.61 (13.80) & 4 & 77.23 (10.08) & 5 & 79.11 (10.34) \\
 & $S_5$ & \cellcolor{green!20}\textbf{1} & \cellcolor{green!20}\textbf{64.20 (13.91)} & 2 & 70.15 (9.71) & 2 & 69.72 (12.48) & 3 & 72.87 (8.86) & 6 & 84.43 (12.41) & 6 & 84.51 (14.40) & 4 & 76.06 (16.27) & 2 & 67.78 (9.68) & 4 & 76.28 (13.41) & 5 & 79.69 (11.05) \\ \hline
\multirow{5}{*}{\textsc{x264}} & $S_1$ & 3 & 10.21 (4.35) & 2 & 8.44 (2.69) & \cellcolor{red!20}\textbf{1} & \cellcolor{red!20}\textbf{8.04 (1.30)} & \cellcolor{red!20}\textbf{1} & \cellcolor{red!20}\textbf{8.04 (3.06)} & 3 & 10.40 (2.15) & 2 & 8.77 (2.89) & 3 & 10.43 (3.40) & 4 & 14.36 (4.34) & 4 & 14.39 (2.64) & 5 & 25.72 (2.91) \\
 & $S_2$ & \cellcolor{green!20}2 & \cellcolor{green!20}2.72 (1.52) & \textbf{1} & \textbf{2.77 (1.07)} & 3 & 3.34 (1.90) & 2 & 3.17 (1.00) & 5 & 6.57 (1.67) & 5 & 6.50 (1.63) & 4 & 3.51 (0.98) & 6 & 8.01 (1.23) & 7 & 9.32 (2.16) & 8 & 18.83 (5.32) \\
 & $S_3$ & 2 & 1.57 (1.52) & \cellcolor{red!20}\textbf{1} & \cellcolor{red!20}\textbf{1.37 (0.46)} & 2 & 1.97 (1.47) & 2 & 2.23 (0.90) & 4 & 5.01 (1.42) & 4 & 4.72 (0.73) & 3 & 2.84 (0.52) & 5 & 7.61 (0.76) & 5 & 8.25 (2.68) & 6 & 13.66 (2.32) \\
 & $S_4$ & 2 & 1.30 (1.17) & \cellcolor{red!20}\textbf{1} & \cellcolor{red!20}\textbf{0.81 (0.28)} & 2 & 1.39 (0.56) & 3 & 1.74 (0.98) & 5 & 3.52 (1.33) & 5 & 3.65 (1.03) & 4 & 2.60 (0.52) & 7 & 7.34 (0.39) & 6 & 7.05 (4.00) & 8 & 10.24 (1.41) \\
 & $S_5$ & 2 & 0.96 (0.79) & \cellcolor{red!20}\textbf{1} & \cellcolor{red!20}\textbf{0.58 (0.15)} & 2 & 1.05 (0.55) & 3 & 1.44 (0.90) & 6 & 2.54 (1.04) & 5 & 2.37 (1.11) & 4 & 2.10 (1.04) & 8 & 7.17 (0.60) & 7 & 7.72 (4.47) & 9 & 9.19 (0.76) \\ \hline
\multirow{5}{*}{\textsc{Dune}} & $S_1$ & 4 & 9.37 (1.15) & 2 & 7.98 (0.98) & 3 & 8.40 (1.61) & \cellcolor{red!20}\textbf{1} & \cellcolor{red!20}\textbf{7.86 (0.93)} & 4 & 9.04 (1.13) & 4 & 9.13 (0.99) & 5 & 11.21 (0.88) & 6 & 13.06 (1.90) & 7 & 14.46 (0.51) & 8 & 14.76 (0.64) \\
 & $S_2$ & \textbf{1} & \textbf{5.95 (0.54)} & 2 & 6.15 (0.40) & \cellcolor{red!20}\textbf{1} & \cellcolor{red!20}\textbf{5.78 (0.41)} & \textbf{1} & \textbf{6.01 (0.21)} & 3 & 6.25 (0.42) & 3 & 6.26 (0.40) & 4 & 11.13 (0.57) & 5 & 13.02 (1.06) & 5 & 13.37 (0.25) & 6 & 13.47 (0.32) \\
 & $S_3$ & \cellcolor{green!20}\textbf{1} & \cellcolor{green!20}\textbf{4.81 (0.48)} & 3 & 5.48 (0.39) & 2 & 5.21 (0.39) & 3 & 5.52 (0.34) & 3 & 5.48 (0.42) & 3 & 5.49 (0.37) & 4 & 11.09 (0.54) & 6 & 13.13 (0.66) & 5 & 12.95 (0.32) & 6 & 13.15 (0.41) \\
 & $S_4$ & \cellcolor{green!20}\textbf{1} & \cellcolor{green!20}\textbf{4.23 (0.35)} & 4 & 5.12 (0.38) & 2 & 4.79 (0.49) & 5 & 5.26 (0.31) & 3 & 5.06 (0.32) & 3 & 5.01 (0.32) & 6 & 11.04 (0.43) & 9 & 13.14 (0.55) & 7 & 12.57 (0.47) & 8 & 12.80 (0.52) \\
 & $S_5$ & \cellcolor{green!20}\textbf{1} & \cellcolor{green!20}\textbf{3.98 (0.19)} & 3 & 4.98 (0.64) & 2 & 4.69 (0.50) & 4 & 5.15 (0.34) & 2 & 4.69 (0.35) & 2 & 4.66 (0.36) & 5 & 10.93 (0.48) & 8 & 13.21 (0.57) & 6 & 12.30 (0.46) & 7 & 12.58 (0.69) \\ \hline
\multirow{5}{*}{\textsc{HIPA$^{cc}$}} & $S_1$ & \textbf{1} & \textbf{7.50 (1.42)} & \cellcolor{red!20}\textbf{1} & \cellcolor{red!20}\textbf{7.40 (1.05)} & 2 & 9.07 (1.18) & 3 & 9.70 (1.28) & 4 & 12.03 (0.93) & 4 & 12.07 (1.06) & 8 & 12.70 (1.22) & 7 & 20.49 (1.61) & 5 & 13.06 (0.51) & 6 & 14.84 (0.62) \\
 & $S_2$ & \cellcolor{green!20}\textbf{1} & \cellcolor{green!20}\textbf{4.48 (0.38)} & 2 & 4.53 (0.48) & 3 & 5.55 (0.60) & 4 & 6.89 (1.44) & 5 & 8.47 (1.10) & 5 & 8.20 (1.00) & 9 & 11.64 (0.35) & 8 & 19.67 (0.83) & 6 & 11.27 (0.44) & 7 & 14.52 (0.54) \\
 & $S_3$ & \cellcolor{green!20}\textbf{1} & \cellcolor{green!20}\textbf{3.43 (0.34)} & 2 & 3.67 (0.28) & 3 & 4.39 (0.41) & 4 & 4.68 (0.90) & 5 & 6.62 (0.80) & 5 & 6.60 (0.75) & 7 & 11.43 (0.34) & 9 & 19.35 (0.72) & 6 & 10.78 (0.28) & 8 & 14.48 (0.57) \\
 & $S_4$ & \cellcolor{green!20}\textbf{1} & \cellcolor{green!20}\textbf{2.60 (0.15)} & 2 & 2.70 (0.18) & 3 & 3.22 (0.35) & 4 & 3.58 (1.00) & 5 & 4.42 (0.22) & 5 & 4.31 (0.44) & 7 & 11.13 (0.25) & 9 & 18.97 (0.46) & 6 & 10.31 (0.14) & 8 & 12.06 (0.25) \\
 & $S_5$ & \cellcolor{green!20}\textbf{1} & \cellcolor{green!20}\textbf{2.11 (0.08)} & 2 & 2.23 (0.23) & 3 & 2.39 (0.22) & 5 & 2.82 (0.57) & 4 & 2.69 (0.15) & 4 & 2.70 (0.17) & 8 & 11.01 (0.15) & 9 & 18.81 (0.37) & 6 & 10.04 (0.09) & 7 & 10.52 (0.24) \\ \hline
\multirow{5}{*}{\textsc{HSMGP}} & $S_1$ & \cellcolor{green!20}\textbf{1} & \cellcolor{green!20}\textbf{4.14 (1.44)} & 2 & 5.03 (2.56) & \textbf{1} & \textbf{4.66 (1.32)} & 6 & 7.09 (3.04) & 6 & 21.65 (2.31) & 6 & 21.34 (2.41) & 3 & 7.72 (0.80) & 7 & 55.91 (19.98) & 4 & 16.00 (2.48) & 5 & 16.63 (1.38) \\
 & $S_2$ & \cellcolor{green!20}\textbf{1} & \cellcolor{green!20}\textbf{2.06 (0.19)} & 3 & 3.13 (0.49) & 2 & 2.66 (0.68) & 4 & 3.69 (0.61) & 7 & 15.59 (1.41) & 7 & 16.43 (1.52) & 5 & 7.35 (0.47) & 8 & 52.54 (10.70) & 6 & 15.24 (1.32) & 7 & 16.30 (2.16) \\
 & $S_3$ & \cellcolor{green!20}\textbf{1} & \cellcolor{green!20}\textbf{1.42 (0.13)} & 3 & 2.34 (0.38) & 2 & 1.56 (0.20) & 3 & 2.28 (0.37) & 6 & 11.35 (0.63) & 6 & 11.26 (0.98) & 4 & 7.14 (0.35) & 8 & 51.38 (9.85) & 5 & 10.05 (0.86) & 7 & 15.03 (0.66) \\
 & $S_4$ & \cellcolor{green!20}\textbf{1} & \cellcolor{green!20}\textbf{1.33 (0.08)} & 3 & 2.19 (0.23) & 2 & 1.49 (0.15) & 4 & 2.23 (0.39) & 7 & 10.13 (0.57) & 8 & 10.14 (0.72) & 5 & 7.16 (0.17) & 10 & 49.63 (9.43) & 6 & 8.95 (0.90) & 9 & 15.07 (0.56) \\
 & $S_5$ & \textbf{1} & \textbf{1.19 (0.08)} & 2 & 2.09 (0.25) & \cellcolor{red!20}\textbf{1} & \cellcolor{red!20}\textbf{1.16 (0.08)} & 3 & 1.94 (0.63) & 7 & 7.46 (0.46) & 6 & 7.43 (0.34) & 5 & 7.15 (0.20) & 9 & 49.05 (3.00) & 4 & 5.87 (0.52) & 8 & 8.60 (0.38) \\ \hline
\multirow{5}{*}{\textsc{Lrzip}} & $S_1$ & 4 & 31.88 (10.96) & 6 & 41.97 (12.02) & 3 & 26.40 (6.94) & 5 & 35.40 (16.59) & \cellcolor{red!20}\textbf{1} & \cellcolor{red!20}\textbf{17.95 (10.85)} & 2 & 23.50 (10.09) & 8 & 117.37 (24.65) & 9 & 369.45 (116.78) & 7 & 59.16 (13.29) & 7 & 63.04 (23.36) \\
 & $S_2$ & 2 & 10.51 (4.39) & 5 & 24.10 (19.98) & 3 & 15.37 (6.04) & 4 & 21.46 (4.53) & \cellcolor{red!20}\textbf{1} & \cellcolor{red!20}\textbf{9.09 (2.56)} & 2 & 10.47 (4.94) & 7 & 109.11 (20.26) & 8 & 315.34 (66.07) & 6 & 63.84 (11.18) & 6 & 61.82 (34.55) \\
 & $S_3$ & 2 & 8.04 (2.37) & 5 & 23.71 (16.49) & 3 & 11.83 (4.53) & 4 & 18.68 (3.20) & \cellcolor{red!20}\textbf{1} & \cellcolor{red!20}\textbf{7.77 (2.09)} & 2 & 8.62 (4.75) & 8 & 104.20 (17.55) & 9 & 322.55 (50.28) & 7 & 63.91 (13.58) & 6 & 66.09 (41.44) \\
 & $S_4$ & \cellcolor{green!20}\textbf{1} & \cellcolor{green!20}\textbf{5.95 (0.99)} & 5 & 19.82 (16.76) & 3 & 10.10 (3.23) & 4 & 15.55 (1.71) & \textbf{1} & \textbf{6.38 (2.72)} & 2 & 7.09 (2.76) & 7 & 103.74 (14.29) & 8 & 323.59 (56.77) & 6 & 65.19 (12.53) & 6 & 69.07 (5.93) \\
 & $S_5$ & 2 & 4.03 (0.26) & 6 & 10.39 (2.92) & 4 & 6.60 (1.34) & 5 & 10.17 (0.91) & \cellcolor{red!20}\textbf{1} & \cellcolor{red!20}\textbf{3.58 (0.97)} & 3 & 4.47 (1.90) & 9 & 105.83 (7.06) & 10 & 327.20 (27.08) & 7 & 48.80 (2.86) & 8 & 51.27 (2.21) \\ \hline
\multirow{5}{*}{\textsc{nginx}} & $S_1$ & \cellcolor{green!20}\textbf{1} & \cellcolor{green!20}\textbf{4.11 (1.45)} & 3 & 8.08 (2.89) & 2 & 4.95 (2.94) & 4 & 8.76 (4.10) & 5 & 14.44 (2.41) & 5 & 14.44 (2.41) & 6 & 22.02 (2.11) & 8 & 574.23 (25.93) & 7 & 456.24 (15.91) & 9 & 1603.42 (66.06) \\
 & $S_2$ & \cellcolor{green!20}\textbf{1} & \cellcolor{green!20}\textbf{2.60 (1.14)} & 2 & 6.95 (1.74) & \textbf{1} & \textbf{3.07 (0.94)} & 2 & 6.15 (4.17) & 3 & 7.44 (3.14) & 3 & 7.44 (3.14) & 4 & 20.25 (1.60) & 6 & 568.09 (19.68) & 5 & 430.40 (22.39) & 7 & 1337.93 (148.42) \\
 & $S_3$ & \cellcolor{green!20}\textbf{1} & \cellcolor{green!20}\textbf{2.14 (0.40)} & 4 & 6.46 (1.41) & 2 & 2.31 (1.29) & 3 & 4.42 (1.44) & 3 & 4.81 (0.81) & 3 & 4.81 (0.81) & 5 & 19.66 (0.75) & 7 & 568.45 (16.13) & 6 & 389.40 (30.26) & 8 & 942.81 (108.67) \\
 & $S_4$ & \cellcolor{green!20}\textbf{1} & \cellcolor{green!20}\textbf{1.99 (0.47)} & 5 & 6.01 (1.68) & 2 & 2.16 (0.88) & 4 & 4.09 (2.27) & 3 & 4.16 (1.01) & 3 & 4.16 (1.01) & 6 & 19.74 (0.82) & 8 & 571.30 (19.18) & 7 & 364.98 (32.75) & 9 & 690.54 (133.59) \\
 & $S_5$ & \textbf{1} & \textbf{1.98 (0.21)} & 4 & 7.27 (2.31) & \cellcolor{red!20}\textbf{1} & \cellcolor{red!20}\textbf{1.97 (0.52)} & 3 & 3.17 (1.72) & 2 & 3.33 (1.00) & 2 & 3.33 (1.00) & 5 & 19.17 (1.07) & 8 & 564.69 (23.96) & 6 & 327.92 (28.40) & 7 & 450.48 (87.27) \\ \hline
\multirow{5}{*}{\textsc{VP8}} & $S_1$ & \cellcolor{green!20}\textbf{1} & \cellcolor{green!20}\textbf{1.56 (0.18)} & \textbf{1} & \textbf{1.72 (0.25)} & 2 & 1.59 (0.36) & 5 & 4.68 (3.27) & 3 & 2.57 (0.70) & 4 & 2.47 (0.70) & 6 & 12.72 (0.84) & 9 & 43.69 (5.81) & 7 & 13.84 (0.76) & 8 & 22.37 (3.68) \\
 & $S_2$ & \cellcolor{green!20}\textbf{1} & \cellcolor{green!20}\textbf{1.15 (0.05)} & 2 & 1.27 (0.15) & 4 & 1.21 (0.09) & 5 & 2.39 (1.21) & 2 & 1.24 (0.24) & 3 & 1.30 (0.22) & 7 & 12.42 (0.44) & 9 & 42.04 (3.16) & 6 & 10.30 (0.49) & 8 & 23.53 (1.46) \\
 & $S_3$ & \cellcolor{green!20}\textbf{1} & \cellcolor{green!20}\textbf{1.08 (0.04)} & 5 & 1.25 (0.14) & 2 & 1.12 (0.08) & 6 & 1.93 (0.72) & 3 & 1.15 (0.12) & 4 & 1.18 (0.21) & 8 & 12.41 (0.34) & 10 & 42.65 (2.99) & 7 & 10.11 (0.56) & 9 & 25.02 (1.93) \\
 & $S_4$ & \cellcolor{green!20}\textbf{1} & \cellcolor{green!20}\textbf{0.98 (0.05)} & 4 & 1.15 (0.11) & 3 & 1.10 (0.06) & 5 & 1.54 (0.40) & 2 & 1.07 (0.08) & 3 & 1.09 (0.08) & 7 & 12.35 (0.31) & 9 & 42.08 (2.79) & 6 & 9.81 (0.36) & 8 & 26.13 (1.14) \\
 & $S_5$ & \cellcolor{green!20}\textbf{1} & \cellcolor{green!20}\textbf{0.90 (0.06)} & 5 & 1.12 (0.11) & 4 & 1.07 (0.05) & 6 & 1.45 (0.23) & 2 & 0.94 (0.04) & 3 & 0.94 (0.05) & 8 & 12.41 (0.27) & 10 & 42.69 (1.94) & 7 & 9.71 (0.21) & 9 & 15.54 (0.77) \\ \hline
\multicolumn{2}{c}{Average $r$} & 1.63 &  & 2.52 &  & 2.27 &  & 3.35 &  & 3.28 &  & 3.37 &  & 5.12 &  & 6.9 &  & 5.35 &  & 6.12 & 
\\
\bottomrule
\end{tabular}
}

\label{tb:scott-knott}
\end{table*}

\subsubsection{Results}

The results have been illustrated in Table~\ref{tb:vsSOTA}, from which we see that, remarkably, \Model~achieves the best median MRE on 41 out of 60 cases. In particular, \Model~considerably improves the accuracy, i.e., by up to $1.61\times$ better than the second-best one on $S_4$ of \textsc{Lrzip} (5.95 vs. 15.55). The above still holds when looking into the results of the statistical test: \Model~is ranked first for 44 out of the 60 cases, in which \Model~obtain the sole best rank for 31 cases. In particular, among the 19 cases where \Model~does not achieve the best MRE, the inferiority to the best on 6 of them is actually insignificant since it is still equally ranked as the best together with the others. That is to say \Model~is, in 44 cases, similar to (13 cases) or significantly better (31 cases) than the best state-of-the-art approaches for each specific case (which could be a different approach). Overall, \Model~obtain an average rank of 1.47---the smallest among those of the others---indicating that it is much more likely to be ranked the best in terms of MRE. It is also worth noting that, the general model from the machine learning community (e.g., \texttt{PILOT}), although sharing a similar concept to \Model, are even inferior to some state-of-the-art models for configuration performance learning such as \texttt{DeepPerf}. This is because they have not been designed to handle the specific sparsity in configuration data.


When considering different training sample sizes, we see that \Model~performs generally more inferior than the others when the size is too limited, i.e., $S_1$ and $S_2$ for the binary systems. This is expected as when there are too few samples, each local model would have a limited chance to observe the right pattern after the splitting, hence blurring its effectiveness in handling sample sparsity. However, in the other cases (especially for mixed systems that have more data even for $S_1$), \Model~needs far fewer samples to achieve the same accuracy as the best state-of-the-art. For example, on \textsc{Lrzip}, \Model~only needs 295 samples ($S_2$) to achieve an accuracy better than the accuracy of the second best model \texttt{DeepPerf} with 907 samples ($S_5$), saving 67\% effort of data measurements. \textcolor{black}{This is a beneficial outcome of properly handling the sample sparsity. That is, when a global model learns all the configuration data, it is often that case that more data is needed in order to correctly learn the full distribution of the data, as the sparsity causes the data points to spread in the landscape. In contrast, with \Model~where the data can be properly divided, each local model can quickly learn the distribution of data in its local division, since the data is much more condensed to each other. }

Another observation is that the improvements of \Model~is much more obvious in mixed systems than those for binary systems. This is because: (1) the binary systems have fewer training samples as they have a smaller configuration space. Therefore, the data learned by each local model is more restricted. (2) The issue of sample sparsity is more severe on mixed systems, as their configuration landscape is more complex and comes with finer granularity.


As a result, we anticipate that the benefit of \Model~can be amplified under more complex configurable systems and/or when the size of the training data sample increases.




To summarize, we can answer \textbf{RQ1} as:

\begin{quotebox}
   \noindent
   \textit{\textbf{RQ1:} For modeling configuration performance, \Model~performs similar or significantly more accurate than the best state-of-the-art approach in 44 out of 60 cases (73\%), in which 31 cases obtain the sole best results with up to $1.61\times$ improvements. It also needs fewer samples to achieve the same accuracy and the benefits can be amplified with complex systems/more training samples.} 
\end{quotebox}

\subsection{\Model~under Different Local Models}
\label{subsec:local}

\subsubsection{Method}

Since the paradigm of dividable learning is naturally agnostic to the underlying local models, we seek to understand how well \Model~performs with different local models against their global model counterparts (i.e., using them directly to learn the entire training dataset without dividing). To that end, we run experiments on a set of global models available in \texttt{scikit-learn} that are widely used in software engineering tasks to make predictions directly~\cite{DBLP:conf/msr/GongC22}, such as regularized Deep Neural Network (\texttt{DNN}), Decision Tree (\texttt{CART}), Random Forest (\texttt{RF}), Linear Regression (\texttt{LR}), and Support Vector Regression (\texttt{SVR}). We used the same settings as those for \textbf{RQ1}. 
Similarly, we show the ranks $r$ produced by the Scott-Knott test and the median MRE (IQR) of the evaluation results.

\subsubsection{Result}

From Table~\ref{tb:scott-knott}, it is clear that when examining each pair of the counterparts in terms of the average rank, i.e., \texttt{DaL$_X$} and \textit{X}, \Model~can indeed improve the accuracy of the local model via the concept of ``divide-and-learn''. Such an improvement is particularly obvious on some simple models, such as \texttt{LR}, which might lead to $26.08\times$ better MRE on \textsc{NGINX} under $S_1$. This confirms the generality and flexibility of \Model: for example, when \texttt{LR} needs to be used as the local model for the sake of training overhead, \Model~can still significantly improve the results compared with its global counterpart. Interestingly, when pairing \Model~with \texttt{CART} as the local model, it remains slightly better than using \texttt{CART} as the global model alone, even though their learning procedures are similar. This is possible as the actual result of the computed loss can be different in \texttt{CART} when it is presented with different proportions of the data samples. Yet, the resulting MREs do not differ much as can be seen.

It is worth noting that, the default of \Model, which uses the \texttt{HINNPerf} as the local model, still performs significantly better than the others: out of the 60 cases, 33 of them come with the best MRE and there are 37 cases with the best rank, leading to an average rank of 1.63 which is again the best of amongst the others. This aligns with the previous findings that \texttt{HINNPerf} handles the feature sparsity better~\cite{DBLP:journals/tosem/ChengGZ23}. 

Therefore, for \textbf{RQ2}, we say:

    
    
    
    

\begin{quotebox}
   \noindent
   \textit{\textbf{RQ2:} Thanks to the paradigm of dividable learning, \Model~is model-agnostic, being able to significantly improve a range of global models when using them as the underlying local model in configuration performance learning. We also reveal that pairing with \Model~with \texttt{HINNPerf} can lead to considerably better accuracy in this work.}
\end{quotebox}

\subsection{Comparing with Other Ensemble Approaches}
\label{subsec:rq3}
\subsubsection{Method}
Since \Model~works with multiple local models, it could naturally be compared with the ensemble learning approaches that also involve a set of local models. Here, we compare \Model~with the most common ensemble learning approaches such as \texttt{Bagging} and \texttt{Boosting}. \textcolor{black}{For the local model, we employ \texttt{HINNPerf} (the default for \Model) as well as two commonly used machine learning models \texttt{LR} and \texttt{CART}.}




Specifically, \textcolor{black}{we examine \texttt{Bagging$_{HINN}$}~\cite{breiman1996bagging}  (an aggregated ensemble of \texttt{HINNPerf} based on random data projection) and \texttt{AdaBoosting$_{HINN}$}~\cite{DBLP:conf/eurocolt/FreundS95} (a sequentially learned ensemble of \texttt{HINNPerf} based on majrotiy voting).} We follow similar settings for \texttt{LR} and \texttt{CART}. Yet, since \texttt{CART} is a tree-like structure, there exist more diverse variants of ensembles. Therefore, in this work, we examine  \Model~that uses \texttt{CART} (denoted as \texttt{\Model$_{CART}$}) against \texttt{RF} (a bagging version that trains multiple \texttt{CART} models), \texttt{XGBoost}~\cite{DBLP:conf/kdd/ChenG16} (a boosting version that builds multiple \texttt{gbtree} booster~\cite{liu2014gb}, which is a variant of \texttt{CART}), and \texttt{AdaBoosting$_{CART}$} (another boosting version of \texttt{CART} that does not rely on optimization of a loss function).

Again, we leverage the \texttt{scikit-learn} package. We use the default parameter settings for all these ensemble approaches, which tends to be ideal for most scenarios. The other experiment settings are the same as \textbf{RQ1} and \textbf{RQ2}.

\begin{table*}[h!]
\centering
\caption{\Model~against the other ensemble approaches. The format is the same as Table~\ref{tb:vsSOTA}.}
\setlength{\tabcolsep}{1.2mm}
\resizebox{\textwidth}{!}{ 
\begin{tabular}{p{1cm}lp{0.2cm}lp{0.2cm}lp{0.2cm}l||p{0.2cm}lp{0.2cm}lp{0.2cm}l||p{0.2cm}lp{0.2cm}lp{0.2cm}lp{0.2cm}l} \toprule
\multirow{2}{*}{System} & \multicolumn{1}{c}{\multirow{2}{*}{Size}} & \multicolumn{2}{c}{\texttt{DaL}} & \multicolumn{2}{c}{\texttt{Bagging$_ {HINN}$}} & \multicolumn{2}{c}{\texttt{AdaBoost$_ {HINN}$}} & \multicolumn{2}{c}{\texttt{DaL$_{LR}$}} & \multicolumn{2}{c}{\texttt{Bagging$_{LR}$}} & \multicolumn{2}{c}{\texttt{AdaBoost$_{LR}$}} & \multicolumn{2}{c}{\texttt{DaL$_{CART}$}} & \multicolumn{2}{c}{\texttt{RF}} & \multicolumn{2}{c}{\texttt{XGBoost}} & \multicolumn{2}{c}{\texttt{AdaBoost$_{CART}$}} \\  \cline{3-22}
 & \multicolumn{1}{c}{} & $r$ & Med (IQR) & $r$ & Med (IQR) & $r$ & Med (IQR) & $r$ & Med (IQR) & $r$ & Med (IQR) & $r$ & Med (IQR) & $r$ & Med (IQR) & $r$ & Med (IQR) & $r$ & Med (IQR) & $r$ & Med (IQR) \\  \hline
\multirow{5}{*}{\textsc{Apache}} & $S_1$ & \cellcolor{green!20}\textbf{1} & \cellcolor{green!20}\textbf{21.86 (7.36)} & 3 & 24.64 (9.80) & 2 & 24.78 (6.66) & \cellcolor{green!20}\textbf{1} & \cellcolor{green!20}\textbf{19.71 (6.80)} & 2 & 20.81 (4.47) & 2 & 20.44 (5.03) & 2 & 21.61 (6.69) & 3 & 21.20 (7.72) & \cellcolor{red!20}\textbf{1} & \cellcolor{red!20}\textbf{19.03 (7.50)} & 4 & 24.06 (7.47) \\
 & $S_2$ & \cellcolor{green!20}\textbf{1} & \cellcolor{green!20}\textbf{12.89 (7.87)} & \textbf{1} & \textbf{14.84 (6.31)} & \textbf{1} & \textbf{15.61 (5.39)} & \cellcolor{green!20}\textbf{1} & \cellcolor{green!20}\textbf{10.77 (7.55)} & 3 & 17.64 (2.12) & 2 & 16.69 (2.45) & \cellcolor{green!20}2 & \cellcolor{green!20}10.12 (5.05) & 2 & 12.02 (6.58) & \textbf{1} & \textbf{10.70 (3.76)} & 3 & 13.82 (4.42) \\
 & $S_3$ & \cellcolor{green!20}\textbf{1} & \cellcolor{green!20}\textbf{6.56 (1.92)} & 2 & 11.20 (5.52) & 3 & 12.19 (10.08) & \cellcolor{green!20}\textbf{1} & \cellcolor{green!20}\textbf{7.42 (2.14)} & 2 & 16.11 (1.34) & 2 & 16.25 (1.68) & 2 & 9.10 (1.31) & \cellcolor{red!20}\textbf{1} & \cellcolor{red!20}\textbf{7.65 (1.46)} & 2 & 9.37 (1.36) & 2 & 9.04 (1.70) \\
 & $S_4$ & \cellcolor{green!20}\textbf{1} & \cellcolor{green!20}\textbf{6.87 (1.95)} & 3 & 9.17 (23.67) & 2 & 9.86 (2.32) & \cellcolor{green!20}\textbf{1} & \cellcolor{green!20}\textbf{6.39 (1.16)} & 2 & 16.20 (0.82) & 2 & 15.96 (1.40) & 3 & 7.97 (1.35) & \cellcolor{red!20}\textbf{1} & \cellcolor{red!20}\textbf{6.63 (1.19)} & 4 & 8.70 (1.14) & 2 & 7.64 (0.99) \\
 & $S_5$ & \cellcolor{green!20}\textbf{1} & \cellcolor{green!20}\textbf{5.78 (1.08)} & 3 & 7.81 (24.03) & 2 & 8.90 (3.19) & \cellcolor{green!20}\textbf{1} & \cellcolor{green!20}\textbf{5.75 (0.55)} & 2 & 16.04 (1.37) & 2 & 16.18 (1.80) & 3 & 7.63 (1.00) & \cellcolor{red!20}\textbf{1} & \cellcolor{red!20}\textbf{6.43 (1.11)} & 4 & 8.17 (1.69) & 2 & 7.22 (0.97) \\ \hline
\multirow{5}{*}{\textsc{BDB-C}} & $S_1$ & \cellcolor{green!20}\textbf{1} & \cellcolor{green!20}\textbf{83.60 (208.75)} & \textbf{1} & \textbf{158.49 (97.21)} & 2 & 239.27 (165.36) & \cellcolor{green!20}\textbf{1} & \cellcolor{green!20}\textbf{65.09 (37.30)} & 3 & 485.41 (235.62) & 2 & 453.92 (139.97) & \textbf{1} & \textbf{47.20 (19.75)} & 4 & 253.78 (93.63) & \cellcolor{red!20}2 & \cellcolor{red!20}40.05 (54.70) & 3 & 211.60 (191.00) \\
 & $S_2$ & \cellcolor{green!20}\textbf{1} & \cellcolor{green!20}\textbf{21.92 (17.91)} & 2 & 87.45 (64.92) & 3 & 175.89 (62.71) & \cellcolor{green!20}\textbf{1} & \cellcolor{green!20}\textbf{49.04 (47.11)} & 3 & 529.58 (158.69) & 2 & 565.28 (157.43) & 2 & 16.81 (8.59) & 3 & 73.41 (118.30) & \cellcolor{red!20}\textbf{1} & \cellcolor{red!20}\textbf{16.30 (4.99)} & 3 & 95.94 (19.68) \\
 & $S_3$ & \cellcolor{green!20}\textbf{1} & \cellcolor{green!20}\textbf{8.31 (7.52)} & 2 & 52.55 (22.07) & 3 & 100.26 (43.36) & \cellcolor{green!20}\textbf{1} & \cellcolor{green!20}\textbf{29.00 (78.81)} & 2 & 460.89 (90.55) & 3 & 496.74 (56.61) & \cellcolor{green!20}2 & \cellcolor{green!20}6.72 (5.82) & 3 & 18.64 (19.42) & \textbf{1} & \textbf{10.23 (1.63)} & 4 & 43.94 (7.78) \\
 & $S_4$ & \cellcolor{green!20}\textbf{1} & \cellcolor{green!20}\textbf{3.60 (4.50)} & 2 & 35.88 (17.56) & 3 & 77.71 (33.23) & \cellcolor{green!20}\textbf{1} & \cellcolor{green!20}\textbf{24.55 (4.18)} & 2 & 440.07 (77.70) & 3 & 475.09 (43.17) & \cellcolor{green!20}\textbf{1} & \cellcolor{green!20}\textbf{4.15 (2.78)} & 3 & 11.32 (6.77) & 2 & 8.51 (0.97) & 3 & 21.47 (5.45) \\
 & $S_5$ & \cellcolor{green!20}\textbf{1} & \cellcolor{green!20}\textbf{1.88 (1.24)} & 2 & 31.49 (17.12) & 3 & 58.63 (19.41) & \cellcolor{green!20}\textbf{1} & \cellcolor{green!20}\textbf{22.88 (3.78)} & 2 & 460.51 (88.56) & 3 & 473.16 (64.47) & \cellcolor{green!20}\textbf{1} & \cellcolor{green!20}\textbf{3.24 (3.09)} & 3 & 6.64 (3.35) & 2 & 7.50 (1.11) & 3 & 11.07 (2.28) \\ \hline
\multirow{5}{*}{\textsc{BDB-J}} & $S_1$ & \cellcolor{green!20}\textbf{1} & \cellcolor{green!20}\textbf{5.37 (4.74)} & 2 & 11.23 (6.65) & 2 & 14.06 (7.13) & \cellcolor{green!20}\textbf{1} & \cellcolor{green!20}\textbf{5.31 (2.96)} & 3 & 43.30 (6.44) & 2 & 40.89 (6.68) & \cellcolor{green!20}\textbf{1} & \cellcolor{green!20}\textbf{2.98 (0.81)} & 3 & 8.36 (10.48) & 2 & 7.39 (1.07) & \textbf{1} & \textbf{3.61 (2.31)} \\
 & $S_2$ & \cellcolor{green!20}\textbf{1} & \cellcolor{green!20}\textbf{1.83 (0.45)} & 3 & 5.24 (33.48) & 2 & 5.17 (3.19) & \cellcolor{green!20}\textbf{1} & \cellcolor{green!20}\textbf{3.77 (0.49)} & 2 & 37.75 (5.70) & 3 & 39.84 (6.41) & 2 & 2.05 (0.33) & \cellcolor{red!20}\textbf{1} & \cellcolor{red!20}\textbf{1.82 (0.29)} & 3 & 6.10 (0.51) & \textbf{1} & \textbf{1.85 (0.20)} \\
 & $S_3$ & \cellcolor{green!20}\textbf{1} & \cellcolor{green!20}\textbf{1.58 (0.26)} & 3 & 3.29 (1.29) & 2 & 3.75 (0.62) & \cellcolor{green!20}\textbf{1} & \cellcolor{green!20}\textbf{3.41 (0.46)} & 2 & 37.72 (3.80) & 2 & 38.85 (4.51) & 3 & 1.92 (0.27) & \cellcolor{red!20}\textbf{1} & \cellcolor{red!20}\textbf{1.57 (0.21)} & 4 & 5.71 (0.61) & 2 & 1.78 (0.25) \\
 & $S_4$ & \cellcolor{green!20}\textbf{1} & \cellcolor{green!20}\textbf{1.45 (0.22)} & 3 & 2.93 (0.80) & 2 & 2.70 (0.59) & \cellcolor{green!20}\textbf{1} & \cellcolor{green!20}\textbf{3.57 (0.66)} & 3 & 37.07 (3.03) & 2 & 36.17 (4.46) & 3 & 1.67 (0.29) & \cellcolor{red!20}\textbf{1} & \cellcolor{red!20}\textbf{1.43 (0.21)} & 4 & 5.48 (0.55) & 2 & 1.66 (0.21) \\
 & $S_5$ & \cellcolor{green!20}\textbf{1} & \cellcolor{green!20}\textbf{1.40 (0.34)} & 3 & 2.48 (0.56) & 2 & 2.39 (0.38) & \cellcolor{green!20}\textbf{1} & \cellcolor{green!20}\textbf{3.23 (0.65)} & 3 & 37.27 (3.01) & 2 & 36.52 (5.25) & 3 & 1.67 (0.43) & \cellcolor{red!20}\textbf{1} & \cellcolor{red!20}\textbf{1.40 (0.23)} & 4 & 5.34 (0.42) & 2 & 1.55 (0.29) \\ \hline
\multirow{5}{*}{\textsc{kanzi}} & $S_1$ & \cellcolor{green!20}\textbf{1} & \cellcolor{green!20}\textbf{196.15 (213.73)} & 2 & 256.87 (151.84) & 3 & 410.20 (187.36) & \cellcolor{green!20}\textbf{1} & \cellcolor{green!20}\textbf{551.03 (1244.41)} & \textbf{1} & \textbf{1158.20 (365.64)} & \textbf{1} & \textbf{1087.10 (248.90)} & \cellcolor{green!20}\textbf{1} & \cellcolor{green!20}\textbf{178.67 (123.16)} & 2 & 359.19 (147.95) & \textbf{1} & \textbf{190.78 (98.22)} & 2 & 426.88 (221.92) \\
 & $S_2$ & \cellcolor{green!20}\textbf{1} & \cellcolor{green!20}\textbf{95.31 (51.34)} & 2 & 184.70 (53.30) & 3 & 285.91 (71.88) & \cellcolor{green!20}\textbf{1} & \cellcolor{green!20}\textbf{383.06 (493.33)} & 3 & 1224.60 (377.56) & 2 & 1259.43 (331.27) & \textbf{1} & \textbf{86.55 (35.63)} & 2 & 155.50 (51.71) & \cellcolor{red!20}\textbf{1} & \cellcolor{red!20}\textbf{84.19 (30.01)} & 3 & 253.92 (74.19) \\
 & $S_3$ & \cellcolor{green!20}\textbf{1} & \cellcolor{green!20}\textbf{59.30 (38.47)} & 2 & 138.52 (75.87) & 3 & 233.73 (39.89) & \cellcolor{green!20}2 & \cellcolor{green!20}177.35 (384.90) & 3 & 1194.11 (1403.01) & \textbf{1} & \textbf{1368.20 (250.10)} & \cellcolor{green!20}\textbf{1} & \cellcolor{green!20}\textbf{62.50 (21.15)} & 2 & 87.44 (21.38) & \textbf{1} & \textbf{65.34 (21.07)} & 3 & 199.28 (45.12) \\
 & $S_4$ & \cellcolor{green!20}\textbf{1} & \cellcolor{green!20}\textbf{42.93 (18.75)} & 2 & 129.60 (73.04) & 3 & 191.77 (38.43) & \cellcolor{green!20}2 & \cellcolor{green!20}134.99 (522.74) & 3 & 1041.55 (230.02) & \textbf{1} & \textbf{1396.16 (189.09)} & \cellcolor{green!20}\textbf{1} & \cellcolor{green!20}\textbf{48.69 (10.31)} & 3 & 65.20 (18.86) & 2 & 54.19 (10.26) & 4 & 158.05 (37.81) \\
 & $S_5$ & \cellcolor{green!20}\textbf{1} & \cellcolor{green!20}\textbf{35.29 (17.13)} & 2 & 114.69 (48.99) & 3 & 172.70 (46.58) & \cellcolor{green!20}\textbf{1} & \cellcolor{green!20}\textbf{605.42 (1213.79)} & 3 & 1014.62 (255.48) & 2 & 1266.23 (184.09) & \cellcolor{green!20}\textbf{1} & \cellcolor{green!20}\textbf{44.15 (8.53)} & 3 & 52.53 (9.96) & 2 & 48.74 (9.81) & 4 & 132.03 (25.97) \\ \hline
\multirow{5}{*}{\textsc{SQLite}} & $S_1$ & 2 & 73.52 (20.40) & \textbf{1} & \textbf{68.63 (16.67)} & \cellcolor{red!20}3 & \cellcolor{red!20}65.45 (19.46) & 2 & 105.75 (83.62) & \cellcolor{red!20}\textbf{1} & \cellcolor{red!20}\textbf{74.67 (20.68)} & \textbf{1} & \textbf{76.22 (21.29)} & 3 & 82.77 (20.99) & \textbf{1} & \textbf{65.16 (12.06)} & \cellcolor{red!20}\textbf{1} & \cellcolor{red!20}\textbf{62.32 (11.64)} & 2 & 70.76 (20.26) \\
 & $S_2$ & 2 & 76.41 (16.66) & \cellcolor{red!20}\textbf{1} & \cellcolor{red!20}\textbf{70.36 (14.36)} & \textbf{1} & \textbf{70.52 (24.47)} & 3 & 161.86 (135.45) & 2 & 95.26 (36.34) & \cellcolor{red!20}\textbf{1} & \cellcolor{red!20}\textbf{89.04 (34.14)} & 3 & 86.76 (20.84) & \textbf{1} & \textbf{69.64 (10.76)} & \cellcolor{red!20}\textbf{1} & \cellcolor{red!20}\textbf{64.04 (13.07)} & 2 & 75.59 (24.63) \\
 & $S_3$ & \textbf{1} & \textbf{71.63 (11.58)} & \cellcolor{red!20}2 & \cellcolor{red!20}67.30 (12.50) & \textbf{1} & \textbf{69.51 (14.85)} & 3 & 103.75 (29.66) & \cellcolor{red!20}\textbf{1} & \cellcolor{red!20}\textbf{76.20 (21.41)} & 2 & 84.61 (35.20) & 3 & 87.65 (24.29) & \textbf{1} & \textbf{69.77 (13.75)} & \cellcolor{red!20}\textbf{1} & \cellcolor{red!20}\textbf{64.25 (11.83)} & 2 & 70.77 (18.66) \\
 & $S_4$ & \cellcolor{green!20}\textbf{1} & \cellcolor{green!20}\textbf{66.94 (10.61)} & 2 & 69.71 (10.60) & 2 & 70.93 (11.85) & 2 & 82.25 (23.16) & \cellcolor{red!20}\textbf{1} & \cellcolor{red!20}\textbf{68.94 (13.83)} & 2 & 82.27 (27.41) & 3 & 86.47 (20.30) & 2 & 69.29 (9.25) & \cellcolor{red!20}\textbf{1} & \cellcolor{red!20}\textbf{62.19 (11.78)} & 2 & 70.01 (16.80) \\
 & $S_5$ & \cellcolor{green!20}\textbf{1} & \cellcolor{green!20}\textbf{64.20 (13.91)} & 2 & 69.41 (10.96) & 2 & 71.51 (14.27) & 2 & 76.06 (16.27) & \cellcolor{red!20}\textbf{1} & \cellcolor{red!20}\textbf{68.03 (10.38)} & 3 & 80.61 (23.78) & 3 & 84.43 (12.41) & 2 & 70.42 (9.55) & \cellcolor{red!20}\textbf{1} & \cellcolor{red!20}\textbf{63.59 (12.77)} & 2 & 71.19 (13.68) \\ \hline
\multirow{5}{*}{\textsc{x264}} & $S_1$ & \cellcolor{green!20}\textbf{1} & \cellcolor{green!20}\textbf{10.21 (4.35)} & 3 & 13.65 (14.17) & 2 & 15.17 (5.23) & \cellcolor{green!20}\textbf{1} & \cellcolor{green!20}\textbf{10.43 (3.40)} & \textbf{1} & \textbf{10.90 (2.21)} & \textbf{1} & \textbf{11.58 (2.30)} & 2 & 10.40 (2.15) & \cellcolor{red!20}\textbf{1} & \cellcolor{red!20}\textbf{8.18 (2.78)} & 3 & 11.19 (2.06) & 4 & 12.22 (5.95) \\
 & $S_2$ & \cellcolor{green!20}\textbf{1} & \cellcolor{green!20}\textbf{2.72 (1.52)} & 3 & 8.44 (24.28) & 2 & 8.68 (2.80) & \cellcolor{green!20}\textbf{1} & \cellcolor{green!20}\textbf{3.51 (0.98)} & 2 & 8.11 (1.06) & 3 & 8.74 (1.21) & \textbf{1} & \textbf{6.57 (1.67)} & \cellcolor{red!20}\textbf{1} & \cellcolor{red!20}\textbf{6.42 (1.20)} & 3 & 8.53 (0.86) & 2 & 6.82 (1.28) \\
 & $S_3$ & \cellcolor{green!20}\textbf{1} & \cellcolor{green!20}\textbf{1.57 (1.52)} & 2 & 5.06 (48.28) & 2 & 6.73 (57.61) & \cellcolor{green!20}\textbf{1} & \cellcolor{green!20}\textbf{2.84 (0.52)} & 2 & 7.57 (0.81) & 3 & 7.91 (0.80) & \textbf{1} & \textbf{5.01 (1.42)} & \cellcolor{red!20}\textbf{1} & \cellcolor{red!20}\textbf{4.95 (0.85)} & 3 & 7.45 (0.85) & 2 & 5.59 (0.64) \\
 & $S_4$ & \cellcolor{green!20}\textbf{1} & \cellcolor{green!20}\textbf{1.30 (1.17)} & 2 & 4.04 (45.08) & 2 & 4.86 (39.30) & \cellcolor{green!20}\textbf{1} & \cellcolor{green!20}\textbf{2.60 (0.52)} & 2 & 7.27 (0.45) & 3 & 7.44 (0.37) & \cellcolor{green!20}\textbf{1} & \cellcolor{green!20}\textbf{3.52 (1.33)} & 2 & 4.04 (0.86) & 4 & 6.61 (0.73) & 3 & 4.53 (0.67) \\
 & $S_5$ & \cellcolor{green!20}\textbf{1} & \cellcolor{green!20}\textbf{0.96 (0.79)} & 2 & 3.80 (28.85) & 2 & 3.98 (3.19) & \cellcolor{green!20}\textbf{1} & \cellcolor{green!20}\textbf{2.10 (1.04)} & 2 & 7.12 (0.67) & 3 & 7.19 (0.46) & \cellcolor{green!20}\textbf{1} & \cellcolor{green!20}\textbf{2.54 (1.04)} & 2 & 2.98 (0.77) & 4 & 6.26 (0.58) & 3 & 3.55 (0.53) \\ \hline
\multirow{5}{*}{\textsc{Dune}} & $S_1$ & \cellcolor{green!20}\textbf{1} & \cellcolor{green!20}\textbf{9.37 (1.15)} & 3 & 13.64 (1.55) & 2 & 12.26 (0.86) & \cellcolor{green!20}\textbf{1} & \cellcolor{green!20}\textbf{11.21 (0.88)} & 2 & 13.01 (1.78) & 3 & 13.76 (5.73) & \cellcolor{green!20}\textbf{1} & \cellcolor{green!20}\textbf{9.04 (1.13)} & \textbf{1} & \textbf{9.27 (0.56)} & 2 & 10.12 (0.71) & \textbf{1} & \textbf{9.29 (0.65)} \\
 & $S_2$ & \cellcolor{green!20}\textbf{1} & \cellcolor{green!20}\textbf{5.95 (0.54)} & 2 & 8.72 (2.48) & 3 & 9.66 (1.10) & \cellcolor{green!20}\textbf{1} & \cellcolor{green!20}\textbf{11.13 (0.57)} & 2 & 12.94 (1.04) & 3 & 27.72 (8.04) & \cellcolor{green!20}\textbf{1} & \cellcolor{green!20}\textbf{6.25 (0.42)} & 3 & 7.27 (0.40) & 4 & 8.39 (0.26) & 2 & 6.62 (0.36) \\
 & $S_3$ & \cellcolor{green!20}\textbf{1} & \cellcolor{green!20}\textbf{4.81 (0.48)} & 2 & 7.87 (1.49) & 3 & 8.77 (1.21) & \cellcolor{green!20}\textbf{1} & \cellcolor{green!20}\textbf{11.09 (0.54)} & 2 & 13.10 (0.50) & 3 & 29.45 (5.73) & \cellcolor{green!20}\textbf{1} & \cellcolor{green!20}\textbf{5.48 (0.42)} & 3 & 6.63 (0.33) & 4 & 8.07 (0.24) & 2 & 5.79 (0.24) \\
 & $S_4$ & \cellcolor{green!20}\textbf{1} & \cellcolor{green!20}\textbf{4.23 (0.35)} & 2 & 7.22 (0.99) & 3 & 8.58 (0.97) & \cellcolor{green!20}\textbf{1} & \cellcolor{green!20}\textbf{11.04 (0.43)} & 2 & 13.15 (0.47) & 3 & 29.21 (5.15) & \cellcolor{green!20}\textbf{1} & \cellcolor{green!20}\textbf{5.06 (0.32)} & 3 & 5.97 (0.45) & 4 & 7.85 (0.26) & 2 & 5.26 (0.19) \\
 & $S_5$ & \cellcolor{green!20}\textbf{1} & \cellcolor{green!20}\textbf{3.98 (0.19)} & 3 & 9.57 (2.40) & 2 & 9.36 (0.92) & \cellcolor{green!20}\textbf{1} & \cellcolor{green!20}\textbf{10.93 (0.48)} & 2 & 13.15 (0.61) & 3 & 29.14 (3.85) & \cellcolor{green!20}\textbf{1} & \cellcolor{green!20}\textbf{4.69 (0.35)} & 3 & 5.68 (0.35) & 4 & 7.66 (0.17) & 2 & 4.90 (0.23) \\ \hline
\multirow{5}{*}{\textsc{HIPA$^{cc}$}} & $S_1$ & \cellcolor{green!20}\textbf{1} & \cellcolor{green!20}\textbf{7.50 (1.42)} & 2 & 10.21 (0.86) & 2 & 10.19 (0.66) & \cellcolor{green!20}2 & \cellcolor{green!20}12.70 (1.22) & 3 & 20.95 (699.46) & \textbf{1} & \textbf{23.50 (2.17)} & 2 & 12.03 (0.93) & \cellcolor{red!20}\textbf{1} & \cellcolor{red!20}\textbf{11.08 (0.66)} & \textbf{1} & \textbf{11.19 (0.60)} & 3 & 14.24 (0.61) \\
 & $S_2$ & \cellcolor{green!20}\textbf{1} & \cellcolor{green!20}\textbf{4.48 (0.38)} & 2 & 7.41 (1.03) & 3 & 7.87 (0.86) & \cellcolor{green!20}3 & \cellcolor{green!20}11.64 (0.35) & \textbf{1} & \textbf{19.70 (0.89)} & 2 & 21.82 (0.68) & \textbf{1} & \textbf{8.47 (1.10)} & \cellcolor{red!20}\textbf{1} & \cellcolor{red!20}\textbf{8.46 (0.70)} & 2 & 9.88 (0.37) & 3 & 10.48 (0.63) \\
 & $S_3$ & \cellcolor{green!20}\textbf{1} & \cellcolor{green!20}\textbf{3.43 (0.34)} & 2 & 6.14 (0.96) & 2 & 6.55 (0.68) & \cellcolor{green!20}\textbf{1} & \cellcolor{green!20}\textbf{11.43 (0.34)} & 2 & 19.34 (0.88) & 3 & 21.82 (0.42) & \cellcolor{green!20}\textbf{1} & \cellcolor{green!20}\textbf{6.62 (0.80)} & \textbf{1} & \textbf{6.69 (0.51)} & 3 & 9.41 (0.33) & 2 & 8.21 (0.47) \\
 & $S_4$ & \cellcolor{green!20}\textbf{1} & \cellcolor{green!20}\textbf{2.60 (0.15)} & 2 & 4.31 (0.77) & 3 & 5.07 (0.74) & \cellcolor{green!20}\textbf{1} & \cellcolor{green!20}\textbf{11.13 (0.25)} & 2 & 18.94 (0.58) & 3 & 21.75 (0.81) & \cellcolor{green!20}\textbf{1} & \cellcolor{green!20}\textbf{4.42 (0.22)} & 2 & 4.47 (0.32) & 4 & 8.81 (0.20) & 3 & 5.37 (0.24) \\
 & $S_5$ & \cellcolor{green!20}\textbf{1} & \cellcolor{green!20}\textbf{2.11 (0.08)} & 2 & 3.04 (0.81) & 3 & 3.86 (0.58) & \cellcolor{green!20}\textbf{1} & \cellcolor{green!20}\textbf{11.01 (0.15)} & 2 & 18.79 (0.35) & 3 & 21.88 (0.76) & \cellcolor{green!20}\textbf{1} & \cellcolor{green!20}\textbf{2.69 (0.15)} & \cellcolor{red!20}\textbf{1} & \cellcolor{red!20}\textbf{2.69 (0.13)} & 3 & 8.30 (0.20) & 2 & 3.10 (0.13) \\ \hline
\multirow{5}{*}{\textsc{HSMGP}} & $S_1$ & \cellcolor{green!20}\textbf{1} & \cellcolor{green!20}\textbf{4.14 (1.44)} & 2 & 7.01 (1.39) & 2 & 8.63 (1.77) & \cellcolor{green!20}\textbf{1} & \cellcolor{green!20}\textbf{7.72 (0.80)} & 2 & 55.90 (20.70) & 3 & 80.26 (38.86) & 3 & 21.65 (2.31) & 2 & 16.63 (2.10) & \cellcolor{red!20}\textbf{1} & \cellcolor{red!20}\textbf{14.63 (1.92)} & 4 & 25.42 (5.90) \\
 & $S_2$ & \cellcolor{green!20}\textbf{1} & \cellcolor{green!20}\textbf{2.06 (0.19)} & 3 & 4.85 (4.52) & 2 & 5.61 (1.17) & \cellcolor{green!20}\textbf{1} & \cellcolor{green!20}\textbf{7.35 (0.47)} & 2 & 51.49 (11.94) & 3 & 81.46 (17.14) & 3 & 15.59 (1.41) & 2 & 11.46 (1.78) & \cellcolor{red!20}\textbf{1} & \cellcolor{red!20}\textbf{10.59 (0.88)} & 4 & 18.59 (1.67) \\
 & $S_3$ & \cellcolor{green!20}\textbf{1} & \cellcolor{green!20}\textbf{1.42 (0.13)} & 2 & 3.60 (0.69) & 3 & 3.87 (0.75) & \cellcolor{green!20}\textbf{1} & \cellcolor{green!20}\textbf{7.14 (0.35)} & 2 & 51.65 (8.15) & 3 & 80.54 (19.98) & 2 & 11.35 (0.63) & \textbf{1} & \textbf{7.58 (0.67)} & \cellcolor{red!20}\textbf{1} & \cellcolor{red!20}\textbf{7.52 (0.46)} & 3 & 12.10 (0.67) \\
 & $S_4$ & \cellcolor{green!20}\textbf{1} & \cellcolor{green!20}\textbf{1.33 (0.08)} & 2 & 3.57 (0.95) & 2 & 3.91 (0.73) & \cellcolor{green!20}\textbf{1} & \cellcolor{green!20}\textbf{7.16 (0.17)} & 2 & 49.96 (10.24) & 3 & 66.32 (36.72) & 3 & 10.13 (0.57) & \cellcolor{red!20}\textbf{1} & \cellcolor{red!20}\textbf{6.88 (0.61)} & 2 & 7.10 (0.38) & 4 & 10.47 (0.65) \\
 & $S_5$ & \cellcolor{green!20}\textbf{1} & \cellcolor{green!20}\textbf{1.19 (0.08)} & 2 & 3.02 (1.31) & 2 & 3.58 (0.78) & \cellcolor{green!20}\textbf{1} & \cellcolor{green!20}\textbf{7.15 (0.20)} & 2 & 49.18 (3.13) & 3 & 85.14 (39.24) & 4 & 7.46 (0.46) & \cellcolor{red!20}\textbf{1} & \cellcolor{red!20}\textbf{5.15 (0.28)} & 2 & 6.11 (0.26) & 3 & 7.04 (0.28) \\ \hline
\multirow{5}{*}{\textsc{Lrzip}} & $S_1$ & \textbf{1} & \textbf{31.88 (10.96)} & \cellcolor{red!20}2 & \cellcolor{red!20}31.78 (29.81) & 3 & 70.36 (28.67) & \cellcolor{green!20}\textbf{1} & \cellcolor{green!20}\textbf{117.37 (24.65)} & 2 & 363.17 (114.48) & 3 & 475.86 (188.56) & 2 & 17.95 (10.85) & 3 & 21.71 (7.45) & \cellcolor{red!20}\textbf{1} & \cellcolor{red!20}\textbf{16.71 (4.27)} & 4 & 76.46 (23.85) \\
 & $S_2$ & \cellcolor{green!20}\textbf{1} & \cellcolor{green!20}\textbf{10.51 (4.39)} & 2 & 46.26 (25.10) & 3 & 74.19 (33.01) & \cellcolor{green!20}\textbf{1} & \cellcolor{green!20}\textbf{109.11 (20.26)} & 2 & 315.54 (57.21) & 3 & 474.88 (196.54) & \cellcolor{green!20}\textbf{1} & \cellcolor{green!20}\textbf{9.09 (2.56)} & 2 & 10.18 (2.12) & \textbf{1} & \textbf{10.19 (1.79)} & 3 & 36.47 (12.40) \\
 & $S_3$ & \cellcolor{green!20}\textbf{1} & \cellcolor{green!20}\textbf{8.04 (2.37)} & 2 & 30.06 (17.51) & 3 & 59.65 (24.60) & \cellcolor{green!20}\textbf{1} & \cellcolor{green!20}\textbf{104.20 (17.55)} & 2 & 321.82 (49.06) & 3 & 360.11 (241.44) & \cellcolor{green!20}\textbf{1} & \cellcolor{green!20}\textbf{7.77 (2.09)} & \textbf{1} & \textbf{7.90 (1.27)} & 2 & 8.43 (0.69) & 3 & 26.46 (7.48) \\
 & $S_4$ & \cellcolor{green!20}\textbf{1} & \cellcolor{green!20}\textbf{5.95 (0.99)} & 2 & 31.16 (20.54) & 3 & 71.33 (22.86) & \cellcolor{green!20}\textbf{1} & \cellcolor{green!20}\textbf{103.74 (14.29)} & 2 & 325.19 (51.13) & 3 & 346.60 (131.69) & \cellcolor{green!20}\textbf{1} & \cellcolor{green!20}\textbf{6.38 (2.72)} & \textbf{1} & \textbf{6.51 (0.69)} & 2 & 7.59 (0.91) & 3 & 19.77 (2.80) \\
 & $S_5$ & \cellcolor{green!20}\textbf{1} & \cellcolor{green!20}\textbf{4.03 (0.26)} & 2 & 23.74 (4.27) & 3 & 58.48 (34.30) & \cellcolor{green!20}\textbf{1} & \cellcolor{green!20}\textbf{105.83 (7.06)} & 2 & 324.82 (32.57) & 3 & 457.93 (299.67) & \cellcolor{green!20}\textbf{1} & \cellcolor{green!20}\textbf{3.58 (0.97)} & 2 & 3.67 (1.31) & 2 & 6.17 (0.43) & 3 & 9.37 (1.21) \\ \hline
\multirow{5}{*}{\textsc{nginx}} & $S_1$ & \cellcolor{green!20}\textbf{1} & \cellcolor{green!20}\textbf{4.11 (1.45)} & 2 & 24.02 (12.20) & 3 & 34.82 (21.61) & \cellcolor{green!20}\textbf{1} & \cellcolor{green!20}\textbf{22.02 (2.11)} & 3 & 576.17 (24.71) & 2 & 550.28 (66.10) & 3 & 14.44 (2.41) & 4 & 84.12 (150.39) & 2 & 6.78 (1.92) & \cellcolor{red!20}\textbf{1} & \cellcolor{red!20}\textbf{4.23 (1.21)} \\
 & $S_2$ & \cellcolor{green!20}\textbf{1} & \cellcolor{green!20}\textbf{2.60 (1.14)} & 2 & 22.99 (9.15) & 3 & 28.21 (2.89) & \cellcolor{green!20}\textbf{1} & \cellcolor{green!20}\textbf{20.25 (1.60)} & 2 & 568.78 (17.19) & 2 & 564.04 (37.05) & 3 & 7.44 (3.14) & 4 & 11.09 (3.33) & 2 & 4.90 (0.44) & \cellcolor{red!20}\textbf{1} & \cellcolor{red!20}\textbf{2.17 (0.48)} \\
 & $S_3$ & \cellcolor{green!20}\textbf{1} & \cellcolor{green!20}\textbf{2.14 (0.40)} & 2 & 23.32 (8.65) & 3 & 27.64 (5.37) & \cellcolor{green!20}\textbf{1} & \cellcolor{green!20}\textbf{19.66 (0.75)} & 2 & 567.70 (19.05) & 2 & 570.41 (23.98) & 2 & 4.81 (0.81) & 3 & 5.68 (1.43) & 2 & 4.56 (0.09) & \cellcolor{red!20}\textbf{1} & \cellcolor{red!20}\textbf{1.85 (0.43)} \\
 & $S_4$ & \cellcolor{green!20}\textbf{1} & \cellcolor{green!20}\textbf{1.99 (0.47)} & 2 & 14.92 (9.09) & 3 & 24.84 (4.19) & \cellcolor{green!20}\textbf{1} & \cellcolor{green!20}\textbf{19.74 (0.82)} & 2 & 566.35 (18.53) & 2 & 566.60 (22.75) & 2 & 4.16 (1.01) & 4 & 4.87 (0.63) & 3 & 4.49 (0.05) & \cellcolor{red!20}\textbf{1} & \cellcolor{red!20}\textbf{1.59 (0.28)} \\
 & $S_5$ & \cellcolor{green!20}\textbf{1} & \cellcolor{green!20}\textbf{1.98 (0.21)} & 2 & 6.11 (6.87) & 3 & 11.91 (11.63) & \cellcolor{green!20}\textbf{1} & \cellcolor{green!20}\textbf{19.17 (1.07)} & 2 & 565.62 (23.39) & 2 & 567.96 (18.51) & 2 & 3.33 (1.00) & 3 & 4.49 (0.56) & 3 & 4.44 (0.06) & \cellcolor{red!20}\textbf{1} & \cellcolor{red!20}\textbf{1.64 (0.30)} \\ \hline
\multirow{5}{*}{\textsc{VP8}} & $S_1$ & \cellcolor{green!20}\textbf{1} & \cellcolor{green!20}\textbf{1.56 (0.18)} & 3 & 6.05 (1.14) & 2 & 8.31 (0.68) & \cellcolor{green!20}\textbf{1} & \cellcolor{green!20}\textbf{12.72 (0.84)} & 3 & 45.51 (7.20) & 2 & 44.69 (5.11) & \cellcolor{green!20}\textbf{1} & \cellcolor{green!20}\textbf{2.57 (0.70)} & 3 & 3.32 (0.79) & 3 & 5.99 (0.31) & 2 & 4.42 (0.46) \\
 & $S_2$ & \cellcolor{green!20}\textbf{1} & \cellcolor{green!20}\textbf{1.15 (0.05)} & 3 & 3.34 (1.09) & 2 & 3.56 (0.43) & \cellcolor{green!20}\textbf{1} & \cellcolor{green!20}\textbf{12.42 (0.44)} & 2 & 42.04 (3.29) & 2 & 42.06 (2.51) & \cellcolor{green!20}\textbf{1} & \cellcolor{green!20}\textbf{1.24 (0.24)} & 2 & 1.45 (0.16) & 3 & 5.25 (0.20) & 2 & 1.44 (0.21) \\
 & $S_3$ & \cellcolor{green!20}\textbf{1} & \cellcolor{green!20}\textbf{1.08 (0.04)} & 3 & 2.91 (0.32) & 2 & 2.66 (0.35) & \cellcolor{green!20}\textbf{1} & \cellcolor{green!20}\textbf{12.41 (0.34)} & 2 & 42.55 (2.55) & 3 & 44.06 (2.66) & 2 & 1.15 (0.12) & 2 & 1.16 (0.16) & 3 & 5.15 (0.15) & \cellcolor{red!20}\textbf{1} & \cellcolor{red!20}\textbf{1.11 (0.09)} \\
 & $S_4$ & \cellcolor{green!20}\textbf{1} & \cellcolor{green!20}\textbf{0.98 (0.05)} & 3 & 2.48 (0.39) & 2 & 2.22 (0.33) & \cellcolor{green!20}\textbf{1} & \cellcolor{green!20}\textbf{12.35 (0.31)} & 2 & 41.90 (3.11) & 3 & 42.98 (2.47) & 3 & 1.07 (0.08) & 2 & 1.03 (0.11) & 4 & 5.07 (0.10) & \cellcolor{red!20}\textbf{1} & \cellcolor{red!20}\textbf{1.02 (0.05)} \\
 & $S_5$ & \cellcolor{green!20}\textbf{1} & \cellcolor{green!20}\textbf{0.90 (0.06)} & 3 & 2.20 (0.33) & 2 & 1.66 (0.16) & \cellcolor{green!20}\textbf{1} & \cellcolor{green!20}\textbf{12.41 (0.27)} & 2 & 42.79 (1.80) & 3 & 43.19 (2.36) & 3 & 0.94 (0.04) & \cellcolor{red!20}\textbf{1} & \cellcolor{red!20}\textbf{0.85 (0.03)} & 4 & 4.89 (0.09) & 2 & 0.88 (0.04) \\ \hline
\multicolumn{2}{c}{Average $r$} & 1.03 &  & 2.22 &  & 2.42 &  & 1.2 &  & 2.1 &  & 2.38 &  & 1.85 &  & 2 &  & 2.35 &  & 2.43 &   \\ 
\bottomrule
\end{tabular}
}

\label{tb:ensemble}
\end{table*}

\subsubsection{Results}
The results can be seen in Table~\ref{tb:ensemble}, in which we compare the approaches within each type of local model, e.g., only comparing those use \texttt{HINNPerf} as the local model. 
\textcolor{black}{Clearly, when using \texttt{HINNPerf} as the local model, \Model~has the best median MRE in 56 out of 60 cases, and is ranked the first in 58 out of all the cases. Moreover, comparing the second best model ensembling method, \Model~achieves up to $9.89\times$ improvements (2.14 vs. 23.32 for $S_{3}$ of \textsc{nginx}). This leads to a remarkable 1.03 average Scott-Knott rank.}

When using \texttt{LR} as the local model, \texttt{\Model$_{LR}$} achieves considerably better accuracy than the other ensemble learning approaches, leading to the best MRE in 55 out of 60 cases with up to $28.50\times$ improvement with respect to the second best counterpart (19.17 vs. 565.62 for $S_{5}$ of \textsc{nginx}). In particular, it obtains the best Scott-Knott ranks on 51 out of 60 cases, within which only 2 cases \texttt{\Model$_{LR}$} does not achieve the sole best rank. This has resulted in the best average rank among the counterparts, i.e., 1.2. 


Since \Model~leverages \texttt{CART} to divide the data samples, one can easily expect that it would benefit less when using \texttt{CART} again as the local model. As a 
result, the relative improvement of \texttt{\Model$_{CART}$} over the counterparts becomes blurred compared with that of \texttt{\Model$_{LR}$}. However, in Table~\ref{tb:ensemble}, we see that \texttt{\Model$_{CART}$} still achieve highly competitive outcomes in relation to the others: amongst the 60 cases, it is ranked as the best for 28 cases, which is higher than the 24, 17, and 10 cases for \texttt{RF}, \texttt{XGBoost}, and \texttt{AdaBoosting$_{CART}$}, respectively. Notably, \texttt{\Model$_{CART}$} also more frequently achieves the best MRE. All the above has led to its best average rank of 1.85. Looking at the detailed MRE differences, compared with \texttt{RF}, the largest improvement and degradation of \texttt{\Model$_{CART}$} is $4.83\times$ (14.44 vs. 84.12 for $S_{1}$ of \textsc{nginx}) and $0.50\times$ (11.35 vs. 7.58 for $S_{3}$ of \textsc{HSMGP}), respectively; compared with \texttt{XGBoost} it is $4.20\times$ (0.94 vs. 4.89 for $S_{5}$ of \textsc{VP8}) and $1.13\times$ (14.44 vs. 6.78 for $S_{1}$ of \textsc{nginx}), respectively; in contrast to \texttt{AdaBoost$_{CART}$}, it is respectively $5.54\times$ (6.72 vs. 43.94 for $S_{3}$ of \textsc{BDB-C}) and $2.43\times$ (7.44 vs. 2.17 for $S_{2}$ of \textsc{nginx}). All the above reveals the superior benefit of \Model~against the others.

\textcolor{black}{With different local models including \texttt{HINNPerf}, \texttt{LR} and \texttt{CART}}, we see that the benefit of \Model~is more obvious for mixed systems: this is again due to the fact that those systems often come with a more sparse and complex configuration landscape, which is precisely what \Model~can cope with.


In conclusion, we can state the following for \textbf{RQ3}:

\begin{quotebox}
   \noindent
   \textit{\textbf{RQ3:} Compared with existing ensemble learning approaches, \Model~generally has a better ability to utilize the local models for predicting configuration performance. The benefits are particularly obvious under complex local models and systems.}
\end{quotebox}

\begin{table*}[h!]
\centering
\caption{Ablation study of \Model~with respect to the effectiveness of using \texttt{CART} for dividing samples, adapting $d$, and $\mu$HV. The format is the same as Table~\ref{tb:vsSOTA}.}
\setlength{\tabcolsep}{1.2mm}
\resizebox{\textwidth}{!}{ 
\begin{tabular}{clllllllll||llll||llll} \toprule
\multirow{2}{*}{System} & \multicolumn{1}{c}{\multirow{2}{*}{Size}} & \multicolumn{2}{c}{\texttt{DaL}} & \multicolumn{2}{c}{\texttt{DBSCAN}} & \multicolumn{2}{c}{\texttt{Agglomerative}} & \multicolumn{2}{c}{\texttt{$k$Means}} & \multicolumn{2}{c}{\texttt{DaL}} & \multicolumn{2}{c}{\texttt{DaL-FSE}} & \multicolumn{2}{c}{\texttt{DaL}} & \multicolumn{2}{c}{\texttt{DaL$^{HV}$}} \\ \cline{3-18}
 & \multicolumn{1}{c}{} & $r$ & Med (IQR) & $r$ & Med (IQR) & $r$ & Med (IQR) & $r$ & Med (IQR) & $r$ & Med (IQR) & $r$ & Med (IQR) & $r$ & Med (IQR) & $r$ & Med (IQR) \\ \hline
\multirow{5}{*}{\textsc{Apache}} & $S_1$ & \cellcolor{green!20}\textbf{1} & \cellcolor{green!20}\textbf{21.86 (7.36)} & 2 & 26.29 (15.23) & 3 & 30.99 (20.01) & 3 & 40.02 (22.13) & \cellcolor{green!20}\textbf{1} & \cellcolor{green!20}\textbf{21.86 (7.36)} & \cellcolor{red!20}\textbf{1} & \cellcolor{red!20}\textbf{21.86 (7.36)} & \cellcolor{green!20}\textbf{1} & \cellcolor{green!20}\textbf{21.86 (7.36)} & \cellcolor{red!20}\textbf{1} & \cellcolor{red!20}\textbf{21.86 (7.36)} \\
 & $S_2$ & \cellcolor{green!20}2 & \cellcolor{green!20}12.89 (7.87) & 2 & 16.24 (10.00) & \textbf{1} & \textbf{15.45 (5.76)} & \textbf{1} & \textbf{14.85 (5.21)} & \textbf{1} & \textbf{12.89 (7.87)} & \cellcolor{red!20}\textbf{1} & \cellcolor{red!20}\textbf{11.87 (9.22)} & \cellcolor{green!20}\textbf{1} & \cellcolor{green!20}\textbf{12.89 (7.87)} & \cellcolor{red!20}\textbf{1} & \cellcolor{red!20}\textbf{12.89 (7.87)} \\
 & $S_3$ & \cellcolor{green!20}\textbf{1} & \cellcolor{green!20}\textbf{6.56 (1.92)} & 3 & 13.24 (9.29) & 2 & 10.62 (3.24) & 2 & 10.60 (6.77) & \cellcolor{green!20}\textbf{1} & \cellcolor{green!20}\textbf{6.56 (1.92)} & \cellcolor{red!20}\textbf{1} & \cellcolor{red!20}\textbf{6.56 (1.92)} & \cellcolor{green!20}\textbf{1} & \cellcolor{green!20}\textbf{6.56 (1.92)} & 2 & 7.00 (2.29) \\
 & $S_4$ & \cellcolor{green!20}\textbf{1} & \cellcolor{green!20}\textbf{6.87 (1.95)} & 4 & 10.44 (4.15) & 3 & 8.63 (2.49) & 2 & 7.39 (2.84) & \cellcolor{green!20}\textbf{1} & \cellcolor{green!20}\textbf{6.87 (1.95)} & \cellcolor{red!20}\textbf{1} & \cellcolor{red!20}\textbf{6.87 (1.95)} & \cellcolor{green!20}\textbf{1} & \cellcolor{green!20}\textbf{6.87 (1.95)} & \cellcolor{red!20}2 & \cellcolor{red!20}6.87 (1.95) \\
 & $S_5$ & \cellcolor{green!20}\textbf{1} & \cellcolor{green!20}\textbf{5.78 (1.08)} & 4 & 9.18 (3.33) & 3 & 7.82 (2.20) & 2 & 6.08 (2.53) & \cellcolor{green!20}\textbf{1} & \cellcolor{green!20}\textbf{5.78 (1.08)} & \cellcolor{red!20}\textbf{1} & \cellcolor{red!20}\textbf{5.78 (1.08)} & \cellcolor{green!20}\textbf{1} & \cellcolor{green!20}\textbf{5.78 (1.08)} & \textbf{1} & \textbf{5.82 (1.77)} \\ \hline
\multirow{5}{*}{\textsc{BDB-C}} & $S_1$ & \cellcolor{green!20}\textbf{1} & \cellcolor{green!20}\textbf{83.60 (208.75)} & 2 & 438.69 (291.46) & 2 & 361.68 (382.62) & 2 & 309.58 (359.65) & \cellcolor{green!20}\textbf{1} & \cellcolor{green!20}\textbf{83.60 (208.75)} & \cellcolor{red!20}\textbf{1} & \cellcolor{red!20}\textbf{83.60 (208.75)} & \cellcolor{green!20}\textbf{1} & \cellcolor{green!20}\textbf{83.60 (208.75)} & \cellcolor{red!20}\textbf{1} & \cellcolor{red!20}\textbf{83.60 (208.75)} \\
 & $S_2$ & \cellcolor{green!20}\textbf{1} & \cellcolor{green!20}\textbf{21.92 (17.91)} & 3 & 170.67 (133.10) & 2 & 117.38 (99.35) & 3 & 145.13 (135.06) & \cellcolor{green!20}\textbf{1} & \cellcolor{green!20}\textbf{21.92 (17.91)} & \cellcolor{red!20}\textbf{1} & \cellcolor{red!20}\textbf{21.92 (17.91)} & \cellcolor{green!20}\textbf{1} & \cellcolor{green!20}\textbf{21.92 (17.91)} & \textbf{1} & \textbf{22.67 (35.19)} \\
 & $S_3$ & \cellcolor{green!20}\textbf{1} & \cellcolor{green!20}\textbf{8.31 (7.52)} & 2 & 43.57 (70.53) & 2 & 39.10 (72.34) & 3 & 38.29 (56.01) & \cellcolor{green!20}\textbf{1} & \cellcolor{green!20}\textbf{8.31 (7.52)} & \cellcolor{red!20}\textbf{1} & \cellcolor{red!20}\textbf{8.31 (7.52)} & \cellcolor{green!20}\textbf{1} & \cellcolor{green!20}\textbf{8.31 (7.52)} & 2 & 12.82 (25.48) \\
 & $S_4$ & \cellcolor{green!20}\textbf{1} & \cellcolor{green!20}\textbf{3.60 (4.50)} & 2 & 29.43 (29.08) & 2 & 25.32 (26.66) & 3 & 23.07 (43.38) & \cellcolor{green!20}\textbf{1} & \cellcolor{green!20}\textbf{3.60 (4.50)} & \cellcolor{red!20}\textbf{1} & \cellcolor{red!20}\textbf{3.60 (4.50)} & \cellcolor{green!20}\textbf{1} & \cellcolor{green!20}\textbf{3.60 (4.50)} & 2 & 10.45 (23.90) \\
 & $S_5$ & \cellcolor{green!20}\textbf{1} & \cellcolor{green!20}\textbf{1.88 (1.24)} & 3 & 33.50 (18.42) & 2 & 11.67 (13.49) & 4 & 19.27 (22.25) & \cellcolor{green!20}\textbf{1} & \cellcolor{green!20}\textbf{1.88 (1.24)} & \cellcolor{red!20}\textbf{1} & \cellcolor{red!20}\textbf{1.88 (1.24)} & \cellcolor{green!20}\textbf{1} & \cellcolor{green!20}\textbf{1.88 (1.24)} & 2 & 3.81 (7.39) \\ \hline
\multirow{5}{*}{\textsc{BDB-J}} & $S_1$ & \cellcolor{green!20}\textbf{1} & \cellcolor{green!20}\textbf{5.37 (4.74)} & 2 & 14.51 (9.69) & 3 & 14.57 (22.68) & 3 & 11.03 (13.62) & \cellcolor{green!20}\textbf{1} & \cellcolor{green!20}\textbf{5.37 (4.74)} & \cellcolor{red!20}\textbf{1} & \cellcolor{red!20}\textbf{5.37 (4.74)} & \cellcolor{green!20}\textbf{1} & \cellcolor{green!20}\textbf{5.37 (4.74)} & \textbf{1} & \textbf{6.21 (5.94)} \\
 & $S_2$ & \cellcolor{green!20}\textbf{1} & \cellcolor{green!20}\textbf{1.83 (0.45)} & 2 & 6.58 (6.33) & 3 & 7.12 (20.32) & 2 & 3.08 (4.46) & \cellcolor{green!20}\textbf{1} & \cellcolor{green!20}\textbf{1.83 (0.45)} & \cellcolor{red!20}\textbf{1} & \cellcolor{red!20}\textbf{1.83 (0.45)} & \cellcolor{green!20}\textbf{1} & \cellcolor{green!20}\textbf{1.83 (0.45)} & 2 & 2.23 (1.59) \\
 & $S_3$ & \cellcolor{green!20}\textbf{1} & \cellcolor{green!20}\textbf{1.58 (0.26)} & 2 & 3.48 (0.94) & 3 & 2.24 (4.85) & 2 & 2.05 (0.47) & \cellcolor{green!20}\textbf{1} & \cellcolor{green!20}\textbf{1.58 (0.26)} & \cellcolor{red!20}\textbf{1} & \cellcolor{red!20}\textbf{1.58 (0.26)} & \cellcolor{green!20}\textbf{1} & \cellcolor{green!20}\textbf{1.58 (0.26)} & 2 & 1.60 (0.32) \\
 & $S_4$ & \cellcolor{green!20}\textbf{1} & \cellcolor{green!20}\textbf{1.45 (0.22)} & 3 & 2.53 (0.87) & 4 & 1.82 (0.68) & 2 & 1.76 (0.34) & \cellcolor{green!20}\textbf{1} & \cellcolor{green!20}\textbf{1.45 (0.22)} & \cellcolor{red!20}\textbf{1} & \cellcolor{red!20}\textbf{1.45 (0.22)} & \cellcolor{green!20}\textbf{1} & \cellcolor{green!20}\textbf{1.45 (0.22)} & 2 & 1.53 (0.26) \\
 & $S_5$ & \cellcolor{green!20}\textbf{1} & \cellcolor{green!20}\textbf{1.40 (0.34)} & 3 & 2.61 (0.58) & 2 & 1.68 (0.26) & 2 & 1.62 (0.39) & \cellcolor{green!20}\textbf{1} & \cellcolor{green!20}\textbf{1.40 (0.34)} & \cellcolor{red!20}\textbf{1} & \cellcolor{red!20}\textbf{1.40 (0.34)} & \cellcolor{green!20}\textbf{1} & \cellcolor{green!20}\textbf{1.40 (0.34)} & 2 & 1.41 (0.35) \\ \hline
\multirow{5}{*}{\textsc{kanzi}} & $S_1$ & \cellcolor{green!20}\textbf{1} & \cellcolor{green!20}\textbf{196.15 (213.73)} & 3 & 545.20 (418.67) & 3 & 553.75 (502.84) & 2 & 531.18 (334.99) & \cellcolor{green!20}\textbf{1} & \cellcolor{green!20}\textbf{196.15 (213.73)} & \cellcolor{red!20}\textbf{1} & \cellcolor{red!20}\textbf{196.15 (213.73)} & \cellcolor{green!20}\textbf{1} & \cellcolor{green!20}\textbf{196.15 (213.73)} & 2 & 292.60 (385.37) \\
 & $S_2$ & \cellcolor{green!20}\textbf{1} & \cellcolor{green!20}\textbf{95.31 (51.34)} & 2 & 204.18 (138.71) & 3 & 196.73 (275.40) & 2 & 190.64 (148.79) & \cellcolor{green!20}\textbf{1} & \cellcolor{green!20}\textbf{95.31 (51.34)} & \cellcolor{red!20}\textbf{1} & \cellcolor{red!20}\textbf{95.31 (51.34)} & \cellcolor{green!20}\textbf{1} & \cellcolor{green!20}\textbf{95.31 (51.34)} & 2 & 216.02 (346.90) \\
 & $S_3$ & \cellcolor{green!20}\textbf{1} & \cellcolor{green!20}\textbf{59.30 (38.47)} & 2 & 160.72 (82.07) & 2 & 118.26 (106.12) & 2 & 119.90 (76.87) & \cellcolor{green!20}\textbf{1} & \cellcolor{green!20}\textbf{59.30 (38.47)} & \cellcolor{red!20}\textbf{1} & \cellcolor{red!20}\textbf{59.30 (38.47)} & \cellcolor{green!20}\textbf{1} & \cellcolor{green!20}\textbf{59.30 (38.47)} & 2 & 94.25 (64.94) \\
 & $S_4$ & \cellcolor{green!20}\textbf{1} & \cellcolor{green!20}\textbf{42.93 (18.75)} & 3 & 103.18 (113.16) & 2 & 81.27 (28.15) & 2 & 81.88 (26.24) & \cellcolor{green!20}\textbf{1} & \cellcolor{green!20}\textbf{42.93 (18.75)} & \cellcolor{red!20}\textbf{1} & \cellcolor{red!20}\textbf{42.93 (18.75)} & \cellcolor{green!20}\textbf{1} & \cellcolor{green!20}\textbf{42.93 (18.75)} & 2 & 134.47 (149.91) \\
 & $S_5$ & \cellcolor{green!20}\textbf{1} & \cellcolor{green!20}\textbf{35.29 (17.13)} & 4 & 91.90 (48.67) & 2 & 60.45 (19.99) & 3 & 65.93 (25.27) & \cellcolor{green!20}\textbf{1} & \cellcolor{green!20}\textbf{35.29 (17.13)} & \cellcolor{red!20}\textbf{1} & \cellcolor{red!20}\textbf{35.29 (17.13)} & \cellcolor{green!20}\textbf{1} & \cellcolor{green!20}\textbf{35.29 (17.13)} & 2 & 78.13 (86.63) \\ \hline
\multirow{5}{*}{\textsc{SQLite}} & $S_1$ & \cellcolor{green!20}\textbf{1} & \cellcolor{green!20}\textbf{73.52 (20.40)} & 2 & 87.19 (30.88) & 2 & 90.37 (22.58) & 2 & 98.63 (64.82) & \cellcolor{green!20}\textbf{1} & \cellcolor{green!20}\textbf{73.52 (20.40)} & \cellcolor{red!20}\textbf{1} & \cellcolor{red!20}\textbf{73.52 (20.40)} & \cellcolor{green!20}\textbf{1} & \cellcolor{green!20}\textbf{73.52 (20.40)} & \cellcolor{red!20}\textbf{1} & \cellcolor{red!20}\textbf{73.52 (20.40)} \\
 & $S_2$ & \cellcolor{green!20}\textbf{1} & \cellcolor{green!20}\textbf{76.41 (16.66)} & 3 & 84.27 (33.49) & 2 & 81.43 (21.75) & 3 & 83.39 (34.67) & \cellcolor{green!20}\textbf{1} & \cellcolor{green!20}\textbf{76.41 (16.66)} & \cellcolor{red!20}\textbf{1} & \cellcolor{red!20}\textbf{76.41 (16.66)} & \cellcolor{green!20}\textbf{1} & \cellcolor{green!20}\textbf{76.41 (16.66)} & 2 & 79.72 (19.44) \\
 & $S_3$ & \cellcolor{green!20}\textbf{1} & \cellcolor{green!20}\textbf{71.63 (11.58)} & 3 & 86.71 (25.06) & 2 & 75.22 (13.79) & 2 & 73.63 (18.36) & \cellcolor{green!20}\textbf{1} & \cellcolor{green!20}\textbf{71.63 (11.58)} & \cellcolor{red!20}\textbf{1} & \cellcolor{red!20}\textbf{71.63 (11.58)} & \cellcolor{green!20}\textbf{1} & \cellcolor{green!20}\textbf{71.63 (11.58)} & 2 & 72.29 (11.70) \\
 & $S_4$ & \cellcolor{green!20}\textbf{1} & \cellcolor{green!20}\textbf{66.94 (10.61)} & 3 & 81.04 (15.76) & 2 & 71.66 (12.67) & 2 & 75.77 (19.71) & \cellcolor{green!20}\textbf{1} & \cellcolor{green!20}\textbf{66.94 (10.61)} & \cellcolor{red!20}\textbf{1} & \cellcolor{red!20}\textbf{66.94 (10.61)} & \cellcolor{green!20}\textbf{1} & \cellcolor{green!20}\textbf{66.94 (10.61)} & 2 & 70.51 (16.58) \\
 & $S_5$ & \cellcolor{green!20}\textbf{1} & \cellcolor{green!20}\textbf{64.20 (13.91)} & 4 & 77.13 (17.36) & 2 & 71.44 (12.01) & 3 & 72.32 (13.70) & \cellcolor{green!20}\textbf{1} & \cellcolor{green!20}\textbf{64.20 (13.91)} & \cellcolor{red!20}\textbf{1} & \cellcolor{red!20}\textbf{64.20 (13.91)} & \cellcolor{green!20}\textbf{1} & \cellcolor{green!20}\textbf{64.20 (13.91)} & 2 & 67.09 (16.08) \\ \hline
\multirow{5}{*}{\textsc{x264}} & $S_1$ & \cellcolor{green!20}\textbf{1} & \cellcolor{green!20}\textbf{10.21 (4.35)} & 2 & 15.09 (9.65) & 2 & 13.36 (12.90) & 3 & 15.72 (26.44) & \cellcolor{green!20}\textbf{1} & \cellcolor{green!20}\textbf{10.21 (4.35)} & \cellcolor{red!20}\textbf{1} & \cellcolor{red!20}\textbf{10.21 (4.35)} & \cellcolor{green!20}\textbf{1} & \cellcolor{green!20}\textbf{10.21 (4.35)} & \textbf{1} & \textbf{10.49 (5.06)} \\
 & $S_2$ & \cellcolor{green!20}\textbf{1} & \cellcolor{green!20}\textbf{2.72 (1.52)} & 3 & 8.49 (6.80) & 2 & 5.44 (3.96) & 3 & 4.48 (3.51) & \cellcolor{green!20}\textbf{1} & \cellcolor{green!20}\textbf{2.72 (1.52)} & \textbf{1} & \textbf{2.74 (1.36)} & \textbf{1} & \textbf{2.72 (1.52)} & \cellcolor{red!20}\textbf{1} & \cellcolor{red!20}\textbf{2.64 (1.51)} \\
 & $S_3$ & \cellcolor{green!20}\textbf{1} & \cellcolor{green!20}\textbf{1.57 (1.52)} & 3 & 6.71 (4.70) & 2 & 2.15 (0.75) & 2 & 1.91 (1.69) & 2 & 1.57 (1.52) & \cellcolor{red!20}\textbf{1} & \cellcolor{red!20}\textbf{1.56 (0.70)} & \cellcolor{green!20}\textbf{1} & \cellcolor{green!20}\textbf{1.57 (1.52)} & \textbf{1} & \textbf{1.93 (2.10)} \\
 & $S_4$ & \textbf{1} & \textbf{1.30 (1.17)} & 2 & 7.28 (6.34) & \textbf{1} & \textbf{1.29 (0.80)} & \cellcolor{red!20}\textbf{1} & \cellcolor{red!20}\textbf{1.05 (0.86)} & 2 & 1.30 (1.17) & \cellcolor{red!20}\textbf{1} & \cellcolor{red!20}\textbf{1.00 (0.56)} & \cellcolor{green!20}\textbf{1} & \cellcolor{green!20}\textbf{1.30 (1.17)} & \textbf{1} & \textbf{1.41 (1.13)} \\
 & $S_5$ & 3 & 0.96 (0.79) & 4 & 4.46 (4.53) & 2 & 0.83 (0.49) & \cellcolor{red!20}\textbf{1} & \cellcolor{red!20}\textbf{0.61 (0.33)} & 2 & 0.96 (0.79) & \cellcolor{red!20}\textbf{1} & \cellcolor{red!20}\textbf{0.60 (0.36)} & \cellcolor{green!20}\textbf{1} & \cellcolor{green!20}\textbf{0.96 (0.79)} & \textbf{1} & \textbf{0.98 (0.77)} \\ \hline
\multirow{5}{*}{\textsc{Dune}} & $S_1$ & \textbf{1} & \textbf{9.37 (1.15)} & 3 & 12.87 (3.21) & \cellcolor{red!20}\textbf{1} & \cellcolor{red!20}\textbf{9.21 (1.43)} & 2 & 9.42 (1.44) & \textbf{1} & \textbf{9.37 (1.15)} & \cellcolor{red!20}\textbf{1} & \cellcolor{red!20}\textbf{9.19 (0.89)} & \cellcolor{green!20}\textbf{1} & \cellcolor{green!20}\textbf{9.37 (1.15)} & 2 & 9.39 (1.15) \\
 & $S_2$ & \cellcolor{green!20}\textbf{1} & \cellcolor{green!20}\textbf{5.95 (0.54)} & 4 & 13.56 (3.20) & 2 & 6.60 (0.64) & 3 & 6.92 (0.61) & \cellcolor{green!20}\textbf{1} & \cellcolor{green!20}\textbf{5.95 (0.54)} & \cellcolor{red!20}\textbf{1} & \cellcolor{red!20}\textbf{5.95 (0.54)} & \cellcolor{green!20}\textbf{1} & \cellcolor{green!20}\textbf{5.95 (0.54)} & \textbf{1} & \textbf{5.98 (0.58)} \\
 & $S_3$ & \cellcolor{green!20}\textbf{1} & \cellcolor{green!20}\textbf{4.81 (0.48)} & 3 & 15.17 (3.61) & 2 & 6.14 (0.46) & 2 & 6.09 (0.56) & \cellcolor{green!20}\textbf{1} & \cellcolor{green!20}\textbf{4.81 (0.48)} & \cellcolor{red!20}\textbf{1} & \cellcolor{red!20}\textbf{4.81 (0.48)} & \cellcolor{green!20}\textbf{1} & \cellcolor{green!20}\textbf{4.81 (0.48)} & 2 & 4.82 (0.48) \\
 & $S_4$ & \cellcolor{green!20}\textbf{1} & \cellcolor{green!20}\textbf{4.23 (0.35)} & 3 & 20.52 (3.50) & 2 & 5.61 (0.63) & 2 & 5.52 (0.51) & \cellcolor{green!20}\textbf{1} & \cellcolor{green!20}\textbf{4.23 (0.35)} & \cellcolor{red!20}\textbf{1} & \cellcolor{red!20}\textbf{4.23 (0.35)} & \cellcolor{green!20}\textbf{1} & \cellcolor{green!20}\textbf{4.23 (0.35)} & \cellcolor{red!20}\textbf{1} & \cellcolor{red!20}\textbf{4.23 (0.35)} \\
 & $S_5$ & \cellcolor{green!20}\textbf{1} & \cellcolor{green!20}\textbf{3.98 (0.19)} & 4 & 26.16 (4.58) & 2 & 4.98 (0.25) & 3 & 5.00 (0.31) & \cellcolor{green!20}\textbf{1} & \cellcolor{green!20}\textbf{3.98 (0.19)} & \cellcolor{red!20}\textbf{1} & \cellcolor{red!20}\textbf{3.98 (0.19)} & \cellcolor{green!20}\textbf{1} & \cellcolor{green!20}\textbf{3.98 (0.19)} & \cellcolor{red!20}\textbf{1} & \cellcolor{red!20}\textbf{3.98 (0.19)} \\ \hline
\multirow{5}{*}{\textsc{HIPA$^{cc}$}} & $S_1$ & \cellcolor{green!20}\textbf{1} & \cellcolor{green!20}\textbf{7.50 (1.42)} & 3 & 14.44 (7.83) & 2 & 9.95 (1.94) & 2 & 9.46 (1.37) & \cellcolor{green!20}\textbf{1} & \cellcolor{green!20}\textbf{7.50 (1.42)} & \cellcolor{red!20}\textbf{1} & \cellcolor{red!20}\textbf{7.50 (1.42)} & \cellcolor{green!20}\textbf{1} & \cellcolor{green!20}\textbf{7.50 (1.42)} & \cellcolor{red!20}\textbf{1} & \cellcolor{red!20}\textbf{7.50 (1.42)} \\
 & $S_2$ & \cellcolor{green!20}\textbf{1} & \cellcolor{green!20}\textbf{4.48 (0.38)} & 4 & 14.76 (3.76) & 3 & 5.61 (0.72) & 2 & 5.50 (0.63) & \cellcolor{green!20}\textbf{1} & \cellcolor{green!20}\textbf{4.48 (0.38)} & \cellcolor{red!20}\textbf{1} & \cellcolor{red!20}\textbf{4.48 (0.38)} & \cellcolor{green!20}\textbf{1} & \cellcolor{green!20}\textbf{4.48 (0.38)} & \cellcolor{red!20}2 & \cellcolor{red!20}4.48 (0.57) \\
 & $S_3$ & \cellcolor{green!20}\textbf{1} & \cellcolor{green!20}\textbf{3.43 (0.34)} & 4 & 10.33 (1.09) & 3 & 4.26 (0.44) & 2 & 4.18 (0.41) & \cellcolor{green!20}\textbf{1} & \cellcolor{green!20}\textbf{3.43 (0.34)} & \cellcolor{red!20}\textbf{1} & \cellcolor{red!20}\textbf{3.43 (0.34)} & \cellcolor{green!20}\textbf{1} & \cellcolor{green!20}\textbf{3.43 (0.34)} & 2 & 3.65 (1.43) \\
 & $S_4$ & \cellcolor{green!20}\textbf{1} & \cellcolor{green!20}\textbf{2.60 (0.15)} & 4 & 7.31 (0.65) & 3 & 2.95 (0.36) & 2 & 2.96 (0.18) & \cellcolor{green!20}\textbf{1} & \cellcolor{green!20}\textbf{2.60 (0.15)} & \cellcolor{red!20}\textbf{1} & \cellcolor{red!20}\textbf{2.60 (0.15)} & \cellcolor{green!20}\textbf{1} & \cellcolor{green!20}\textbf{2.60 (0.15)} & 2 & 4.06 (2.09) \\
 & $S_5$ & \cellcolor{green!20}\textbf{1} & \cellcolor{green!20}\textbf{2.11 (0.08)} & 4 & 6.92 (0.60) & 3 & 2.27 (0.14) & 2 & 2.20 (0.09) & \cellcolor{green!20}\textbf{1} & \cellcolor{green!20}\textbf{2.11 (0.08)} & \cellcolor{red!20}\textbf{1} & \cellcolor{red!20}\textbf{2.11 (0.08)} & \cellcolor{green!20}\textbf{1} & \cellcolor{green!20}\textbf{2.11 (0.08)} & 2 & 2.64 (0.56) \\ \hline
\multirow{5}{*}{\textsc{HSMGP}} & $S_1$ & \cellcolor{green!20}\textbf{1} & \cellcolor{green!20}\textbf{4.14 (1.44)} & 4 & 21.00 (94.17) & 2 & 10.21 (14.94) & 3 & 12.27 (47.67) & \cellcolor{green!20}\textbf{1} & \cellcolor{green!20}\textbf{4.14 (1.44)} & \cellcolor{red!20}\textbf{1} & \cellcolor{red!20}\textbf{4.14 (1.44)} & \cellcolor{green!20}\textbf{1} & \cellcolor{green!20}\textbf{4.14 (1.44)} & 2 & 7.80 (10.46) \\
 & $S_2$ & \cellcolor{green!20}\textbf{1} & \cellcolor{green!20}\textbf{2.06 (0.19)} & 4 & 26.68 (25.59) & 2 & 3.42 (3.02) & 3 & 5.47 (17.62) & \cellcolor{green!20}\textbf{1} & \cellcolor{green!20}\textbf{2.06 (0.19)} & \cellcolor{red!20}\textbf{1} & \cellcolor{red!20}\textbf{2.06 (0.19)} & \cellcolor{green!20}\textbf{1} & \cellcolor{green!20}\textbf{2.06 (0.19)} & 2 & 2.49 (4.01) \\
 & $S_3$ & \cellcolor{green!20}\textbf{1} & \cellcolor{green!20}\textbf{1.42 (0.13)} & 4 & 99.71 (84.77) & 2 & 2.31 (0.41) & 3 & 2.59 (0.72) & \cellcolor{green!20}\textbf{1} & \cellcolor{green!20}\textbf{1.42 (0.13)} & \cellcolor{red!20}\textbf{1} & \cellcolor{red!20}\textbf{1.42 (0.13)} & \cellcolor{green!20}\textbf{1} & \cellcolor{green!20}\textbf{1.42 (0.13)} & 2 & 1.48 (0.48) \\
 & $S_4$ & \cellcolor{green!20}\textbf{1} & \cellcolor{green!20}\textbf{1.33 (0.08)} & 4 & 144.33 (83.45) & 2 & 2.15 (0.51) & 3 & 2.30 (0.37) & \cellcolor{green!20}\textbf{1} & \cellcolor{green!20}\textbf{1.33 (0.08)} & \cellcolor{red!20}\textbf{1} & \cellcolor{red!20}\textbf{1.33 (0.08)} & \cellcolor{green!20}\textbf{1} & \cellcolor{green!20}\textbf{1.33 (0.08)} & 2 & 1.40 (0.35) \\
 & $S_5$ & \cellcolor{green!20}\textbf{1} & \cellcolor{green!20}\textbf{1.19 (0.08)} & 4 & 51.04 (24.94) & 2 & 1.80 (0.29) & 3 & 1.89 (0.22) & \cellcolor{green!20}\textbf{1} & \cellcolor{green!20}\textbf{1.19 (0.08)} & \cellcolor{red!20}\textbf{1} & \cellcolor{red!20}\textbf{1.19 (0.08)} & \cellcolor{green!20}\textbf{1} & \cellcolor{green!20}\textbf{1.19 (0.08)} & 2 & 1.20 (0.15) \\ \hline
\multirow{5}{*}{\textsc{Lrzip}} & $S_1$ & \cellcolor{green!20}\textbf{1} & \cellcolor{green!20}\textbf{31.88 (10.96)} & 3 & 72.91 (33.63) & 2 & 38.43 (19.53) & 3 & 66.44 (35.18) & \cellcolor{green!20}\textbf{1} & \cellcolor{green!20}\textbf{31.88 (10.96)} & \cellcolor{red!20}\textbf{1} & \cellcolor{red!20}\textbf{31.88 (10.96)} & \cellcolor{green!20}\textbf{1} & \cellcolor{green!20}\textbf{31.88 (10.96)} & 2 & 32.32 (12.02) \\
 & $S_2$ & \cellcolor{green!20}\textbf{1} & \cellcolor{green!20}\textbf{10.51 (4.39)} & 3 & 209.93 (186.42) & 2 & 23.70 (17.77) & 3 & 48.56 (93.35) & \cellcolor{green!20}\textbf{1} & \cellcolor{green!20}\textbf{10.51 (4.39)} & \cellcolor{red!20}\textbf{1} & \cellcolor{red!20}\textbf{10.51 (4.39)} & \cellcolor{green!20}\textbf{1} & \cellcolor{green!20}\textbf{10.51 (4.39)} & 2 & 16.23 (4.19) \\
 & $S_3$ & \cellcolor{green!20}\textbf{1} & \cellcolor{green!20}\textbf{8.04 (2.37)} & 4 & 288.22 (299.90) & 2 & 14.75 (16.26) & 3 & 32.91 (20.58) & \cellcolor{green!20}\textbf{1} & \cellcolor{green!20}\textbf{8.04 (2.37)} & \cellcolor{red!20}\textbf{1} & \cellcolor{red!20}\textbf{8.04 (2.37)} & \cellcolor{green!20}\textbf{1} & \cellcolor{green!20}\textbf{8.04 (2.37)} & 2 & 13.28 (6.10) \\
 & $S_4$ & \cellcolor{green!20}\textbf{1} & \cellcolor{green!20}\textbf{5.95 (0.99)} & 3 & 124.12 (186.64) & 2 & 12.59 (7.86) & 3 & 34.68 (46.30) & \cellcolor{green!20}\textbf{1} & \cellcolor{green!20}\textbf{5.95 (0.99)} & \cellcolor{red!20}\textbf{1} & \cellcolor{red!20}\textbf{5.95 (0.99)} & \cellcolor{green!20}\textbf{1} & \cellcolor{green!20}\textbf{5.95 (0.99)} & 2 & 10.15 (3.76) \\
 & $S_5$ & \cellcolor{green!20}\textbf{1} & \cellcolor{green!20}\textbf{4.03 (0.26)} & 3 & 110.41 (46.00) & 2 & 5.56 (4.32) & 4 & 22.75 (136.13) & \cellcolor{green!20}\textbf{1} & \cellcolor{green!20}\textbf{4.03 (0.26)} & \cellcolor{red!20}\textbf{1} & \cellcolor{red!20}\textbf{4.03 (0.26)} & \cellcolor{green!20}\textbf{1} & \cellcolor{green!20}\textbf{4.03 (0.26)} & 2 & 6.31 (1.86) \\ \hline
\multirow{5}{*}{\textsc{nginx}} & $S_1$ & \cellcolor{green!20}\textbf{1} & \cellcolor{green!20}\textbf{4.11 (1.45)} & 3 & 15.18 (10.27) & 2 & 7.30 (3.27) & 2 & 7.89 (4.12) & \cellcolor{green!20}\textbf{1} & \cellcolor{green!20}\textbf{4.11 (1.45)} & \cellcolor{red!20}\textbf{1} & \cellcolor{red!20}\textbf{4.11 (1.45)} & \cellcolor{green!20}\textbf{1} & \cellcolor{green!20}\textbf{4.11 (1.45)} & 2 & 9.44 (5.40) \\
 & $S_2$ & \cellcolor{green!20}\textbf{1} & \cellcolor{green!20}\textbf{2.60 (1.14)} & 3 & 7.83 (4.25) & 2 & 5.87 (1.00) & 2 & 5.52 (1.26) & \cellcolor{green!20}\textbf{1} & \cellcolor{green!20}\textbf{2.60 (1.14)} & \cellcolor{red!20}\textbf{1} & \cellcolor{red!20}\textbf{2.60 (1.14)} & \cellcolor{green!20}\textbf{1} & \cellcolor{green!20}\textbf{2.60 (1.14)} & 2 & 3.15 (1.94) \\
 & $S_3$ & \cellcolor{green!20}\textbf{1} & \cellcolor{green!20}\textbf{2.14 (0.40)} & 3 & 4.94 (3.79) & 2 & 5.23 (1.55) & 2 & 4.49 (1.57) & \cellcolor{green!20}\textbf{1} & \cellcolor{green!20}\textbf{2.14 (0.40)} & \cellcolor{red!20}\textbf{1} & \cellcolor{red!20}\textbf{2.14 (0.40)} & \cellcolor{green!20}\textbf{1} & \cellcolor{green!20}\textbf{2.14 (0.40)} & 2 & 2.34 (0.79) \\
 & $S_4$ & \cellcolor{green!20}\textbf{1} & \cellcolor{green!20}\textbf{1.99 (0.47)} & 2 & 3.75 (3.14) & 3 & 4.95 (1.01) & 3 & 4.19 (1.15) & \cellcolor{green!20}\textbf{1} & \cellcolor{green!20}\textbf{1.99 (0.47)} & \cellcolor{red!20}\textbf{1} & \cellcolor{red!20}\textbf{1.99 (0.47)} & \cellcolor{green!20}\textbf{1} & \cellcolor{green!20}\textbf{1.99 (0.47)} & 2 & 2.00 (0.77) \\
 & $S_5$ & \cellcolor{green!20}\textbf{1} & \cellcolor{green!20}\textbf{1.98 (0.21)} & 4 & 5.83 (2.09) & 2 & 4.10 (1.83) & 3 & 4.32 (1.20) & \cellcolor{green!20}\textbf{1} & \cellcolor{green!20}\textbf{1.98 (0.21)} & \cellcolor{red!20}\textbf{1} & \cellcolor{red!20}\textbf{1.98 (0.21)} & \cellcolor{green!20}\textbf{1} & \cellcolor{green!20}\textbf{1.98 (0.21)} & 2 & 2.10 (0.96) \\ \hline
\multirow{5}{*}{\textsc{VP8}} & $S_1$ & \cellcolor{green!20}\textbf{1} & \cellcolor{green!20}\textbf{1.56 (0.18)} & 4 & 7.58 (2.89) & 2 & 2.16 (1.08) & 3 & 2.86 (1.04) & \cellcolor{green!20}\textbf{1} & \cellcolor{green!20}\textbf{1.56 (0.18)} & \cellcolor{red!20}\textbf{1} & \cellcolor{red!20}\textbf{1.56 (0.18)} & \cellcolor{green!20}\textbf{1} & \cellcolor{green!20}\textbf{1.56 (0.18)} & \cellcolor{red!20}\textbf{1} & \cellcolor{red!20}\textbf{1.56 (0.18)} \\
 & $S_2$ & \cellcolor{green!20}\textbf{1} & \cellcolor{green!20}\textbf{1.15 (0.05)} & 4 & 7.39 (2.36) & 3 & 1.45 (0.54) & 2 & 1.40 (0.16) & \cellcolor{green!20}\textbf{1} & \cellcolor{green!20}\textbf{1.15 (0.05)} & \cellcolor{red!20}\textbf{1} & \cellcolor{red!20}\textbf{1.15 (0.05)} & \cellcolor{green!20}\textbf{1} & \cellcolor{green!20}\textbf{1.15 (0.05)} & \cellcolor{red!20}\textbf{1} & \cellcolor{red!20}\textbf{1.15 (0.05)} \\
 & $S_3$ & \cellcolor{green!20}\textbf{1} & \cellcolor{green!20}\textbf{1.08 (0.04)} & 4 & 9.02 (3.35) & 3 & 1.32 (0.32) & 2 & 1.25 (0.13) & \cellcolor{green!20}\textbf{1} & \cellcolor{green!20}\textbf{1.08 (0.04)} & \cellcolor{red!20}\textbf{1} & \cellcolor{red!20}\textbf{1.08 (0.04)} & \cellcolor{green!20}\textbf{1} & \cellcolor{green!20}\textbf{1.08 (0.04)} & \cellcolor{red!20}\textbf{1} & \cellcolor{red!20}\textbf{1.08 (0.04)} \\
 & $S_4$ & \cellcolor{green!20}\textbf{1} & \cellcolor{green!20}\textbf{0.98 (0.05)} & 4 & 9.65 (3.27) & 3 & 1.31 (0.31) & 2 & 1.17 (0.09) & \cellcolor{green!20}\textbf{1} & \cellcolor{green!20}\textbf{0.98 (0.05)} & \cellcolor{red!20}\textbf{1} & \cellcolor{red!20}\textbf{0.98 (0.05)} & \cellcolor{green!20}\textbf{1} & \cellcolor{green!20}\textbf{0.98 (0.05)} & \cellcolor{red!20}\textbf{1} & \cellcolor{red!20}\textbf{0.98 (0.05)} \\
 & $S_5$ & \cellcolor{green!20}\textbf{1} & \cellcolor{green!20}\textbf{0.90 (0.06)} & 4 & 10.23 (5.77) & 2 & 1.03 (0.12) & 3 & 1.05 (0.10) & \cellcolor{green!20}\textbf{1} & \cellcolor{green!20}\textbf{0.90 (0.06)} & \cellcolor{red!20}\textbf{1} & \cellcolor{red!20}\textbf{0.90 (0.06)} & \cellcolor{green!20}\textbf{1} & \cellcolor{green!20}\textbf{0.90 (0.06)} & \cellcolor{red!20}\textbf{1} & \cellcolor{red!20}\textbf{0.90 (0.06)} \\ \hline
\multicolumn{2}{c}{Average $r$} & 1.05 &  & 3.15 &  & 2.25 &  & 2.43 &  & 1.05 &  & 1 &  & 1 &  & 1.65 & 
\\
\bottomrule
\end{tabular}
}

\label{tb:clustering}
\end{table*}

\subsection{The Effectiveness of Components in \Model}
\label{subsec:component}

\subsubsection{The Benefits of \texttt{CART}}

As discussed in Section~\ref{subsec:modeify-cart}, we modify \texttt{CART} to divide the configuration data samples since its training procedure naturally fits our needs better. To confirm this point, here we experimentally examine its effectiveness over several alternative clustering algorithms that can serve as the replacement of \texttt{CART} in the \textit{dividing} phase of \Model, namely: \texttt{DBSCAN}~\cite{DBLP:conf/kdd/EsterKSX96} (a density-based clustering algorithm), \texttt{Agglomerative clustering}~\cite{Inchoate:Ward63} (a hierarchical clustering algorithm), and \texttt{$k$Means}~\cite{DBLP:journals/tit/Lloyd82} (a centroid-based clustering algorithm). We use the most common Euclidean distance while both the configuration options and performance values are combined as the features of clustering.



Again, we directly use the implementations from the \texttt{scikit-learn} package while the other experiment settings are the same as the previous RQs. Among these, \texttt{DBSCAN} adaptively determines the best number of divisions, while for \texttt{Agglomerative clustering} and \texttt{$k$Means}, we tuned the division numbers for each case and use the overall best setting throughout the runs. The other parameters are set as the default values.

As can be seen from Table~\ref{tb:clustering} (left), the original design of \Model~that extends \texttt{CART} achieves significantly and overwhelmingly better results: among the 60 cases, there are 56 cases of the solely best rank and 57 cases of the best MRE with up to $5.41\times$ improvement over the second-best counterpart. The main reason is that, while the clustering algorithms do divide the configuration data, they often fail to take the prediction loss into account---a key feature within the training procedure of \texttt{CART}.



\subsubsection{The Necessity of Adapting $d$}

To investigate the necessity of adapting $d$, we compare \Model~with the version from our previous FSE work~\cite{DBLP:conf/sigsoft/JChen2023} where an overall best $d$ value for each system-size pair is pre-defined via profiling, denoted as \texttt{DaL}-\texttt{FSE}. Pragmatically, the procedure that one would need to perform without $d$ adaptation is as below:

\begin{enumerate}
    \item Choose a system and a training size considered.
    \item Pick a $d=k$ where $k \in\{1,2,..,d_{max}\}$ and train \texttt{DaL}-\texttt{FSE} using the corresponding training size, where $d_{max}$ is the largest depth that the \texttt{CART} can ever reach under a training size of the system.
    \item Repeat the experiment for 30 runs under $d=k$.
    \item Calculate the MREs of \texttt{DaL}-\texttt{FSE} under $d=k$ using all the remaining data as the testing set. 
    \item Repeat from (2) if not all possible $d$ value has been examined.
    \item Calculate the Scott-Knott ranks for MRE of \texttt{DaL}-\texttt{FSE} under different $d$ values.
    \item The $d$ value with the best rank is used; if there are multiple $d$ values that lead to the best rank, we select the one with the best average MRE therein.
    \item Repeat from (1) if not all system and size pairs have been covered.
\end{enumerate}

All other settings are the same as the previous RQs.

The comparisons between \texttt{DaL}-\texttt{FSE} and the proposed \Model~are illustrated in Table~\ref{tb:clustering} (middle). Clearly, both approaches perform very similarly across the cases. In particular, \Model~reaches identical results to that of \texttt{DaL}-\texttt{FSE}, which is pre-tuned, in 54 out of 60 cases. Interestingly, it is even possible for \Model~to achieve better accuracy, e.g., in $S_{2}$ of x264. This is because \Model~adapts $d$ for each individual run while \texttt{DaL}-\texttt{FSE} uses an overall best setting, as such \Model~will be able to control the $d$ in a finer granularity, hence further pushing the full potential of the dividable learning paradigm.


While in general \Model~performs very similar to \Model-\texttt{FSE}, it is worth noting that it does so without incurring the large overhead that would otherwise be required to find the optimal $d$ under \Model-\texttt{FSE}. Figure~\ref{fig:time_select_d} illustrates the time taken of \Model-\texttt{FSE} to find the best $d$ setting (where $d \in [1,4]$) beforehand for a given system, under the samples and largest training data size considered in this work. Clearly, across the systems, finding the optimal $d$ between 1 and 4 (with 30 repeats) needs at least 13.2 hours up to 65 hours for the smallest training size; this can turn into between 45.9 and 78.2 hours under the biggest training size, respectively. In contrast, since the newly proposed $d$ adaptation does not involve any additional training process, it only takes the magnitude of seconds as we will show in Section~\ref{subsec:overhead}. We believe such a resource-saving is remarkable. 



\begin{figure}[!t]
\centering
\footnotesize

\begin{subfigure}{.5\columnwidth}
  \centering
  \begin{tikzpicture}
\begin{axis}[
height=8cm, width=10cm,
 axis x line  = bottom,
    axis y line  = left  ,
grid style = {dashed},
legend pos=outer north east,
ymin=0, ymax=70,
enlarge x limits=0.05,
ybar=0pt,
bar width=10pt,
nodes near coords,
nodes near coords style={font=\footnotesize},
xtick={1,2,3,4,5,6,7,8,9,10,11,12},
xticklabels={
\textsc{nginx},
\textsc{Dune},
\textsc{HIPA$^{cc}$},
\textsc{BDB-J},
\textsc{kanzi},
\textsc{SQLite},
\textsc{Lrzip},
\textsc{VP8},
\textsc{BDB-C},
\textsc{Apache},
\textsc{HSMGP},
\textsc{x264},
},
y label style={at={(0.04, 0.5)}},
ylabel=Time taken (hours),
xticklabel style={rotate=45},
legend columns=1,
legend cell align={left},
    legend style={nodes={scale=1},fill=none,at={(0.9,0.95)}},
        extra y tick labels={},
        extra y tick style={grid=major,major grid style={thick,draw=black}}
]
\addplot[fill=beaublue,draw=black]
coordinates {
(1,65.0)
(2,53.5)
(3,49.2)
(4,43.5)
(5,40.9)
(6,39.4)
(7,37.6)
(8,34.5)
(9,33.7)
(10,33.1)
(11,28.6)
(12,13.2)
};


\end{axis}
\end{tikzpicture} 
  \caption{Total overhead using $S_1$}
\end{subfigure}
~\hspace{-0.3cm}
\begin{subfigure}{.5\columnwidth}
  \centering
  \begin{tikzpicture}
\begin{axis}[
height=8cm, width=10cm,
grid style = {dashed},
 axis x line  = bottom,
    axis y line  = left  ,
legend pos=outer north east,
ymin=0, ymax=85,
enlarge x limits=0.05,
ybar=0pt,
bar width=10pt,
nodes near coords,
nodes near coords style={font=\footnotesize},
y label style={at={(0.04, 0.5)}},
xtick={1,2,3,4,5,6,7,8,9,10,11,12},
xticklabels={
\textsc{nginx},
\textsc{HIPA$^{cc}$},
\textsc{Apache},
\textsc{Dune},
\textsc{HSMGP},
\textsc{BDB-C},
\textsc{Lrzip},
\textsc{VP8},
\textsc{kanzi},
\textsc{x264},
\textsc{BDB-J},
\textsc{SQLite},
},
ylabel=Time taken (hours),
xticklabel style={rotate=45},
legend columns=1,
legend cell align={left},
    legend style={nodes={scale=1},fill=none,at={(0.9,0.95)}},
        extra y tick labels={},
        extra y tick style={grid=major,major grid style={thick,draw=black}}
]

\addplot[fill=beaublue,draw=black]
coordinates {
(1,78.2)
(2,72.6)
(3,71.7)
(4,70.1)
(5,68.5)
(6,62.4)
(7,58.4)
(8,58.3)
(9,56.5)
(10,54.2)
(11,53.6)
(12,45.9)
};
\end{axis}
\end{tikzpicture}
  \caption{Total overhead using $S_5$}
\end{subfigure}
   \caption{The total overhead to select the optimal $d$ (from $d=1$ to $d=4$) for 30 runs under the smallest ($S_1$) and biggest ($S_5$) training data size considered.}
     \label{fig:time_select_d}
\end{figure}

\subsubsection{The Usefulness of $\mu$HV}

Recall from Section~\ref{subsubsec:adapting_depth}, we have theoretically analyzed why the newly proposed indicator $\mu$HV is needed for adapting $d$. Here, to confirm its usefulness, we compare the results of \Model~with its variant where the standard HV is used instead, which is denoted as \Model$^{HV}$. To that end, we leverage the HV implementation from the Python library \texttt{pymoo}. All other settings are the same as the previous RQs.

From Table~\ref{tb:clustering} (right), we see that using $\mu$HV clearly achieves considerably better accuracy than using the standard HV: the former is overwhelmingly ranked better (47 cases) or similar (12 cases) out of the 60 cases, which is a remarkable result.





Overall, for \textbf{RQ4}, we conclude that:

\begin{quotebox}
   \noindent
   \textit{\textbf{RQ4:}} The designed components in \Model~is effective:

   \begin{enumerate}
       \item \textbf{RQ4.1:} \texttt{CART} is ranked the solely best for 93.3\% (56/60) of the cases amongst the other clustering counterparts with up to $5.41\times$ MRE improvement over the second-best.
       \item \textbf{RQ4.2:} Adapting $d$ achieves almost identical result to the version of \Model~that relies on a pre-tuned $d$ value while saving an extensive amount of effort.
       \item \textbf{RQ4.3:} $\mu$HV is more suitable than the standard HV for determining the optimal and most balanced $d$ value on the configuration data, leading to 98\% (59/60) cases of better (47/60=78\%) or similar (12/60=20\%) accuracy.
   \end{enumerate}
   
\end{quotebox}



\subsection{Sensitivity and Adaptivity to the Depth $d$}
\label{subsec:sen}

\subsubsection{Method}
To understand the sensitivity of \Model~to the depth, we examine different $d$ values. Since the number of divisions (and hence the possible depth) is sample size-dependent, for each system, we use 80\% of the full dataset for training and the remaining for testing. This has allowed us to achieve up to $d=4$ with 16 divisions as the maximum possible bound for all systems studied. For different $d$ values, we report on the median MRE together over 30 runs, as well as counting the number of times that a $d$ value leads to the best MRE. 

Recall that in \textbf{RQ4}, we examine \Model-\texttt{FSE} which serves as the variant that takes a generally optimal $d$ (over all repeated runs) obtained via pre-tuning. Yet, while \Model-\texttt{FSE} mimics the practical scenario, it does not cater to the optimal $d$ for each individual run in which the training/testing data differ. Therefore, here we further examine to what extent \Model~can indeed reach the optimal $d$, considering individual run under each system-size pair, which serves as the ground truth.

All other settings are the same as the previous RQs.



\begin{figure*}[!t]
    \centering
    \footnotesize
    \input{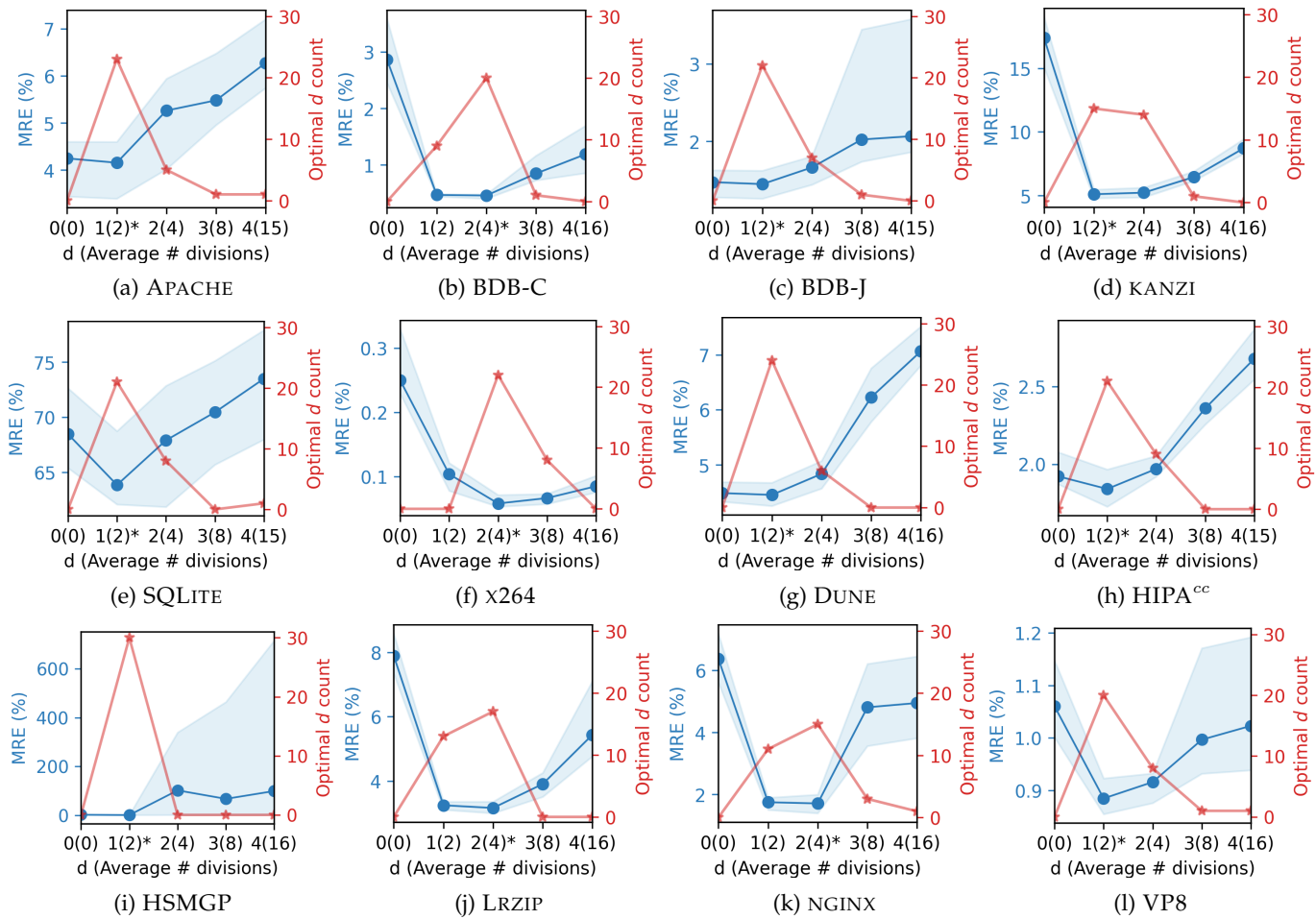}
    \caption{The median MRE (\markone{0}{20}{10}{20}), its IQR (area), and the number of runs that a $d$ value leads to the best MRE (\marktwo{0}{20}{7.5}{20}) over all systems and 30 runs. The better or worse is identified by the Scott-Knott test and average MRE. The generally optimal $d$, in terms of the overall MRE, is marked as $*$.}
    \label{fig:depths}
\end{figure*}

\begin{figure*}[!t]
    \centering
    \input{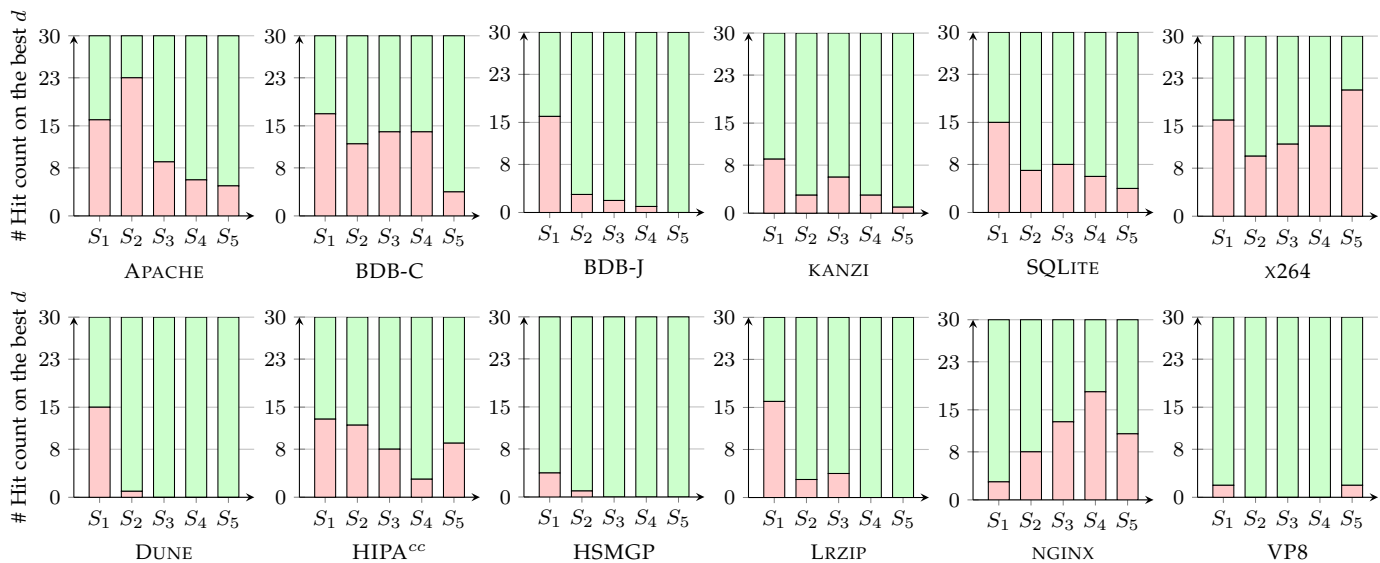}
    \caption{Hit count of reaching the optimal $d$ at each run (which leads to the best MRE) by the adaptive mechanism in \Model. The \setlength{\fboxsep}{1.5pt}\colorbox{red!20}{red bars} and \setlength{\fboxsep}{1.5pt}\colorbox{green!20}{green bars} show the miss hit counts and correct hit counts, respectively.}
    \label{fig:hit-count}
\end{figure*}

\begin{figure*}[!t]
\centering
    \input{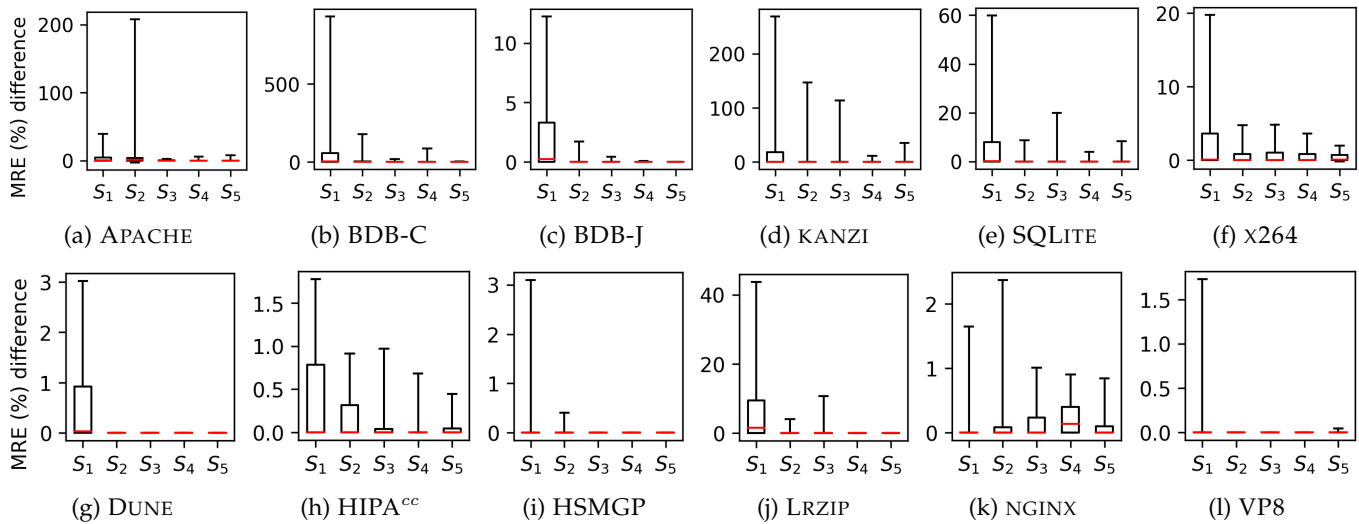}
\caption{MRE difference between the $d$ adapted by \Model~and the optimal $d$ for each of the 30 runs.}
\label{fig:box-plot}
\end{figure*}

\subsubsection{Results}

In Figure~\ref{fig:depths}, the best $d$ of each subject system is marked as $*$, which has the best Scott-Knott rank, and if more than one $d$ are ranked first, we mark the one with the smallest median MRE.

As can be seen from Figure~\ref{fig:depths}, we see that the correlation between the MRE (blue axis) of \Model~and $d$ value is close to quadratic: \Model~reaches its best MRE with $d=1$ (2 divisions) or $d=2$ (4 divisions). Since $d$ controls the trade-off between the ability to handle sample sparsity and ensuring sufficient data samples to train all local models, $d=1$ or $d=2$ tends to be the ``sweet points'' that reach a balance for the systems studied. After the point of $d=1$ or $d=2$, the MRE will worsen, as the local models' training size often drops dramatically. This is a clear sign that, from that point, the side-effect of having too less samples to train a local model has started to surpass the benefit that could have been brought by dealing with sample sparsity using more local models. When $d=0$, which means only one division and hence \Model~is reduced to \texttt{HINNPerf} that ignores sample sparsity, the resulted MRE is the worst on 6 out of 12 systems; a similar result can be observed for the cases when $d=4$. This suggests that neither too small $d$ (e.g., $d=0$ with only one division) nor too large $d$ (e.g., $d=4$ with up to 16 divisions, i.e., too many divisions) are ideal, which matches our theoretical analysis in Section~\ref{subsubsec:adapting_depth}.

Clearly, another important observation is that there is not a single optimal $d$ that can be applied to all systems: four of them favor $d=2$ in general while the remaining reveal that $d=1$ is the overall optimal setting. This becomes even more obvious when we take a closer look at each individual run with different training/testing data: as from the count of optimal $d$ (red axis), for the case where $d=1$ leads to the generally best MRE, there are a considerable number of runs under which $d=2$ is the optimal, even occasionally $d=3$, e.g., for system \textsc{kanzi}. The above suggests that \Model~is highly sensitive to $d$, depending on the systems and training/testing data---the key motivation to the extension of adapting $d$ in this work.


To compare the ability of \Model~on adapting $d$ against the ground truth for each individual run, we plot the hit/miss hit counts in Figure~\ref{fig:hit-count}. As can be seen, in the majority of the cases, \Model~can accurately adapt to the optimal $d$ of a run. There are a good amount of cases where the hit count is 100\% and only in 5 out of 60 cases, the miss hit counts are more than 50\% of the runs. We also observe a clear pattern where the hit count often increases as the training size expands, e.g., for systems \textsc{BDB-J} and \textsc{Dune}: this is as expected as the conflicts between the ability to handle sample sparsity and the necessary amount of data to learn are relieved when the amount of data increases.

Overall, the average hit count for all the cases is 22.93 out of 30, which means in 76.43\% of the runs, \Model~can indeed adapt to the optimal $d$. For mixed systems, \Model~generally reaches the optimal $d$ more frequently, as the average hit count for all the mixed systems is 25.13 against the 20.73 for binary systems. This is because the complex configuration landscape for mixed systems can provide more information when assessing the $d$ values during its adaptation process. To understand the severity of miss hitting the optimal $d$ in a run, we also report on the MRE difference between the results of the optimal $d$ and what is achieved by \Model~under an adapted $d$. It is clear that, from Figure~\ref{fig:box-plot}, the MRE difference is marginal in most of the cases according to their distributions: only very occasionally there are one or two extreme cases of more than 50\% MRE difference, the majority of them are within 10\%. This implies that, even for the cases/runs where \Model~does miss hit, a promising $d$ value can still be chosen, leading to minimal accuracy loss.

Therefore, for \textbf{RQ5}, we conclude that:

\begin{quotebox}
   \noindent
   \textit{\textbf{RQ5:} The error of \Model~has a (upward) quadratic correlation to $d$, the optimal value of which varies depending on the actual systems and training/testing data. \Model~can adapt to the optimal $d$ in 76.43\% of the individual runs while even when it misses hit, a promising $d$ value that leads to generally marginal MRE degradation can still be selected.}
   
\end{quotebox}

\subsection{Overhead of Model Building}
\label{subsec:overhead}


\subsubsection{Method}

To study \textbf{RQ6}, we examine the overall time required for training and building the models, including those \textcolor{black}{state-of-the-art configuration performance learning} approaches compared in \textbf{RQ1}, \textcolor{black}{and ensemble learning methods with different local models compared in \textbf{RQ3}}.

\subsubsection{Result}


From Table~\ref{tb:max_min}, which shows the range of training time required across all 60 cases, \Model~incurs an overall overhead from 4 to 56 minutes. Yet, from the breakdown, we see that the majority of the overhead comes from the \textit{training phase} that trains the local models, especially when a deep learning model like the \texttt{HINNPerf} is used (which is the default in \Model). We also note that the entire adaptation process of adapting $d$ is lightweight and it is independent of the local model; the only computation overhead lies in the calculation of the two functions in Equation (6), as well as the $\mu$HV. The time cost for adaptation ranges only from $10^{-6}$ to 0.02 minutes, which is marginal considering that it is an offline process.


Compared the default of \Model~(using \texttt{HINNPerf} as the local model) with other \textcolor{black}{state-of-the-art configuration performance learning} approaches, i.e., \texttt{HINNPerf} (3 to 54 minutes) and \texttt{DeepPerf} (3 to 48 minutes), \Model~have slightly higher training overhead, but the difference is merely in a matter of a few minutes, even though it trains multiple local models. This is because (1) each local model has less data to train and (2) those local models can be trained in parallel which speeds up the process. However, it is worth noting that, from \textbf{RQ1}, \Model~leads to considerably better accuracy than those two. In contrast to \texttt{Perf-AL} (a few seconds to two minutes), \Model~appears to be rather slow as the former does not use hyperparameter tuning but fixed-parameter values~\cite{DBLP:conf/esem/ShuS0X20}. Yet, as we have shown for \textbf{RQ1}, \Model~achieves up to a few magnitudes of accuracy improvement. Although \texttt{SPLConqueror} and \texttt{DECART} have an overhead of less than a minute, again their accuracy is much inferior. Further, \texttt{DECART} does not work on mixed systems. It is also worth noting that, when using with cheaper model like \texttt{LR} in \Model, the overhead can be greatly reduced.

\textcolor{black}{Compared with other ensemble learning methods, it can be seen that: (1) when using \texttt{HINNPerf} as the local model, \Model~has significantly lower overhead compared with \texttt{AdaBoost}$_{HINN}$ and \texttt{Bagging}$_{HINN}$, as these ensemble learning methods rely on training a large number of local models to get better results, whereas \Model~only trains a handful number of local models in most of the cases; (2) meanwhile, with \texttt{LR} and \texttt{CART} as the local model, \Model~is slightly slower than \texttt{Bagging} and \texttt{AdaBoost}. For example, \Model$_{CART}$ can take up to 0.32 minutes compared with the maximum overhead of 0.02 minutes for \texttt{AdaBoost}$_{CART}$. This is as expected, considering the time required for adapting depth and dividing, which could take up to 0.2 minutes. } 



\begin{table}[t!]
\caption{The overhead ranges across all systems and sizes.}
\centering

\begin{adjustbox}{width=0.78\columnwidth,center}
\hspace{-1.2cm}
\begin{subtable}[t]{0.5\columnwidth}
\centering
\begin{tabular}{ll}
\toprule
\textbf{Approach} & \textbf{Overhead (minutes)} \\ \hline

\texttt{DaL$_{LR}$} & \textcolor{black}{2$\times 10^{-4}$ to 0.32} \\
--- \Model~(\textit{adapting $d$}) & 1$\times 10^{-6}$ to 0.02 \\
--- \Model~(\textit{dividing}) & 9$\times 10^{-4}$ to 0.18 \\
--- \Model~(\textit{training}) & 3$\times 10^{-}4$ to 0.21 \\
--- \Model~(\textit{predicting}) & 4$\times 10^{-}5$ to 0.09 \\

\texttt{DaL$_{CART}$} & \textcolor{black}{1$\times 10^{-3}$ to 0.55} \\
--- \Model~(\textit{adapting $d$}) & 1$\times 10^{-6}$ to 0.02 \\
--- \Model~(\textit{dividing}) & 9$\times 10^{-4}$ to 0.18 \\
--- \Model~(\textit{training}) & 2$\times 10^{-4}$ to 0.31 \\
--- \Model~(\textit{predicting}) & 5$\times 10^{-4}$ to 0.18 \\

\Model & 4 to 56 \\
--- \Model~(\textit{adapting $d$}) & 1$\times 10^{-6}$ to 0.02 \\
--- \Model~(\textit{dividing}) & 9$\times 10^{-4}$ to 0.18 \\
--- \Model~(\textit{training}) & 3 to 53 \\
--- \Model~(\textit{predicting}) & 0.3 to 3 \\

\bottomrule
\end{tabular}
\end{subtable}
\hspace{1.3cm}
\begin{subtable}[t]{0.5\columnwidth}
\centering
\begin{tabular}{ll}
\toprule
\textbf{Approach} & \textbf{Overhead (minutes)} \\ \hline~\vspace{0.1cm}
\texttt{SPLConqueror} & 4$\times 10^{-4}$ to 5$\times 10^{-3}$ \\~\vspace{0.093cm}
\texttt{AdaBoost$_{LR}$} & \textcolor{black}{5$\times 10^{-6}$ to 8$\times 10^{-3}$} \\~\vspace{0.093cm}
\texttt{RF} & \textcolor{black}{1$\times 10^{-4}$ to 8$\times 10^{-3}$} \\~\vspace{0.093cm}
\texttt{XGBoost} & \textcolor{black}{3$\times 10^{-4}$ to 8$\times 10^{-3}$} \\~\vspace{0.093cm}
\texttt{Adaboost$_{CART}$} & \textcolor{black}{5$\times 10^{-5}$ to 0.02} \\~\vspace{0.093cm}
\texttt{Bagging$_{LR}$} & \textcolor{black}{1$\times 10^{-3}$ to 0.02} \\~\vspace{0.093cm}
\texttt{DECART} & 0.07 to 0.5 \\~\vspace{0.093cm}
\texttt{Perf-AL} & 0.08 to 2.4 \\~\vspace{0.093cm}
\texttt{DeepPerf} & 3 to 48 \\~\vspace{0.093cm}
\texttt{HINNPerf} & 3 to 54 \\~\vspace{0.093cm}
\texttt{AdaBoost$_{HINN}$} & \textcolor{black}{2.28 to 92.51} \\
~\texttt{Bagging$_{HINN}$} & \textcolor{black}{2.43 to 92.63} \\

\bottomrule
\end{tabular}
\end{subtable}
\end{adjustbox}

\label{tb:max_min}
\end{table}

In summary, we say that:

\begin{quotebox}
   \noindent
   \textit{\textbf{RQ6:} \Model, when using \texttt{HINNPerf} as the local model, has competitive model building time with respect to \texttt{HINNPerf} and \texttt{DeepPerf}, and has higher overhead than the other state-of-the-art approaches, but this can be acceptable considering its improvement in accuracy. When using other cheaper local models such as \texttt{LR}, the overhead is marginal.}
\end{quotebox}





\section{Discussion}
\label{sec:discussion}

Here, we discuss a few insights and pointers observed from the experiment results.

\subsection{Why Does \Model~Work?}

\begin{figure}[!t]
\centering
\footnotesize

\begin{subfigure}{.48\columnwidth}
  \centering
  \includegraphics[width=\linewidth]{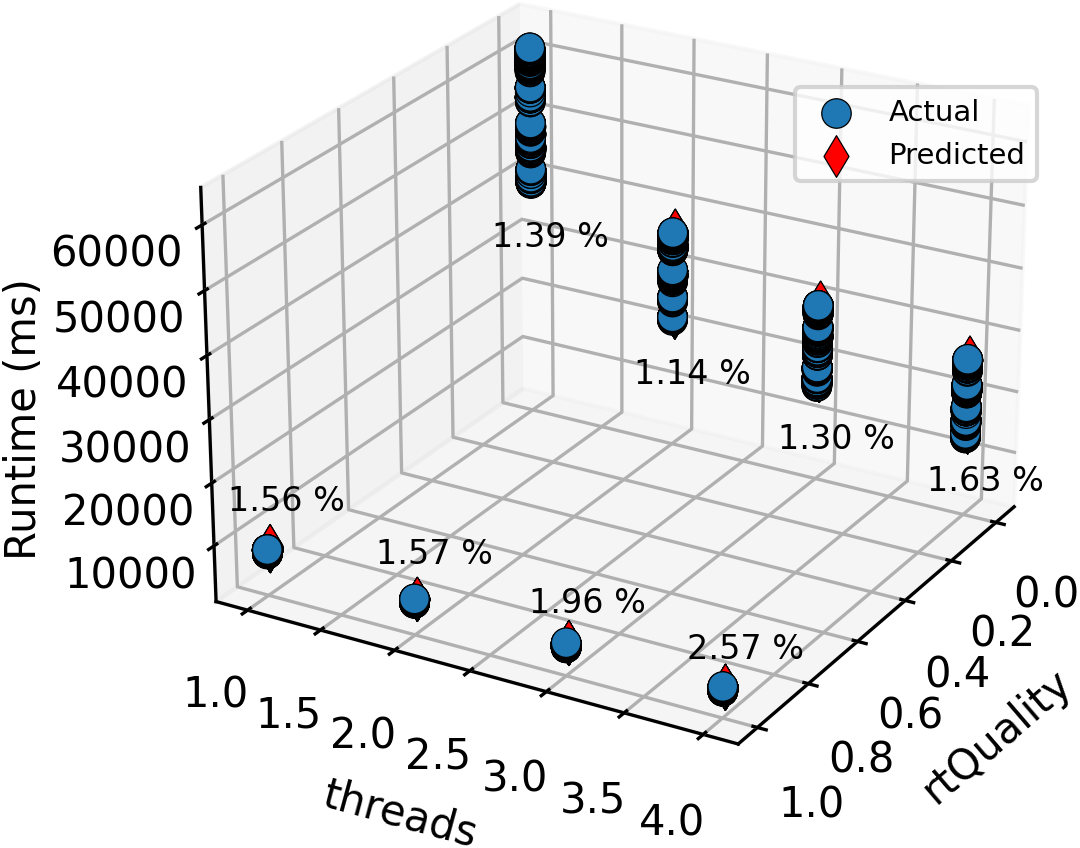} 
  \caption{\Model}
\end{subfigure}
~\hfill
\begin{subfigure}{.48\columnwidth}
  \centering
  \includegraphics[width=\linewidth]{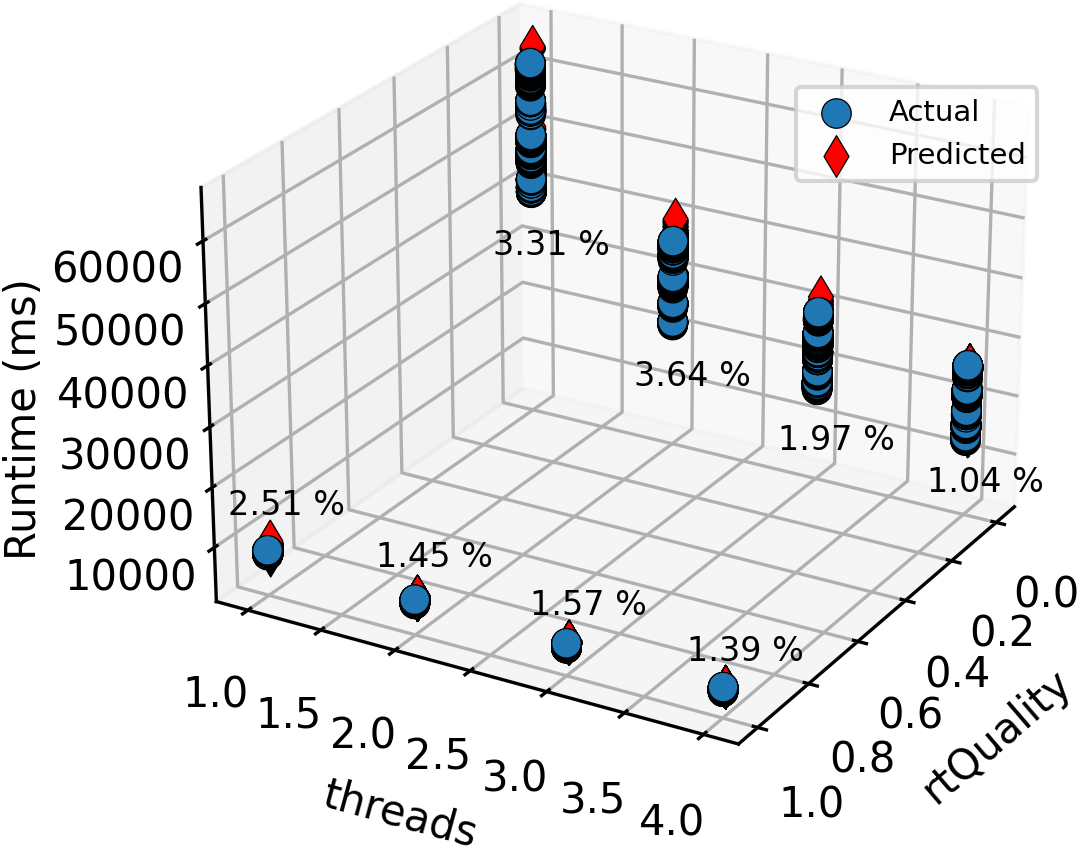} 
  \caption{\texttt{HINNPerf}}
\end{subfigure}

\begin{subfigure}{.48\columnwidth}
  \centering
  \includegraphics[width=\linewidth]{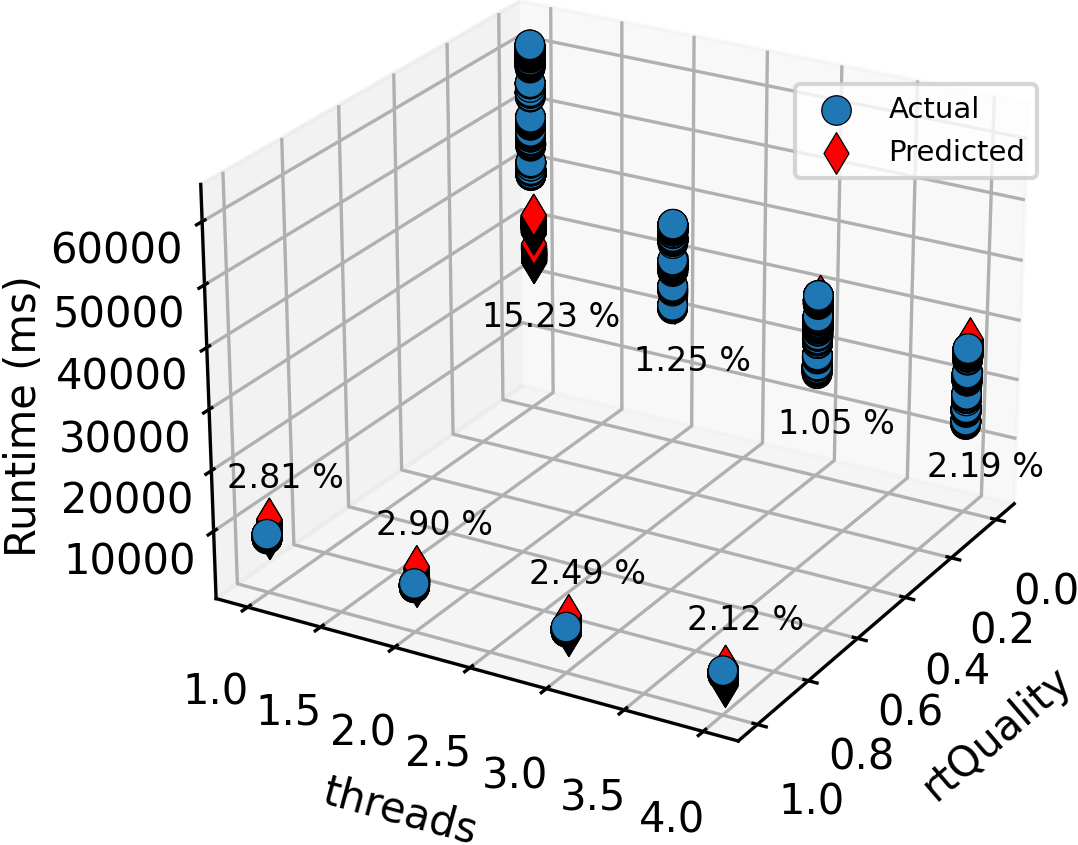} 
  \caption{\texttt{DeepPerf}}
\end{subfigure}
~\hfill
\begin{subfigure}{.48\columnwidth}
  \centering
  \includegraphics[width=\linewidth]{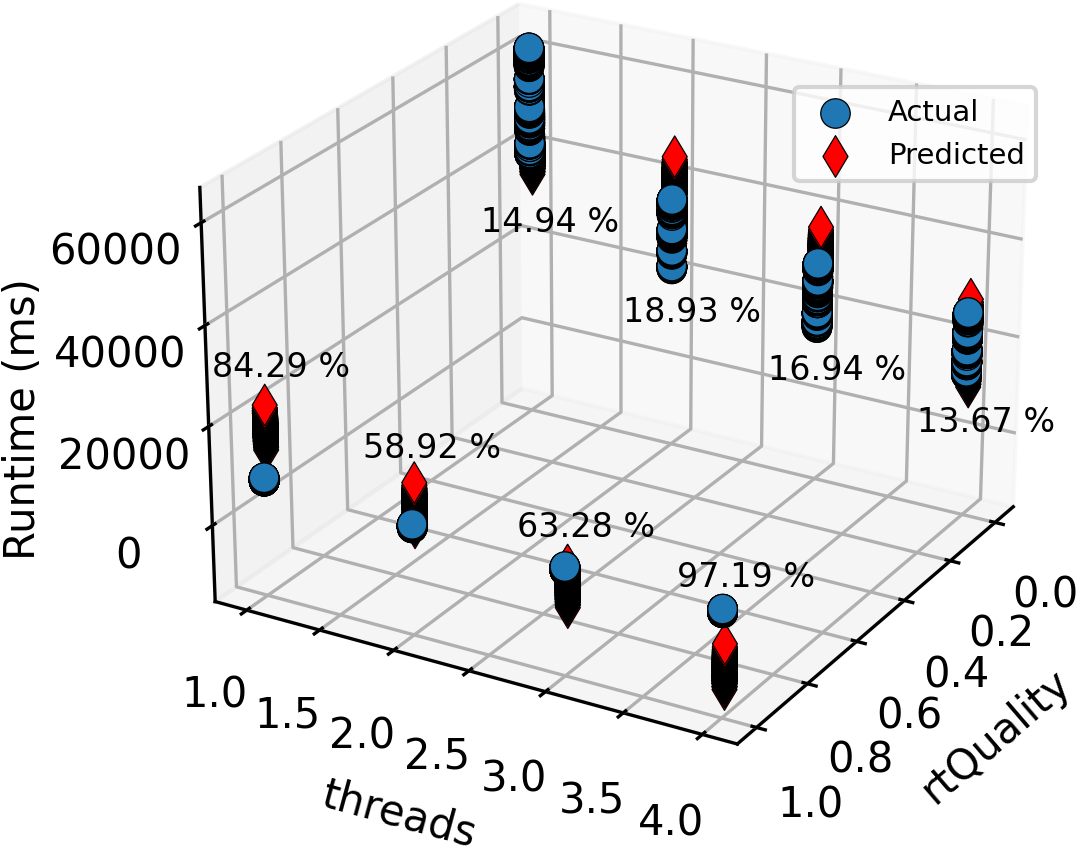} 
  \caption{\texttt{SPLConqueror}}
\end{subfigure}

\caption{\textcolor{black}{Example run of the actual and predicted performance by \Model, \texttt{HINNPerf}, \texttt{DeepPerf}, and \texttt{SPLConqueror} for $S_1$ of \textsc{VP8}. The numbers are the fine-grained MRE over each chunk of the samples.}}
\label{fig:discussion_scatter}
\end{figure}

To provide a more detailed understanding of why \Model~performs better than the state-of-the-art approaches, in Figure~\ref{fig:discussion_scatter}, we showcase the most common run of the predicted performance by \Model, \texttt{HINNPerf}, \texttt{DeepPerf} and \texttt{SPLConqueror} against the actual performance. Clearly, we note that the sample sparsity is rather obvious where there are two distant divisions. 

Those approaches that rely on a single local model have been severely affected by such highly sparse samples: we see that the models try to cover points in both divisions, but fail to do so as they tend to overfit/memorize the points in one or the other. For example, on \texttt{DeepPerf} (Figure~\ref{fig:discussion_scatter}c), its predictions on some configurations, which should result in high runtime, tend to have much lower values (e.g., when \texttt{rtQuality=0} and \texttt{threads=1}) since it overfits the points with some drastically lower runtime, i.e., when \texttt{rtQuality=1}. Conversely, \texttt{HINNPerf} (Figure~\ref{fig:discussion_scatter}b) and \texttt{SPLConqueror} (Figure~\ref{fig:discussion_scatter}d) estimate high runtime on some configurations that should lead to excellent performance (e.g., \texttt{rtQuality=1} and \texttt{threads=1}), which is, again, due to the fact that it memorizes those points with much higher runtime (\texttt{rtQuality=0}). Further, it also creates additional noises that make the predictions too high against what they should be, e.g., when \texttt{rtQuality=0} and \texttt{threads=2}. \Model, in contrast, handles such a sample sparsity well as it contains different local models that particularly cater to each division identified, hence leading to high accuracy (Figure~\ref{fig:discussion_scatter}a).


\subsection{Why Adapting $d$ Helps?}

\begin{figure}[!t]
  \centering
   \begin{subfigure}[t]{0.46\columnwidth}
        \centering
\includegraphics[width=\columnwidth]{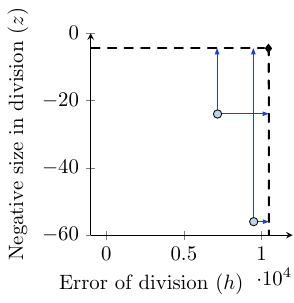}
    \subcaption{$d=1$ ($\mu$HV=$56411.36$)}
   \end{subfigure}
~\hfill
      \begin{subfigure}[t]{0.46\columnwidth}
        \centering
\includegraphics[width=\columnwidth]{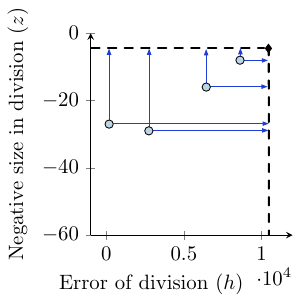}
 \subcaption{$d=2$ ($\mu$HV=$118013.45$)}
   \end{subfigure}

      \begin{subfigure}[t]{0.46\columnwidth}
        \centering
\includegraphics[width=\columnwidth]{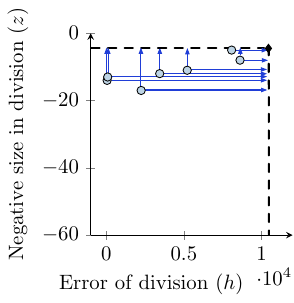}
  \subcaption{$d=3$ ($\mu$HV=$54755.69$)}
   \end{subfigure}
~\hfill
      \begin{subfigure}[t]{0.46\columnwidth}
        \centering
\includegraphics[width=\columnwidth]{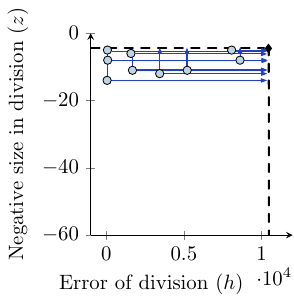}
     \subcaption{$d=4$ ($\mu$HV=$33827.60$)}
   \end{subfigure}

    \caption{The snapshot of the objective values of $h$ and $z$ on divisions when adapting $d$ for a run under system \textsc{x264}. The black diamond denotes the reference point. The testing MRE for $d=1$, $d=2$, $d=3$, and $d=4$ are $0.9676$, $0.7876$, $1.7858$, and $3.1285$, respectively.}
       \label{fig:changing-d}
  \end{figure}

We have showcased that the proposed adaptive mechanism is effective in dynamically setting the suitable $d$ in \Model. In order to take a deeper look at why it helps, Figure~\ref{fig:changing-d} illustrates the process of selecting the $d$ value at a run under \textsc{x264}. Here, we observed clear traces of how the sample size and mean square error in the divisions change along different $d$ values:

\begin{itemize}
    \item The overall sample size of the divisions ($z$) decreases as the $d$ value increases since more divisions are created to divide the configuration data. This implies that there is less chance for a local model to learn and generalize.
    \item The overall mean square error of the divisions ($h$) tends to decrease with larger $d$, as the additionally created divisions contain some much-reduced errors. This means that the overall ability to handle sample sparsity is improved, which is expected because the larger $d$, the smaller the sample size, hence it is more likely to have more dense points in a division, leading to lower mean square error.
\end{itemize}

The above observations match with our theoretical analysis in Section~\ref{subsubsec:adapting_depth}, and therefore adapting the $d$ value requires finding a good trade-off between $h$ and $z$. For the example in Figure~\ref{fig:changing-d}, the proposed $\mu$HV reflects the collective contributions of all possible divisions on both objectives, including those that are dominated, by calculating the average area that each $d$ value covers with respect to the reference point. In this way, we identify that $d=2$ is the best and most balanced value, which is subsequently confirmed to be indeed the case when validating their MRE.

\subsection{Strengths and Limitations}

The first strength of \Model~is that the concept of ``divide-and-learn'' under the paradigm of dividable learning, paired with an appropriately chosen local model, can handle both sample sparsity and feature sparsity well. As from Section~\ref{subsec:rq1} for \textbf{RQ1}, this has led to better accuracy and better utilization of the sample data than the state-of-the-art approaches.

The second strength is that while using \texttt{HINNPerf} has been shown to be the best option for accuracy in our cases, one can also easily replace it with others such as LR for faster training and better interoperability, hence offering great flexibility. As such, \Model~permits software engineers to make trade-offs based on different concerns depending on the practical scenarios. In particular, as from Section~\ref{subsec:local} for \textbf{RQ2}, \Model~can additionally improve different local models compared with when they are used alone as a global model. Further, when compared with other model-agnostic frameworks from the paradigm of ensemble learning, \Model~also produces more effective and accurate predictions, as demonstrated at \textbf{RQ3}.

The third strength of \Model~is that it is now less parameter-dependent (apart from those of the local model) given the mechanism that adapts $d$. This, as from \textbf{RQ5}, is necessary and the adaptation leads to a high hit rate with effective results. More importantly, adapting $d$ would have saved a great amount of effort to tailor and understand the approach together with its customized parameters, which is not uncommon for a model-agnostic framework~\cite{DBLP:conf/eurocolt/FreundS95}. Meanwhile, at \textbf{RQ4}, we illustrate that each proposed component in \Model~can individually contributes to its superiority. 


From \textbf{RQ6}, a limitation of \Model~is that it can incur slightly larger computational overhead when paired with a deep learning model: it could take a longer time to build the model than some state-of-the-art approaches. Using a typical server with a CPU of 2.6GHz and 256GB RAM, \Model~requires between 4 minutes and 56 minutes for systems with up to 33 configuration options and more than 2,000 samples. While this time effort might be relieved by implementing parallel training of the local models, the computational resources consumed are unavoidable. However, if efficiency is more important than accuracy, one could opt for a cheaper local model such as linear regression, which has trivial overhead as we have shown.

The other factor of \Model~is that it needs to be paired with a sampling method and in this work we use random sampling. This would work if we randomly sample configurations from the valid configuration space, which has already been measured by datasets from previous runs. This is also a widely used evaluation method in the field of configuration performance learning~\cite{DBLP:conf/icse/HaZ19,DBLP:journals/tosem/ChengGZ23,DBLP:conf/esem/ShuS0X20,DBLP:conf/msr/GongC22}. In practice, the vast majority of the dependency/constraints analysis tools~\cite{DBLP:conf/sigsoft/ChenWLLX20,DBLP:conf/sosp/XuZHZSYZP13} that are based on taint analysis would also be able to provide such information for sampling when there is a new system without any previously measured data. We see sampling as a complementary technique to the model building. As such, \Model~is certainly not random sampling-dependent; in fact, similar to \texttt{SPLConqueror}, it can be seamlessly paired with any sampling method, such as option-wise/pair-wise sampling, depending on the cases.


\subsection{\textcolor{black}{How would Invalid Configurations Affect \Model?}}

\textcolor{black}{While applying datasets from existing studies as shown in Table~\ref{tb:subject-system}, we only randomly
sample configurations from the valid configuration space. However, it does not affect the effectiveness of \Model~even if those datasets are not available in advance.} This is because, when collecting the configuration data via profiling/measuring, there are two major consequences related to the invalid configurations as demonstrated by a recent study~\cite{DBLP:conf/sigsoft/ChenWLLX20}: 

\begin{itemize}
    \item \textbf{Case 1:} The system crashes or throws runtime exceptions.
    \item \textbf{Case 2:} The system runs as usual but the violated dependency/constraint is automatically corrected by the system via some internal mechanisms or the relevant settings of the invalid options are simply ignored by some system-level checkers.
\end{itemize}

For \textbf{Case 1}, one can simply filter out those configurations that lead to invalid measurements, and hence all the remaining data contain the valid configurations only; this is similar to the data cleaning step in the common machine learning pipeline. \textbf{Case 2} is difficult to be certain about, but it is one of the typical causes of the sparsity issue (as those options that cause dependency/constraint violation would simply be non-influential to the performance), which is the exact challenge that we design \Model~for. 

As a result, even when collecting data for a new system for which we do not have any information about the dependency/constraint, we would still be able to easily deal with \textbf{Case 1} during the data collection process and the models in \Model~would be able to handle the sparsity caused by \textbf{Case 2}. Therefore, we believe that \Model~would not be weakened in such a circumstance.

\subsection{\textcolor{black}{How Many Options Can \Model~Handle?}}


\Model~can adapt well to any new subject systems with a large number of options, thanks to a unique structure that operates at two levels---the \textit{Dividing} and \textit{Training}:

\begin{itemize}
    \item \textbf{Dividing:} Here the goal is to divide the data into divisions using CART.
    \item \textbf{Training:} Here is where the actual learning and modeling happens, using a paired local model for each individual division.
\end{itemize}

The CART at the \textit{Dividing} would build a tree that partitions the data based on the sorted importance of the options. In the meantime, the parameter $d$ determines at which option we start to identify the divisions, e.g., $d=n$ basically means that we divide the data according to the top $n$ options produced by CART (please see Figure~\ref{fig:DT_example}). As such, the most important options are jointly determined by CART and $d$ and, given the characteristics of the learning procedure in CART, \Model~can handle a high number of options at the \textit{Dividing}.

Since the actual learning at the \textit{Training} is achieved by a local model, the extent to which it can handle the number of options entirely depends on the local model used. In our work, by default, we use \texttt{HINNPerf} in which the high dimensionality issue is handled by the hierarchical regularization (it is also its mechanism to deal with feature sparsity).

Therefore, \Model~can handle a large number of options if an appropriate local model has been chosen. Since the above has made no assumption about the system's characteristics, \Model~can work on new systems too.

\subsection{\textcolor{black}{What is the Typical and Best Size of Divisions?}}

As shown in Figure~\ref{fig:depths}, \Model~often performs the best commonly when $d=1$ (2 divisions) or $d=2$ (3-4 divisions), or sometimes, $d=3$ (4-8 divisions). However, this can vary depending on the systems and sample sizes, which motivates this extension work. 

As for what division size is most appropriate, our observation indicates that the accuracy does improve with more data used for training, i.e., $S_5$ in our case. It is difficult to confirm the best training size required as this is system/case dependent and the fundamental question itself is still an open problem in the machine learning community. However, if there is a given preferred level of accuracy for learning configuration performance for a system, then our results can provide some insights. For example, if we expect the average of MRE to be 2\% or better for \textsc{NGINX}, then we know that a training data size of $S_4$ (1,012 samples) or more is needed when using \Model.

\subsection{Threats to Validity}

{\textbf{Internal Threats.}} Internal threats to validity are related to the parameter values used. In this work, we set the same settings as used in state-of-the-art approaches~\cite{DBLP:journals/sqj/SiegmundRKKAS12,DBLP:conf/kbse/GuoCASW13,DBLP:conf/icse/HaZ19,DBLP:conf/esem/ShuS0X20}. Automated and adaptive mechanisms are enabled for those approaches that come with parameter tuning capability, e.g., \texttt{DeepPerf} and \texttt{DBSCAN}, or the best settings are profiled in a trail-and-error manner before the experiments, e.g., for the $k$ value in $k$\texttt{Means} clustering. If no parameter values are specified, we use the default settings. To mitigate stochastic error, we repeat the experiments for 30 runs and use Scott-Knott test for multiple comparisons. Indeed, although we have carefully ensured the fairness of comparisons, it is hard to guarantee that all compared models are in their best possible status.

{\textbf{Construct Threats.}} Threats to construct validity may lie in the metric used. In this study, MRE is chosen for two reasons: (1) it is a relative metric and hence is insensitive to the scale of the performance; (2) MRE has been recommended for performance prediction by many latest studies~\cite{DBLP:conf/icse/HaZ19, DBLP:conf/esem/ShuS0X20, DBLP:journals/ese/GuoYSASVCWY18}. To assess the cost of the modeling and training process, we measure the time taken to build the models. We also compare \Model~against state-of-the-art approaches, other ensemble learning models, and different variants thereof that are paired with different local models and come with alternative components. To confirm that adapting $d$ is indeed effective, we verify the sensitivity of \Model~to $d$ value while examining the adapted $d$ against the optimal $d$ on both the overall results and each individual run. However, programming errors or other small implementation defects are always possible.

{\textbf{External Threats.}} External validity could be raised from the subject systems and training samples used. To mitigate such, we evaluate 12 commonly used subject software systems, which are of diverse domains, scales, and performance metrics, selected from the latest studies~\cite{DBLP:journals/tse/KrishnaNJM21, DBLP:conf/icse/HaZ19,DBLP:journals/jss/CaoBWZLZ23,DBLP:conf/sigsoft/SiegmundGAK15,10172849}. We have also examined different training sample sizes as determined by the number of configuration options and \texttt{SPLConqueror}~\cite{DBLP:journals/sqj/SiegmundRKKAS12}---a typical method. Nevertheless, we agree that using more subject systems and data sizes may be fruitful, especially for examining the adaptive mechanism for $d$.


\section{Related Work}
\label{sec:related}

We now discuss the related work in light of the key contributions in \Model.

\subsection{Analytical Models} 

Predicting software performance can be done by analyzing the code structure and architecture of the systems~\cite{di2004compositional,DBLP:conf/icse/VelezJSAK21}. For example, Marco and Inverardi~\cite{di2004compositional} apply queuing network to model the latency of requests processed by the software. Velez \textit{et al.}~\cite{DBLP:conf/icse/VelezJSAK21} use local measurements and
dynamic taint analysis to build a model that can predict performance for part of the configuration code. However, analytical models require full understanding and access to the software's internal states, which may not always be possible/feasible.  Further, due to the theoretical assumption or limitations of the measurement tool, analytical models are often restricted to certain performance metrics, most commonly time-related ones such as runtime and latency. \Model~is not restricted to those scenarios as it is a data-driven approach while still being designed to cater to the unique sparse nature of configuration data.


\subsection{Statistical Learning-based Models} 

Data-driven learning has relied on various statistical models, such as linear regressions~\cite{DBLP:journals/sqj/SiegmundRKKAS12,DBLP:conf/sigsoft/SiegmundGAK15,DBLP:journals/tc/SunSZZC20,DBLP:journals/is/KangKSGL20}, tree-like models~\cite{DBLP:conf/icdcs/HsuNFM18,DBLP:conf/kbse/SarkarGSAC15,DBLP:journals/corr/abs-1801-02175}, fourier-learning models~\cite{zhang2015performance,DBLP:conf/icsm/Ha019}, and even transfer learning~\cite{DBLP:conf/sigsoft/JamshidiVKS18,DBLP:journals/tse/KrishnaNJM21,DBLP:conf/kbse/JamshidiSVKPA17}, \textit{etc}. Among others, \texttt{SPLConqueror}~\cite{DBLP:journals/sqj/SiegmundRKKAS12} utilizes linear regression
combined with different sampling methods and a step-wise feature selection to capture the interactions between configuration options. \texttt{DECART}~\cite{DBLP:journals/ese/GuoYSASVCWY18} is an improved CART with an efficient sampling method~\cite{zhang2015performance}. Jamshidi and Casale~\cite{DBLP:conf/mascots/JamshidiC16} use Gaussian Process (GP) to model configuration performance, which is also gradually updated via Bayesian Optimization. However, recent work reveals that those approaches do not work well with small datasets~\cite{DBLP:conf/icse/HaZ19}, which is rather common for configurable software systems due to their expensive measurements. This is a consequence of not fully handling the sparsity in configuration data. Further, they come with various restrictions and prerequisites, e.g., \texttt{DECART} does not work on mixed systems while \texttt{SPLConqueror} needs an extensive selection of the right sampling method(s); GP has also been proved to struggle in modeling the sparse configuration data~\cite{DBLP:journals/corr/abs-1801-02175}.

In contrast, we showed that \Model~produces significantly more accurate results while not limited to those restrictions. In addition, the paradigm of dividable learning underpins \Model~prevent it from suffering the particular pitfalls of a single machine learning algorithm, as it can be paired with different local models.

\subsection{Deep Learning-based Models}

A variety of studies apply deep neural networks with multiple layers to predict configuration performance~\cite{DBLP:conf/icse/HaZ19, DBLP:conf/esem/ShuS0X20,DBLP:conf/sbac-pad/NemirovskyAMNUC17,app11083706,DBLP:journals/concurrency/FalchE17,DBLP:conf/iccad/KimMMSR17,DBLP:conf/sc/MaratheAJBTKYRG17,DBLP:conf/im/JohnssonMS19,DBLP:journals/jsa/ZhangLWWZH18}. Among others, \texttt{HINNPerf} uses embedding to encodes the configuration into a latent space after which a deep neural network, combined with the hierarchical regularization, is applied to handle feature sparsity. \texttt{DeepPerf}~\cite{DBLP:conf/icse/HaZ19} is a state-of-the-art DNN model with $L_{1}$ regularization to mitigate feature sparsity for any configurable systems, and it can be more accurate than many other existing approaches. The most recently proposed \texttt{Perf-AL}~\cite{DBLP:conf/esem/ShuS0X20} relied on adversarial learning that consists of a generative network to predict the performance and a discriminator network to distinguish the predictions and the actual labels. Nevertheless, existing deep learning approaches capture only the feature sparsity while ignoring the sample sparsity, causing severe risks of overfitting even with regularization in place. 

Compared with those, we have demonstrated that, by additionally capturing sample sparsity, \Model~is able to improve the accuracy considerably with better efficiency and acceptable overhead.

\subsection{Ensemble Models}

Similar to the paradigm of dividable learning, ensemble learning also leads to different frameworks that can be paired with different local models, some of which have already been adopted for configuration performance learning. For example, Chen and Bahsoon~\cite{DBLP:journals/tse/ChenB17} propose an ensemble approach, paired with feature selection for mitigating feature sparsity, to model software performance. Other ensemble learning such as Bagging~\cite{breiman1996bagging} and Boosting~\cite{schapire2003boosting} can also be similarly applied. However, those ensemble learning approaches allow different local models to share information at one or more phases of the learning pipeline:

\begin{itemize}
    \item Local models can be trained on the same data~\cite{DBLP:journals/tse/ChenB17}; or data that has been processed sequentially~\cite{schapire2003boosting}.
    \item Local models can make predictions collectively as opposed to individually~\cite{breiman1996bagging}.
\end{itemize}

While information sharing may be useful when different models learn data with high similarity; they might not be effective in coping with highly sparse data. \Model, in contrast, produces isolated local models without sharing knowledge at all phases, including training and prediction. This has been shown to be more suitable for dealing with the sample sparsity exhibited in configuration data, preventing overfitting and memorizing largely spread data samples.



\subsection{Hybrid Models}

The analytical models can be combined with data-driven ones to form a hybrid model~\cite{DBLP:conf/wosp/HanYP21,didona2015enhancing,DBLP:conf/icse/WeberAS21}. Among others, Didona \textit{et al.}~\cite{didona2015enhancing} use linear regression and $k$NN to learn certain components of a queuing network. Conversely, Weber \textit{et al.}~\cite{DBLP:conf/icse/WeberAS21} propose to learn the performance of systems based on the parsed source codes from the system to the function level. 

We see \Model~as being complementary to those hybrid models due to its flexibility in selecting the local model: when needed, the local models can be replaced with hybrid ones, making itself a hybrid variant. In case the internal structure of the system is unknown, \Model~can also work in its default as a purely data-driven approach.


\subsection{Data Splitting Models}
\textcolor{black}{From the machine learning community, there are also studies that have explored tree-based learning paradigms to split the training dataset. For example, \texttt{IBMB}~\cite{DBLP:conf/icml/Quinlan93} combines predictions from both instance-based learning and model-based learning techniques, i.e., model trees. Especially, during the construction of a model tree, the recursive algorithm partitions the data based on attribute tests until certain stopping criteria are met (e.g., reaching a maximum tree depth or a minimum number of instances per leaf). At each leaf node, a linear regression model is fitted using the samples that reach that leaf. \texttt{M5}~\cite{Quinlan1992LearningWC} splits the original dataset to minimize the variance within each subset, and subsequently trains a linear model for each leaf node. \texttt{PILOT}~\cite{raymaekers2023fast} divides samples in a similar manner to \texttt{CART} but without pruning. Additionally, it improves the efficiency of \texttt{CART} by utilizing linear models at the leaf nodes. Moreover, \texttt{MOB}~\cite{Zeileis2009partyWT} improves the dividing method of \texttt{CART} by fitting the dataset using a parametric model and performing recursive partitioning by splitting the samples based on the partitioning variable that exhibits the highest parameter instability.}

\textcolor{black}{However, these models are demonstrated to be ill-suited to some extent for configuration performance prediction, as they are not specifically designed for complex and non-linear regression tasks, especially given the fact that most of them utilize linear models at the leaf nodes. Moreover, one of the differences between \Model~and those tree-based data splitting methods is that we allow \Model~to pair with any local model depending on the preferences.}

\section{Conclusion}
\label{sec:conclusion}

This paper proposes a model-agnostic framework dubbed \Model~that effectively handles the issues of both feature and sample sparsity in configuration performance learning. The key novelty of \Model~is that it follows a new paradigm of dividable learning, in which the branches/leaves are extracted from a \texttt{CART} that divides the samples of configuration into a number of distant divisions, which is adaptively adjusted, and trains a dedicated local model for each division thereafter. Prediction of the new configuration is then made by the local model of division inferred based on a Random Forest classifier. Such a theory of ``divide-and-learn'' handles the sample sparsity while the local model used (e.g., one with regularization) deals with the feature sparsity, hence collectively addressing the overall sparsity issue in configuration data.

By means of extensive experiments, we comprehensively evaluate \Model~on 12 real-world systems that are of diverse domains and scales, together with five sets of training data. The results show that \Model~is:

\begin{itemize}
    \item \textbf{effective} as it is competitive to the best state-of-the-art approach on 44 out of 60 cases, in which 31 of them are significantly better with up to $1.61\times$ MRE improvement; the paradigm of dividable learning is also more suitable than the classic ensemble learning to handle the sample sparsity in configuration data;
    \item \textbf{efficient} since it often requires fewer samples to reach the same/better accuracy compared with the others; the adaptation of $d$ also leads to negligible overhead as no additional training is required;
    \item \textbf{flexible} given that it considerably improves various global models when they are used as the local model therein; 
    \item \textbf{robust} because the mechanism that adapted the parameter $d$ can reach the optimal value for 76.43\% of the individual runs while a similarly promising $d$ value can be chosen even when the optimal $d$ is missed.
    
\end{itemize}


Mitigating the issues caused by sparsity is only one step towards more advanced configuration performance learning, hence the possible future work based on \Model~is vast, including multi-task prediction of configuration performance under different environments and merging diverse local models as part of the dividable learning paradigm.

\section*{Acknowledgements}
This work was supported by an NSFC Grant (62372084) and a UKRI Grant (10054084).


\bibliographystyle{IEEEtranS}
\bibliography{IEEEabrv,references}

\begin{IEEEbiography}[{\includegraphics[width=1in,height=1.25in,clip,keepaspectratio]{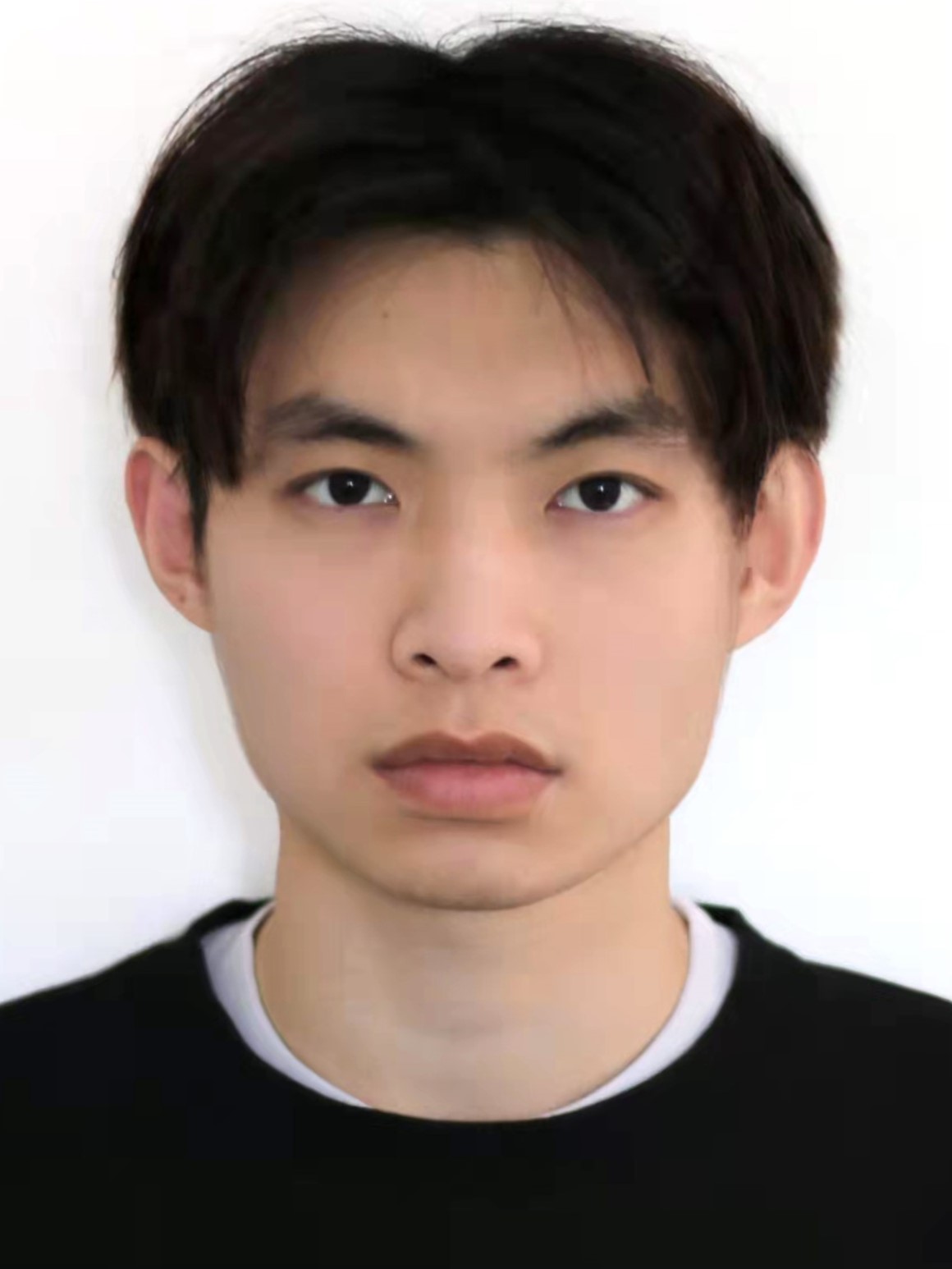}}]{Jingzhi Gong} received the Ph.D. degree from the Loughborough University, U.K., in 2024. As a member of the IDEAS Lab, his doctoral research has been published in major Software Engineering conferences/journals including FSE, TOSEM, and MSR. His research interests mainly include configuration performance modeling, machine/deep learning, and software engineering. He is currently a postdoctoral research fellow at the University of Leeds, U.K., and TurinTech AI, U.K., developing cutting-edge code optimization approaches using large language models and compiler-based techniques. 
\end{IEEEbiography}

\begin{IEEEbiography}[{\includegraphics[width=1in,height=1.25in,clip,keepaspectratio]{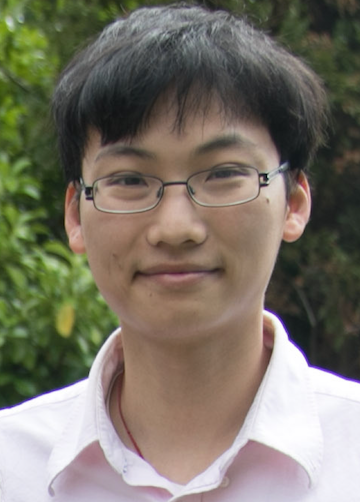}}]{Tao Chen} directs the Intelligent Dependability Engineering for Adaptive Software Laboratory (IDEAS Lab), conducting cutting-edge research on the intersection between AI and Software Engineering. Currently, his research interests include performance engineering, self-adaptive systems, search-based software engineering, data-driven software engineering, and their interplay with machine learning and computational intelligence. His work has been regularly published in all major Software Engineering conferences/journals (ICSE, FSE, ASE, TOSEM, and TSE) and has been supported by projects worth over $\pounds$1 million from external funding bodies. He currently serves as an AE for ACM Transactions on Autonomous and Adaptive Systems and a PC member for many prestigious conferences.
\end{IEEEbiography}

\begin{IEEEbiography}[{\includegraphics[width=1in,height=1.25in,clip,keepaspectratio]{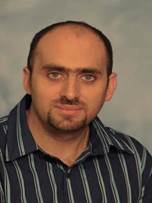}}]{Rami Bahsoon} is a Reader in Autonomous and at the School of Computer Science, University of Birmingham, UK.  He conducts research in the fundamentals of self-adaptive and managed software engineering and its application to emerging paradigms such as cloud, microservices, IoT, Blockchain, Digital Twins and CPS etc. He co-edited four books on Software Architecture. He is a fellow of the Royal Society of Arts, Associate Editor of IEEE Software, Area Editor (Software Engineering)  of Wiley Practice and Experience and EiC of ACM Transactions on Autonomous and Adaptive Systems. He holds a PhD in Software Engineering from University College London (2006).
\end{IEEEbiography}

\end{document}